\documentclass[11pt,leqno]{article}
\usepackage{amssymb,amsfonts}%amsLatex
\usepackage{amsmath,amsthm,amsxtra}
\usepackage[all]{xy}

\newcommand{\scriptpp}{
  \begin{picture}(6,4)(0.5,2)
    \put(0,3.5){$\scriptstyle +$}
    \put(0,0){$\scriptstyle +$}
  \end{picture}
}

\newcommand{\scriptmm}{
  \begin{picture}(6,4)(0.5,2)
    \put(0,3.5){$\scriptstyle -$}
    \put(0,0){$\scriptstyle -$}
  \end{picture}
}

\newcommand{\st}[1]{\ensuremath{^{\scriptstyle \textrm{#1}}}}

\newcommand\bigcheck[1]{#1 \raise1ex\hbox{$\hspace{-1ex}{}^\vee$}}
\newcommand\sucheck[1]{#1 \raise0.5ex\hbox{$\hspace{-1ex}{}^\vee$}}

\newcommand{\alphaparenlist}{% changes enumerate 1st level to (a)...(z)
  \renewcommand{\theenumi}{\alph{enumi}}%
  \renewcommand{\labelenumi}{(\theenumi)}%
}
\newcommand{\Mlist}{% changes enumerate 1st level to (M1). ... (M9).
  \renewcommand{\theenumi}{\arabic{enumi}}%
  \renewcommand{\labelenumi}{(M\theenumi)}%
}
\newcommand{\romanparenlist}{% changes enumerate 1st level to (i)...(ix)
  \renewcommand{\theenumi}{\roman{enumi}}%
  \renewcommand{\labelenumi}{(\theenumi)}%
}
\newcommand{\Romanlist}{% changes enumerate 1st level to I...X.
  \renewcommand{\theenumi}{\Roman{enumi}}%
  \renewcommand{\labelenumi}{\theenumi.}%
}
\newcommand{\ad}{\mathop{\rm ad}\,}
\newcommand{\adv}{{\rm adv}}

\newcommand{\ch}{{\rm ch}}

\newcommand{\even}{\mathop{\rm even \, }}
\newcommand{\End}{\mathop{\rm End }}
\newcommand{\Gr}{{\rm Gr}}
\renewcommand{\Im}{\mathop{\rm Im  \, }}
\newcommand{\Ker}{\mathop{\rm Ker \, }}

\renewcommand{\ne}{\mathop{\rm ne}\,}
\newcommand{\odd}{{\rm odd}}
\renewcommand{\Re}{\mathop{\rm Re  \, }}
\newcommand{\Res}{\mathop{\rm Res  \, }}
\renewcommand{\sl}{s\ell}
\newcommand{\str}{{\rm str}}

\newcommand{\sdim}{\mathop{\rm sdim \, }}

\newcommand{\tr}{\rm tr \, }
\newcommand{\vac}{|0\rangle}

\newcommand{\A}{\mathcal{A}}
\newcommand{\C}{\mathcal{C}}

\newcommand{\CC}{\mathbb{C}}
\newcommand{\NN}{\mathbb{N}}

\newcommand{\RR}{\mathbb{R}}
\newcommand{\ZZ}{\mathbb{Z}}

\newcommand{\fc}{\mathfrak{c}}
\newcommand{\fg}{\mathfrak{g}}
\newcommand{\fh}{\mathfrak{h}}

\newcommand{\fn}{\mathfrak{n}}

\newcommand{\fp}{\mathfrak{p}}

\renewcommand{\hat}{\widehat}
\makeatletter
\renewcommand\section{\@startsection {section}{1}{\z@}%
                                   {-3.5ex \@plus -1ex \@minus -.2ex}%
                                   {2.3ex \@plus.2ex}%
                                   {\normalfont\large\bfseries}}
\renewcommand\subsection{\@startsection{subsection}{2}{\z@}%
                                     {-3.25ex\@plus -1ex \@minus -.2ex}%
                                     {0ex \@plus .0ex}%
                                     {\normalfont\normalsize\bfseries}}

\@addtoreset{equation}{section}
\makeatother

\newtheorem{theorem}{Theorem}[section]

\newtheorem{lemma}{Lemma}[section]
\newtheorem{corollary}{Corollary}[section]

\newtheorem*{lemma*}{Lemma}

\theoremstyle{remark}
\newtheorem{remark}{Remark}[section]
\newtheorem{example}{Example}[section]

\makeatletter
\def\@maketitle{\newpage
 \null
 \vskip 2em
 \begin{center}%
  \vskip 3em
  {\Large\bf \@title \par}%
  \vskip 1.5em
  {\normalsize
   \lineskip .5em
   \begin{tabular}[t]{c}\@author
   \end{tabular}\par}%
  \vskip 2em

 \end{center}%
 \par
 \vskip 2.5em}
\makeatother
\begin{document}

\title{Quantum Reduction and Representation Theory of
  Superconformal Algebras}

\author{Victor G. Kac\thanks{Department of Mathematics, M.I.T.,
    Cambridge, MA 02139, USA.~~kac@math.mit.edu}~~\thanks{Supported in part by NSF grants
    DMS-9970007 and DMS-0201017.}~~and Minoru Wakimoto\thanks{Graduate School
    of Mathematics, Kyushu University, Fukuoka 812-8581,
    Japan~~wakimoto@math.kyushu-u.ac.jp}~~\thanks{Supported by Grant-in-aid 13440012 for
scientific research Japan.}\\[4ex] \text{To Ivan Todorov on his 70th birthday}}

\maketitle

\begin{abstract}
  We study the structure and representations of a family of
  vertex algebras obtained from affine superalgebras by quantum
  reduction.  As an application, we obtain in a unified way free field
  realizations and determinant formulas for all superconformal
  algebras.

\end{abstract}

\section{Introduction}
\label{sec:intro}

In this paper we study the structure and representations of the
vertex algebras $W_k(\fg ,x,f)$ introduced in \cite{KRW}.
Let us briefly recall the construction of these vertex algebras.
The datum we
begin with is a quadruple $(\fg ,x,f,k)$, where  $\fg$ is a
simple finite-dimensional Lie superalgebra with a non-zero even
invariant supersymmetric bilinear form $(. \, | \, .)$, $x$ is an
$\ad$-diagonalizable element of $\fg$ with eigenvalues in
$\tfrac{1}{2}\ZZ$, $f$ is an even element of $\fg$ such that
$[x,f]=-f$ and the eigenvalues of $\ad x$ on the centralizer
$\fg^f$ of $f$ in $\fg$ are non-positive, and $k \in \CC$.
Recall that a bilinear form  $(. \, | \, .)$ on $\fg$ is called
even if $(\fg_{\bar{0}}| \fg_{\bar{1}})=0$, supersymmetric if
$(. \, | \, .)$ is symmetric (resp. skewsymmetric) on
$\fg_{\bar{0}}$ (resp. $\fg_{\bar{1}}$), invariant if
$([a,b]|c)=(a|[b,c])$ for all $a,b,c \in \fg$.  Note also that a
pair $(x,f)$ satisfying the above properties can be obtained by
taking a non-zero  nilpotent element $f \in \fg_{\bar{0}}$ and
including it in an $\sl_2$-triple $\{ e,x,f\}$, so that
$[x,e]=e$, $[x,f]=-f$, $[e,f]=x$ (then, up to conjugacy, $x$ is
determined by $f$).

We associate to the quadruple $(\fg ,x,f,k)$ a homology complex
\begin{equation}
  \label{eq:0.1}
  \C (\fg ,x,f,k)= (V_k (\fg)\otimes F^{\ch}\otimes F^{\ne}
     \, , \, d_0) \, ,
\end{equation}
where $V_k (\fg)$ is the universal affine vertex algebra of level
$k$ associated to $\fg$, $F^{\ch}$ is the vertex algebra of free
charged superfermions based on $\fg_+ +\fg^*_+$ with reversed
parity, $F^{\ne}$ is the vertex algebra of free neutral
superfermions based on $\fg_{1/2}$ (see \cite{K4} for definitions),
and $d_0$ is an explicitly
constructed odd derivation of the vertex algebra $\C (\fg
,x,f,k)$ whose square is $0$.  Here $\fg_+$ (resp. $\fg_{1/2}$)
denote the sum of eigenspaces with positive eigenvalues
(resp. eigenspace with eigenvalue $1/2$) of $\ad \, x$ (see
Section~\ref{sec:1}).  The vertex algebra $W_k (\fg ,x,f)$ is the
homology of the complex (\ref{eq:0.1}).  In the case when the
pair $(x,f)$ can be included in an $\sl_2$-triple, we denote this
vertex algebra by $W_k (\fg ,f)$.

Our main result on the structure of the vertex algebras $W_k(\fg,x,f)$
is Theorem~\ref{th:4.1} which states that the $j$\st{th} homology
of the complex (\ref{eq:0.1}) is zero if $j \neq 0$ and the
$0$\st{th} homology vertex algebra (which is $W_k (\fg ,x,f)$) is
strongly generated by fields $J^{\{ a_i \}}$, where $a_1 , \ldots
  ,a_s$ is a basis of $\fg^f$ consisting of eigenvectors of $\ad
  \, x$.  This theorem also gives a rather explicit form of these
  fields.  Furthermore, provided that $k \neq -h\spcheck$, we
  have an explicitly constructed energy-momentum field $L(z)$
  (see (\ref{eq:2.2})), with respect to which the fields
  $J^{\{a_i\}}$ have conformal  weight $1-m_i$, where
  $[x,a_i]=m_ia_i$.  Theorem~\ref{th:2.1} of the present paper
  gives explicit formulas for the fields $J^{\{ a_i \}}$ in the
  cases when $m_i=0$ and $-1/2$, and the commutation relations
  between these fields.

We study in more detail the simplest, but a very interesting
subclass of vertex algebras in question, namely $W_k (\fg
,e_{-\theta})$ that correspond to $f=e_{-\theta}$, the lowest root
vector of $\fg$ (for some choice of positive roots),
see Theorem~\ref{th:5.1}.  These vertex algebras are
characterized among all vertex algebras $W_k (\fg ,x,f)$ by the
property that they are strongly generated by the energy-momentum
field and the fields of conformal weight $1$ and $3/2$.
Moreover, it turns out that, under some natural assumptions, any
vertex algebra satisfying this property is one of the vertex
algebras $W_k (\fg ,e_{-\theta})$\cite{FL}, \cite{D}.  Thus, all well
known superconformal algebras are covered by this construction:

\begin{list}{}{}
\item $W_k (s\ell_2,e_{-\theta})$ is the Virasoro vertex
  algebra,

\item $W_k (s\ell_3 , e_{-\theta})$ is the
  Bershadsky--Polyakov algebra \cite{B},

\item  $W_k (spo (2|1), e_{-\theta})$ is the Neveu--Schwarz
  algebra,

\item  $W_k (spo (2|m), e_{-\theta})$ for $m \geq 3$ are the
  Bershadsky--Knizhnik algebras \cite{BeK},

\item $W_k (s\ell (2|1) =spo (2|2), e_{-\theta})$ is the $N=2$
  superconformal algebra,

\item $W_k (s\ell (2|2)/\CC I, e_{-\theta})$ is the $N=4$
  superconformal algebra,

\item  $W_k (spo (2|3), e_{-\theta})$ tensored with one fermion
  is the $N=3$ superconformal algebra (cf. \cite{GS}),

\item $W_k (D(2,1;a), e_{-\theta})$ tensored with four
  fermions and one boson is the big $N=4$ superconformal algebra (cf.~\cite{GS}).

\end{list}

Theorem~\ref{th:5.1} also provides a construction of the vertex
algebra $W_k (\fg ,x,f)$ as a subalgebra of $V_{\alpha_k}
(\fg_{\leq}) \otimes F^{\ne}$, where $\fg_{\leq}$ is the sum of
eigenspaces of $\ad \, x$ with non-positive eigenvalues and
$\alpha_k$ is a ``shifted'' $2$-cocycle on
$\fg_{\leq}[t,t^{-1}]$.  Therefore the homomorphism $\fg_{\leq}
\to \fg_0$, where $\fg_0$ is the centralizer of $x$, induces a
realization of $W_k(\fg ,x,f)$ inside $V_{\alpha_k} (\fg_0)
\otimes F^{\ne}$.  In particular, this construction gives an
explicit free field realization of all vertex algebras $W_k (\fg
,e_{-\theta})$ (Theorem~\ref{th:5.2}), recovering thereby all
previously known free field realizations of superconformal
algebras (see \cite{K4} and references there).

Given a highest weight module $P$ of level $k$ over the affine
Lie superalgebra $\hat{\fg}$ associated to $\fg$, we have the
associated $\C (\fg ,x,f,k)$-module $\C (P)=P\otimes F^{\ch}
\otimes F^{\ne}$ with the differential $d^P_0$.  The homology
$H(P)$ of the complex $(\C (P) , \, d^P_0)$ is a $W_k (\fg
,x,f)$-module.
We show that if $P$ is a generalized Verma module over
$\hat{\fg}$, then $H_j (P) =0$ for $j \neq 0$
(Theorem~\ref{th:6.2}), and that $H_0(P)$ is a Verma module over
$W_k (\fg ,x,f)$ if $P$ is a Verma module over $\hat{\fg}$
(Theorem~\ref{th:6.3}).  Since the singular weights of a Verma
module over $\hat{\fg}$ are known \cite{K2}, this allows us to
construct sufficiently many singular weights for a Verma module
over $W_k (\fg ,x,f)$.
As a result, we get an explicit determinant formula for all
vertex algebras $W_k (\fg ,e_{-\theta})$ (Theorem~\ref{th:7.2}),
which includes as special cases all previously known determinant
formulas for superconformal algebras (in the ``untwisted' sector;
the twisted sector will be treated in \cite{KW5}).

The special cases of free field realization and determinant
formulas for the Virasoro, $N=1, 2, 3, 4$ and big $N=4$
superconformal algebras are considered in Section~\ref{sec:8}.
(Due to \cite{FK} these are all finite simple formal distribution
Lie superalgebras which admit a central extension containing a Virasoro
subalgebra with a non-trivial center.)

In this paper we use some very simple homological arguments.  The
necessary general facts about homology are collected in
Section~\ref{sec:3}.

Our paper represents further development of the ideas of Feigin
and Frenkel \cite{FF1}, \cite{FF2}, \cite{FB} who considered the very
important case of the principal nilpotent element $f$, and of the
paper \cite{BT}, which treated the case when all eigenvalues of
$\ad \, x$ are integers. In these works the homology of the
corresponding BRST complex is studied and many results of the
present paper are proved in the respective special cases, using
similar methods. After the present paper was send to the archive,
we learned about the paper \cite{ST} where a BRST complex similar
to ours is studied along similar lines, and the paper \cite{BG}
where free field realisations similar to ours are established.

In a forthcoming paper \cite{KW6} we shall develop the theory of characters
of the vertex algebras $W_k (\fg,e_{-\theta})$, building on \cite{KW1}
-\cite{KW4} and \cite{KRW}.

Throughout the paper all vector spaces, algebras and tensor products
are considered over the field of complex numbers $\CC$, unless otherwise
stated. We denote by $\ZZ$, $\ZZ_+$ and $\NN$ the sets of all integers,
all non-negative integers and all positive integers, respectively.

We wish to thank E.~Frenkel for many insights, in
particular, for providing a proof of Lemma~\ref{lem:3.2}.
We also wish to thank D.~Vogan for a suggestion concerning anti-involutions.

\section{The complex $\C (\fg , x, f,k)$ and the associated
  vertex algebra $W_k (\fg ,x,f)$}
\label{sec:1}

We recall here, following \cite{KRW} (see also \cite{K5}), the construction of a
vertex algebra $W_k (\fg ,x,f)$ depending on a complex parameter
$k$, via a differential complex $\C (\fg ,x,f,k)$, associated to
a simple finite-dimensional Lie superalgebra $\fg$ with a
non-degenerate even supersymmetric invariant bilinear form $(. \, |
\, .)$, and a pair $x,f$ of even elements of $\fg$ such that $\ad
x$ is diagonalizable on $\fg$ with half-integer eigenvalues and
$[x,f] =-f$.

We have the eigenspace decomposition of $\fg$ with respect to
$\ad x$:
\begin{equation}
  \label{eq:1.1}
  \fg = \oplus_{j \in\tfrac{1}{2} \ZZ} \fg_j \, ,
\end{equation}
and we let:
\begin{equation}
  \label{eq:1.2}
  \fg_+ =\oplus_{j>0} \fg_j \, , \quad \fg_- = \oplus_{j<0} \fg_j
  \, , \quad \fg_{\leq} = \fg_0 \oplus \fg_- \, .
\end{equation}
The element $f$ defines a skew-supersymmetric even bilinear form
$\langle . \, | \, . \rangle_{\ne}$ on $\fg_{1/2}$ by the
formula:
\begin{equation}
  \label{eq:1.3}
  \langle a , b \rangle_{\ne} = (f | [a,b]) \, .
\end{equation}
This bilinear form is non-degenerate if and only if the map
\begin{equation}
  \label{eq:1.4}
  \ad f : \fg_{1/2} \to \fg_{-1/2}
   \hbox{ is an isomorphism,}
\end{equation}
which we shall assume.  Denote by $\fg^{\natural}$ the centralizer
of $f$ in $\fg_0$.  By (\ref{eq:1.4}) the $\fg^{\natural}$-modules
$\fg_{1/2}$ and $\fg_{-1/2}$ are isomorphic.  The bilinear form $\langle . \, , \,
. \rangle_{\ne}$ is invariant with respect to the representation of
$\fg^{\natural}$ on
$\fg_{1/2}$:
\begin{equation}
  \label{eq:1.5}
  \langle [v,a],b \rangle_{\ne} + (-1)^{p(v)p(a)}
  \langle a,[v,b] \rangle_{\ne} =0 \quad
  (v \in \fg^{\natural}, a,b \in \fg_{1/2}) \, .
\end{equation}

Denote by $A_{\ne}$ the vector superspace $\fg_{1/2}$
with the bilinear form (\ref{eq:1.3}).  Denote by $A$
(resp. $A^*$) the vector superspace $\fg_+$ (resp. $\fg^*_+$)
with the reversed parity, let $A_{\ch} =A \oplus A^*$ and define
an even skew-supersymmetric non-degenerate bilinear form $\langle . \,
, \, . \rangle_{\ch}$ on $A_{\ch}$ by
\begin{eqnarray}
  \label{eq:1.6}
  \langle A,A \rangle_{\ch} &=& 0 = \langle A^* ,A^* \rangle_{\ch}
   \, , \, \\
\nonumber
\langle a,b^* \rangle_{\ch} &=& -(-1)^{p(a)p(b^*)}
   \langle b^* ,a \rangle_{\ch} =b^* (a) \hbox{ for }
   a \in A , b^* \in A^* \, .
\end{eqnarray}
Here and further, $p(a)$ stands for the parity of an
(homogeneous) element of a vector superspace.

Following \cite{KRW}, introduce the differential complex $(\C (\fg
,x,f,k),d_0)$, where $\C (\fg ,x,f,k)$  is a vertex algebra depending on a
complex parameter $k$ and $d_0$ is an odd derivation of all
products of this vertex algebra, such that $d^2_0 =0$.  We have:
\begin{equation}
  \label{eq:1.7}
  \C (\fg , x,f ,k)=V_k (\fg) \otimes F (\fg , x,f) \, ,
\end{equation}
where $V_k (\fg)$ is the \emph{universal affine vertex algebra of level}
$k$ associated to $\fg$ and
\begin{equation}
  \label{eq:1.8}
  F(\fg , x,f) =F (A_{\ch}) \otimes F (A_{\ne}) \, ,
\end{equation}
where $F(A_{\ch})$ (resp. $F(A_{\ne})$) is the \emph{vertex algebra of
charged} (resp. \emph{neutral}) \emph{free superfermions} based on $A_{\ch}$
(resp. $A_{\ne}$).

Recall that the vertex algebra $V_k (\fg)$ is constructed using
the \emph{Kac--Moody affinization} of $\fg$ (cf. \cite{K3}), which is the Lie
superalgebra $\hat{\fg}=\fg [t,t^{-1}] \oplus \CC K \oplus \CC D$
with the commutation relations $(a, b \in \fg\, , \, m,n \in \ZZ)$:
\begin{displaymath}
  [at^m , bt^n] = [a,b] t^{m+n} + m \delta_{m,-n}
  (a|b)K \, , \, [D, at^m] = m at^m \, , \,
  [K, \hat{\fg}]=0 \, .
\end{displaymath}
Let $\hat{\fg}' = \fg [t,t^{-1}] \oplus \CC K$ be its derived
algebra.  Then $V_k (\fg) = U (\hat{\fg}') / U (\hat{\fg}') (\fg
[t] \oplus \CC (K-k))$, as a left $\hat{\fg}'$-module.  Here and
further, $U(\fp)$ stands for the universal enveloping algebra of
the Lie superalgebra $\fp$.  Introduce the \emph{current} attached
to $a \in \fg$:  $a(z) = \sum_{n\in \ZZ} (at^n)z^{-n-1}$.  Recall
that the $\lambda$-bracket of the currents is $[a_{\lambda}b] = [a,b]+\lambda
(a|b)k$, $a,b \in \fg$.  Then, by the existence theorem
\cite{K4}, the space $V_k (\fg)$ carries a unique vertex algebra
structure such that the vacuum vector $\vac =1$, the
infinitesimal translation operator $T=-\tfrac{d}{dt}$, and $Y
(at^{-1} \vac ,z) =a(z)$, $a \in \fg$.

Given a vector superspace $A$ with an even skew-supersymmetric
non-degenerate bilinear form $\langle . \, , \,  . \rangle$, the associated vertex
algebra $F(A)$ of free superfermions is defined by making use of
the \emph{Clifford affinization} of $A$, which is the Lie superalgebra
$\hat{A} =A [t,t^{-1}]+\CC K$ with the commutation relations
$(a,b \in A \, , \,\, m,n \in \ZZ)$:
\begin{displaymath}
  [at^m,bt^n]= \delta_{m,-n-1} \langle a,b\rangle K \, , \,
  [K,\hat{A}]=0 \, .
\end{displaymath}
Then $F(A) =U(\hat{A})/ U(\hat{A}) (A[t] \oplus \CC (K-1))$, as a
left $\hat{A}$-module.  The free superfermion attached to
$\varphi \in A$ is $\varphi (z) = \sum_{n \in \ZZ} (\varphi t^n)
z^{-n-1}$, and the $\lambda$-bracket is:  $[\varphi_{\lambda}
\psi] = \langle \varphi , \psi \rangle , \varphi, \psi \in A$.
Likewise, $F(A)$ carries a unique vertex algebra structure with
$\vac =1$, $T=-\tfrac{d}{dt}$ and $Y(\varphi t^{-1} \vac
,z)=\varphi (z) $, $\varphi \in A$.

The vertex algebra $F(A_{\ch})$ has the charge decomposition:
\begin{displaymath}
  F(A_{\ch}) = \oplus_{m \in \ZZ} F_m (A_{\ch}) \, , \,
  \hbox{ where charge } \varphi (z) =- \hbox{ charge }
  \varphi^* (z)=1 \hbox{ for }\varphi \in A \, , \,
  \varphi^* \in A^* \, .
\end{displaymath}
Letting charge $V_k (\fg) =0$, charge $F(A_{\ne})=0$, this
induces the charge decompositions:
\begin{equation}
  \label{eq:1.9}
  F (\fg ,x,f) =\oplus_{m \in \ZZ} F_m \, , \quad
  \C(\fg ,x,f ,k) =\oplus_{m \in \ZZ} \C_m \, .
\end{equation}

In order to define the differential $d_0$, choose a basis $\{
u_{\alpha} \}_{\alpha \in S_j}$ of each $\fg_j$ in
(\ref{eq:1.1}), and let $S=\coprod_{j \in \tfrac{1}{2} \ZZ} S_j$,
$S_+ = \coprod_{j>0} S_j$.  Let $p(\alpha) \in \ZZ / 2\ZZ$ denote the parity of
$u_{\alpha}$, and let $m_{\alpha} =j$
if $\alpha \in S_j$.  Define the structure constants
$c^{\gamma}_{\alpha \beta}$ by $[u_{\alpha}, u_{\beta}] =
\sum_{\gamma} c^{\gamma}_{\alpha \beta} u_{\gamma}$ $(\alpha,
\beta, \gamma \in S)$.  Denote by $\{ \varphi_{\alpha}\}_{\alpha
  \in S_+}$ the corresponding basis of $A$ and by $\{
\varphi^{\alpha}\}_{\alpha \in S_+}$ the basis of $A^*$ such that
$\langle \varphi_{\alpha}, \varphi^{\beta} \rangle_{\ch}
=\delta_{\alpha \beta}$.  Denote by $\{ \Phi_{\alpha}\}_{\alpha
  \in S_+}$ the corresponding basis of $A_{\ne}$, and by $\{
\Phi^{\alpha} \}_{\alpha \in S_{1/2}}$ the dual basis
with respect to $\langle . \, , \, . \rangle_{\ne}$, i.e.,~$\langle
\Phi_{\alpha} , \Phi^{\beta} \rangle_{\ne} = \delta_{\alpha \beta}$.
  It will also be convenient to define $\Phi_u$ for any $u \in
  \sum_{\alpha \in S} c_{\alpha}u_{\alpha} \in \fg$ by letting
  $\Phi_u =\sum_{\alpha \in S_{1/2}} c_{\alpha}\Phi_{\alpha}$.
Following \cite{KRW}, introduce the following odd field of the vertex
algebra $\C (\fg ,x,f,k)$:
\begin{eqnarray*}
  d(z) &=& \sum_{\alpha \in S_+} (-1)^{p(\alpha)}
       u_{\alpha} (z) \otimes \varphi^{\alpha} (z) \otimes 1
- \frac{1}{2} \sum_{\alpha ,\beta ,\gamma \in S_+}
    (-1)^{p(\alpha) p(\gamma)} c^{\gamma}_{\alpha \beta}
    \otimes \varphi_{\gamma} (z) \varphi^{\alpha} (z)
    \varphi^{\beta} (z) \otimes 1 \\
&& + \sum_{\alpha \in S_+} (f| u_{\alpha}) \otimes \varphi^{\alpha}
    (z) \otimes 1 + \sum_{\alpha \in S_{1/2}}
    1 \otimes \varphi^{\alpha} (z) \otimes \Phi_{\alpha}(z) \, .
\end{eqnarray*}
Then $d_0 := \Res_z \, d(z)$ is an odd derivation of all products
of the vertex algebra $\C (\fg ,x,f,k)$ and $d^2_0 =0$, since by
\cite{KRW}, Theorem~2.1, $[d(z), d(w)] =0$.  Also, $d_0
(\C_m) \subset \C_{m-1}$.  Thus, $(\C (\fg ,x,f,k), d_0)$ is a
$\ZZ$-graded homology complex.  The homology of this complex is a vertex
algebra, denoted by $W_k (\fg ,x,f)$, and called the
\emph{quantum reduction} for the triple $(\fg ,x,f)$.

We shall often drop the tensor product sign $\otimes$, and also
shall often drop $z$ (for example we write $d$ in place of
$d(z)$, $\partial a$ in place of $\frac{d}{dz} a(z)$, etc.) if no
confusion may arise.

The most interesting pair $x,f$ satisfying (\ref{eq:1.4}) comes
from an $s\ell_2$-triple $\{ e,x ,f \}$, where $[x,e] =e$, $[x,f]
=-f$, $[e,f]=x$.  In this case we have a stronger property:
$\fg^f := \{ a \in \fg | [f,a] =0 \} \subset \fg_{\leq}$,
 or, equivalently,
\begin{equation}
  \label{eq:1.10}
\fg^f = \oplus_{j \leq 0}
\fg^f_j \, , \hbox{ where } \fg^f_j = \{ a \in \fg_j| [f,a]=0 \} \, .
\end{equation}
This property is immediate by the
$s\ell_2$-representation theory.  Since a nilpotent even element
$f$ determines uniquely (up to conjugation) the element $x$ of an
$s\ell_2$-triple (by a theorem by Dynkin), we use in this case
the notation $C (\fg ,f,k)$ for the complex and $W_k (\fg ,f)$
for the quantum reduction.  In fact, up to isomorphism, the
vertex algebra $W_k (\fg ,f)$ depends obviously only on the adjoint orbit
of $f$ in the even part of $\fg$.  Note also that in this case the subalgebra $\fg^f_0$ coincides
with the centralizer of $\{ e,x,f \}$ in $\fg$.

Property (\ref{eq:1.10}) of the pair $(x,f)$ will play an
important role in the sequel.  Note that this property is
equivalent to
\begin{equation}
  \label{eq:1.11}
  [f,\fg_j] = \fg_{j-1} \hbox{  if  } j \leq \tfrac{1}{2} \, .
\end{equation}
Indeed, $[f,\fg_j] \neq \fg_{j-1}$ for $j \leq \tfrac{1}{2}$ is
equivalent to existence of a non-zero $a \in \fg_{-j+1}$ such that
$([f,\fg_j] |a)=0$, i.e.,~$([f,a] | \fg_j)=0$, which is
equivalent to $a \in \fg^f_{-j+1}$.

The pair $(x,f)$ and the
corresponding $\tfrac{1}{2} \ZZ$-gradation (\ref{eq:1.1}) are
called \emph{good} if one of the equivalent properties
(\ref{eq:1.10}) or (\ref{eq:1.11}) holds.  In this case the
$\fg^{\natural}$-modules $\fg_{-1/2}$ and $\fg_{1/2}$ are
isomorphic.  Furthermore, we have
the exact sequence of $\fg^{\natural}$-modules:
\begin{displaymath}
  0 \to \fg^f \to \fg_- + \fg_0 + \fg_{1/2} \overset{\ad f}{\longrightarrow}
  \fg_- \to 0 \, .
\end{displaymath}
It follows that $\ad f$ induces the following isomorphism of
$\fg^{\natural}$-modules:
\begin{equation}
  \label{eq:1.12}
\fg^f\overset{ \sim}{\longrightarrow} \fg_0 \oplus \fg_{1/2} \, .
\end{equation}

One can find a detailed study and a classification of good gradations of simple Lie algebras
in \cite{EK}.

\begin{remark}
  \label{rem:1.1}

Property (\ref{eq:1.11}) for $j=0$ gives:  $[\fg_0,f]=\fg_{-1}$.
In particular, it follows that $f$ lies in the dense open orbit
in $\fg_{-1}$ of the algebraic group $G_{0\, \bar{0}}$ whose Lie
algebra is the even part of $\fg_0$.  Thus, up to conjugacy,  $f$
is determined by $x$.  Hence, instead of a good pair $(x,f)$ one
may talk about a good semi-simple element $x$, defined by the
properties that the eigenvalues of $\ad \, x$ lie in
$\tfrac{1}{2}\ZZ$ and the centralizer of any element $f$ from the
open orbit of $G_{0\, \bar{0}}$ in $\fg_{-1}$ lies in $\fg_{\leq}$.
\end{remark}

\section{Fields of the vertex algebra $W_k (\fg ,x,f)$}
\label{sec:2}

The dual Coxeter number $h\spcheck$, defined as the half of the
eigenvalue of the Casimir operator on $\fg$, is given by
\begin{equation}
  \label{eq:2.1}
  h\spcheck =(\rho | \theta) + \tfrac{1}{2}(\theta | \theta) \, ,
\end{equation}
where $\rho$ is the half of the difference of the sum of positive
even roots and the sum of positive odd roots of $\fg$, and
$\theta$ is the highest root (for any choice of a set of
positive roots).

Provided that $k \neq -h\spcheck$, define the energy--momentum
(or Virasoro)
field
\begin{equation}
  \label{eq:2.2}
  L(z) = L^{\fg} (z) + \tfrac{d}{dz} x(z) + L^{\ch}(z)
         +  L^{\ne} (z) \, .
\end{equation}
Here
\begin{displaymath}
L^{\fg} = \tfrac{1}{2(k+h\spcheck)} \sum_{\alpha \in S}
(-1)^{p(\alpha)} : u_{\alpha} u^{\alpha}:
\end{displaymath}
is the Sugawara
construction, where $\{ u^{\alpha}\}_{\alpha \in
  S}$ is the dual basis, i.e.,~$(u_{\alpha}|u^{\beta})
=\delta_{\alpha \beta}$,
\begin{displaymath}
  L^{\ch} = - \sum_{\alpha \in S_+} m_{\alpha} : \varphi^{\alpha}
    \partial \varphi_{\alpha} : + \sum_{\alpha \in S_+}(1-m_{\alpha})
    : (\partial \varphi^{\alpha}) \varphi_{\alpha }: \, \, ,\, \,
  L^{ne} = \tfrac{1}{2} \sum_{\alpha \in S_{1/2}} :
    (\partial \Phi^{\alpha}) \Phi_{\alpha}: \, .
\end{displaymath}
The central charge of $L(z)$ equals \cite{KRW}:
\begin{equation}
  \label{eq:2.3}
  c(\fg ,x,f,k) =\tfrac{k\sdim \fg}{k+h\spcheck}
  -12 k (x|x)
- \sum_{\alpha \in S_+} (-1)^{p(\alpha)}
   (12 m^2_{\alpha}-12m_{\alpha}+2)-1/2
   \sdim \fg_{1/2} \, .
\end{equation}
(Here and further $\sdim$ stands for the super-dimension of a
super vector space.)

With respect to $L(z)$ the fields $\varphi_{\alpha}$
(resp.~$\varphi^{\alpha}$) are primary of conformal weight
$1-m_{\alpha}$ (resp.~$m_{\alpha}$), the fields $\Phi_{\alpha}$
are primary of conformal weight $\tfrac{1}{2}$, and the fields
$a(z)$ for $a \in \fg_j$ have conformal weight $1-j$ and are
primary unless $j=0$ and $(x|a) \neq 0$.  Furthermore, it was
shown in \cite{KRW} that the field $d(z)$ is primary of conformal
weight $1$, hence $[d_{\lambda}L]=\lambda d$ and $d_0 (L) =
[d_{\lambda}L]|_{\lambda =0} =0$.  Thus, the homology class of
$L$ defines the energy--momentum field of $W_k (\fg ,x,f)$,
which we again denote by $L$.

In order to construct some other fields of the vertex algebra
$W_k (\fg ,x,f)$, for $v \in \fg$ denote by $c_{\alpha \beta}
(v)$ the matrix of $\ad v$ in the basis $\{ u_{\alpha}\}_{\alpha
  \in S}$, i.e.,~$[v,u_{\beta}]=\sum_{\alpha \in S}
c^{\alpha}_{\beta}(v) u_{\alpha}$.  Given $v \in \fg_j$,
introduce the following field of conformal weight $1-j$:
\begin{equation}
  \label{eq:2.4}
  J^{(v)} (z)= v(z) + \sum_{\alpha ,\beta \in S_+}
  (-1)^{p(\alpha)} c^{\alpha}_{\beta} (v): \varphi_{\alpha}
  (z) \varphi^{\beta} (z) : \, .
\end{equation}

The fields $J^{(v)}$ will be the main building blocks for the
vertex algebra $W_k (\fg ,x,f)$.  They obey the following
$\lambda$-brackets \cite{KRW}:
\begin{equation}
  \label{eq:2.5}
  [{J^{(v)}}{}_{\lambda} J^{(v')}] = J^{([v,v'])} +
   \lambda (k (v|v') + \tfrac{1}{2}
  (\kappa_{\fg} (v,v') - \kappa_{\fg_0} (v,v')))
\end{equation}
if $v \in \fg_i$, $v' \in \fg_j$ and $ij\geq 0$, where $\kappa_{\fg}$
(resp. $\kappa_{\fg_0}$) denotes the Killing form on $\fg$
(resp.~$\fg_0$).

The following important formula is established via the
$\lambda$-bracket calculus, using formulas (2.4) from \cite{KRW}:
\begin{eqnarray}
    \label{eq:2.6}
    d_0 (J^{(v)}) = \sum_{\beta \in S_+} ([f,v]|u_{\beta})
    \varphi^{\beta}
    + \sum_{ \beta \in S_+}
    (-1)^{p(v)(p(\beta)+1)}
    :\varphi^{\beta} \Phi_{[v,u_{\beta}]}:\\
\nonumber
    -\sum_{\substack{\beta \in S_+:\\ [v,u_{\beta}] \in \fg_{\leq}}}
    (-1)^{p(\beta) (p(v)+1)} :
    \varphi^{\beta} J^{([v,u_{\beta}])} :
    + \sum_{\beta \in S_+} (k(v|u_{\beta})
    +\str_{\fg_+} p_+ (\ad v)  \,  (\ad u_{\beta}))
    \partial \varphi^{\beta} \, ,
\end{eqnarray}
where $p_+$ is the projection of $\fg$ on $\fg_+$.

Furthermore, introduce the following fields of conformal weight
$1$ and $\tfrac{3}{2}$ for $v \in \fg_0$ and $v \in
\fg_{-1/2}$, respectively:
\begin{eqnarray}
  \label{eq:2.7}
  v^{\ne} &=&
  \tfrac{(-1)^{p(v)}}{2}
        \sum_{\alpha \in S_{1/2}}
        : \Phi^{\alpha}\Phi_{[u_{\alpha},v]}: (v \in \fg_0)\\
  \label{eq:2.8}
  v^{\ne}
      &=& -\tfrac{(-1)^{p(v)}}{3}
         \sum_{\alpha ,\beta  \in S_{1/2}} :
         \Phi^{\alpha} \Phi^{\beta}\Phi_{[u_{\beta}, [u_{\alpha},v]]}:
         (v \in \fg_{-1/2}) \, .
\end{eqnarray}

\begin{theorem}
  \label{th:2.1}\alphaparenlist
  \begin{enumerate}
  \item %%%a
    For  $v \in \fg_0$ let $
      J^{\{ v \}} = J^{(v)} + v^{\ne}$.
Then, provided that $v \in \fg^{\natural}$, we have $d_0 (J^{\{ v \}})
=0$, hence the homology class of $J^{\{ v \}}$ defines a field of
the vertex algebra $W_k (\fg ,x,f)$ of conformal weight $1$.

\item %%b
  $[L_{\lambda}J^{(v)}] = (\partial + (1-j)\lambda) J^{(v)} +
  \delta_{j0} \lambda^2 (\tfrac{1}{2} \str_{\fg_+} (\ad v)
  -(k+h\spcheck) (v|x))$ if $v \in \fg_j$, and the same formula
  holds for $J^{\{ v \}}$ if $v \in \fg_0$.

\item %%c
  $[{J^{\{ v \}}}{}_{\lambda} J^{\{ v' \}}] =J^{\{ [v,v']\}}+ \lambda
  (k(v|v') + \tfrac{1}{2} (\kappa_{\fg} (v,v') -\kappa_{\fg_0} (v,v')
  -\kappa_{1/2} (v,v')))$
provided that  $v,v' \in \fg^f_0$, where $\kappa_{1/2}$ is
the supertrace form of $\fg_0$ on $\fg_{1/2}$.

\item %%d
  For $v \in \fg_{-1/2}$ let
  \begin{eqnarray*}
    G^{\{ v\}} &=& J^{(v)} + v^{\ne} +
       \sum_{ \beta \in S_{1/2}}
      :J^{([v,u_{\beta}])} \Phi^{\beta}: \\
     &&  -\sum_{\beta \in S_{1/2}}(k(v |u_{\beta})
         +\str_{\fg_+} (\ad v)\, (\ad u_{\beta}))
         \partial \Phi^{\beta}\, .
  \end{eqnarray*}
Then provided that $v \in \fg^f_{-1/2}$, we have $d_0
(G^{\{ v \}})=0$, hence the homology class of $G^{\{ v \}}$
defines a field of the vertex algebra $W_k (\fg ,x,f)$ of
conformal weight $\tfrac{3}{2}$.  This field is primary.

\item %%e
Provided that $a \in \fg^{\natural}$ and $v \in \fg^f_{-1/2}$,
we have:
\begin{displaymath}
  [{J^{\{ a \}}}{}_{\lambda} G^{\{ v \}}] = G^{\{ [a,v]\}} \, .
\end{displaymath}

  \end{enumerate}
\end{theorem}

\begin{proof}
  (a), (b) and (c) were proved in \cite{KRW}, Theorem~2.4, by
  making use of the $\lambda$-bracket calculus.  The proof of (d)
  and (e) is similar.  In the calculation of the $\lambda$-bracket of $L$ with $G^{\{ v
  \}}$ $(v \in \fg^f_{-1/2})$ we find:
$  [L_{\lambda} G^{\{ v \}}] = (\partial + \tfrac{3}{2} \lambda)
  G^{\{ v \}} + \lambda^2 \sum_{\alpha} c_{\alpha} \Phi^{\alpha}
  \, , \quad c_{\alpha} \in \CC$.
Since $d_0 (L) =0$, $d_0 (G^{\{ v \}}) =0$ and $d_0
(\Phi^{\alpha})=\varphi^{\alpha}$, we conclude that all
$c_{\alpha} =0$.

\end{proof}

\section{A digression to homology}
\label{sec:3}

In this section we collect some general facts about homology that
will be used in the sequel.

Let $V$ be a vector superspace with an odd endomorphism $d$ such
that $d^2=0$, and let $H(V,d)= \Ker d /\Im d$ denote the
\emph{homology} of the \emph{complex} $(V,d)$.  The following
 well known lemma is very useful.
\vspace{1ex}

\begin{lemma}[K\"unneth lemma]
  \label{lem:3.1}    Let $(V,d)$ be a complex, and
suppose that $V=V_1 \otimes V_2$ and $d=d_1 \otimes 1 + 1 \otimes
d_2$.  Then the canonical map $H(V_1,d_1) \otimes H(V_2 ,d_2) \to
H(V,d)$ is a vector space isomorphism.
\end{lemma}

The next lemma is probably well known too.

\begin{lemma}
  \label{lem:3.2}
Let $\fg$ be a  Lie superalgebra and let $d$ be an odd derivation
of $\fg$ such that $d^2 =0$.  Extend $d$ to an odd derivation of
$U(\fg)$.  Then $H:=H (\fg ,d)$ is a Lie superalgebra too and the
canonical homomorphism of $U(H)$ to $H(U (\fg), d)$ is an
isomorphism of associative algebras.  In particular, if $H=0$,
then $H(U(\fg),d)=\CC 1$.
\end{lemma}

\begin{proof}
First, it follows from the K\"unneth lemma that
\begin{equation}
  \label{eq:3.1}
  H (S^n (\fg),d) = S^n (H(\fg)) \hbox{  for any  } n \in \ZZ \, .
\end{equation}
Indeed, by the K\"unneth lemma, (\ref{eq:3.1}) holds if $S$ is
replaced by $T$  .  But since the action of the symmetric group
$S_n$ on $\fg^{\otimes n}$ commutes with the action of $d$,
(\ref{eq:3.1}) holds too.

Next, consider the exact sequence of vector spaces:
\begin{displaymath}
  0 \to U_{n-1} (\fg) \to U_n (\fg) \to S^n(\fg)\to 0 \, ,
\end{displaymath}
where $\{ U_n (\fg) \}_n$ is the increasing PBW filtration of
$U(\fg)$.  This exact sequence induces a long exact sequence of
homology:
\begin{displaymath}
  H(S^n (\fg),d)\overset{\delta}{\longrightarrow} H (U_{n-1}(\fg),d) \to
  H(U_n(\fg),d) \to H (S^n (\fg),d) \overset{\delta}{\longrightarrow}
  H (U_{n-1} (\fg)\, , d) \, ,
\end{displaymath}
where $\delta$ is the boundary map.  Recall the construction of
$\delta$.  Take $w \in H (S^n (\fg),d)$.  By (\ref{eq:3.1}), we
can choose a representative $\tilde{w}$ of $w$, which is a
polynomial in the closed elements of $\fg$.  Take the preimage
$\tilde{\tilde{w}}$ of $\tilde{w}$ in $U_n (\fg)$ via the
supersymmetrization map.  Then the class of $d\tilde{\tilde{w}}$ lies
in $H (U_{n-1} (\fg),d)$ and we let $\delta (w) = d
\tilde{\tilde{w}}$.  But $d \tilde{\tilde{w}} =0$, hence $\delta
=0$ and we have the exact sequence
\begin{displaymath}
  0 \to H (U_{n-1} (\fg),d) \to H (U_n (\fg) ,d) \to
  H(S^n (\fg),d) \to 0 \, .
\end{displaymath}
By induction on $n$, this proves the lemma using (\ref{eq:3.1}).
\end{proof}

\begin{lemma}
  \label{lem:3.3}
Let $0 \to \fc \to \tilde{\fg} \to \fg \to 0$ be an extension of
Lie superalgebras and let $d$ be an odd derivation of
$\tilde{\fg}$ such that $d (\fc) =0$, so that $d$ induces a
derivation $d$ of $\fg$, and suppose that $d (\tilde{\fg})$ is
contained in an ideal $J$ of $\tilde{\fg}$ such that $J \cap \fc
=0$.  Let $\lambda : \fc \to \CC$ be a Lie superalgebra
homomorphism and let $\fc_{\lambda} \subset \tilde{\fg} + \CC
\cdot 1$ denote the subspace $\{ c-\lambda (c)$\break $| c \in \fc \}$.
Then
\begin{displaymath}
  H (U (\tilde{\fg}),d))/(\fc_{\lambda})= U(H(\tilde{\fg},d))/
  (\fc_{\lambda}) \, .
\end{displaymath}

\end{lemma}

\begin{proof}
Include $J$ in a complementary subspace $\fg$ to $\fc$ in
$\tilde{\fg}$, choose a basis of $J$, extend it to a basis of
$\fg$ and denote by $U(\fg)$ the span of the PBW monomials in
this basis.  Then $U(\fg)$ is a $d$-invariant subspace of
$U(\fg)$ and we have a vector space decomposition:
$
  U(\tilde{\fg}) = U (\fc) \otimes U (\fg)$.
  Hence, by the K\"unneth lemma we have:
\begin{displaymath}
  H (U (\tilde{\fg}),d) = U (\fc) \otimes H (U (\fg),d)
  (\hbox{  tensor product of vector spaces}) \, ,
\end{displaymath}
and the lemma follows.

\end{proof}

\begin{lemma}
  \label{lem:3.4}
\noindent{(a)\quad   Let $R$ be a Lie conformal superalgebra with an odd derivation $
 d$ (i.e.,~$d[a_{\lambda}\, b]$ $ = [da_{\lambda} b] + (-1)^{p(a)}
 [a_{\lambda}\, db]$) such that $d^2=0$.  Extend $d$ to a derivation
 of the universal enveloping vertex algebra $V(R)$.  Then $H:= H
 (R,d)$ is a Lie conformal superalgebra too,  and the canonical
 homomorphism of its universal enveloping vertex algebra $V(H)$
 to $H(V(R), \, d)$ is an isomorphism of vertex algebras.  In particular, if $H=0$, then
 $H (V(R),d)=\CC \vac$.}

(b)\quad
  Lemma~\ref{lem:3.3} holds if we replace Lie superalgebras by
  Lie conformal superalgebras and universal enveloping algebras
  by universal enveloping vertex algebras.

\end{lemma}

\begin{proof}
Since $V(R)$ is identified with $U(R_-)$, where $R_-$ is the
space $R$ with the Lie superalgebra bracket $[a,b] =
\int^0_{-\partial} [a_{\lambda} b] \, d \lambda$
\cite{GMS}, \cite{BK}, Lemma~\ref{lem:3.4} follows from
  Lemmas~\ref{lem:3.2} and \ref{lem:3.3}.

\end{proof}

\section{The structure of the vertex algebra $W_k (\fg ,x,f)$}
\label{sec:4}

Denote by $\C^+$ the vertex subalgebra of the vertex algebra $\C
(\fg ,x,f,k)$ generated by the fields $\varphi_{\alpha}$ and $d_0
(\varphi_{\alpha})$ for all $\alpha \in S_+$.  From \cite{KRW},
formula~(2.4), we have:

\begin{displaymath}
  d_0 (\varphi_{\alpha}) = \left\{
      \begin{array}{ll}
        J^{(u_{\alpha})} + (-1)^{p(\alpha)} \Phi_{\alpha}
           & \hbox{if  } \alpha \in S_{1/2} \,  \\
       J^{(u_{\alpha})} + (f|u_{\alpha})
           & \hbox{if  }\alpha \in S_+\backslash
           S_{1/2}\, .
      \end{array}\right.
\end{displaymath}
Using this and \cite{KRW}, Lemma~2.1(a) we get
\begin{displaymath}
  [d_0 (\varphi_{\alpha})_{\lambda} \varphi_{\beta}] =
  (-1)^{p(\alpha)} \sum_{\gamma} c^{\alpha}_{\beta}
  (u_{\alpha}) \varphi_{\gamma} \, .
\end{displaymath}
It follows that
\begin{displaymath}
  R=\sum_{\alpha \in S_+} \CC [\partial] \varphi_{\alpha}+
\sum_{\alpha \in S_+} \CC [\partial] \, d_0 (\varphi_{\alpha})
\end{displaymath}
is closed under the $\lambda$-bracket and is $d_0$-invariant.
Hence we have a Lie conformal algebra complex $(R, d_0)$, and the
homology of this complex is obviously zero.  It is also clear
that $\C^+$ is the universal enveloping vertex algebra of $R$.
Applying Lemma~\ref{lem:3.2}, we obtain:
\begin{equation}
  \label{eq:4.1}
  H (\C^+ , d_0) = \CC \vac \, .
\end{equation}

Next, denote by $\C^-$ the vertex subalgebra of the vertex
algebra $\C (\fg ,x,f,k)$  generated by the fields
$J^{(u)}$ for all $u \in \fg_{\leq}$, the fields
$\varphi^{\alpha}$ for all $\alpha \in S_+$ and the fields
$\Phi_{\alpha}$ for all $\alpha \in S_{1/2}$.  Then
obviously we have:
\begin{equation}
  \label{eq:4.2}
  \C (\fg ,x,f,k) = \C^+ \otimes \C^- \hbox{  (as vector spaces).}
\end{equation}
Recall that (\cite{KRW}, formula~(2.4)):
\begin{eqnarray}
  \label{eq:4.3}
  d_0 (\varphi^{\alpha}) &=& -\tfrac{1}{2}
     \sum_{\beta,\gamma \in S_+} (-1)^{p(\alpha)p(\beta)}
     c^{\alpha}_{\beta \gamma} \varphi^{\beta}\varphi^{\gamma}\,
     , \\[2ex]
\label{eq:4.4}
  d_0 (\Phi_{\beta}) &=& \sum_{\alpha \in S_{1/2}}
  \langle u_{\alpha}, u_{\beta} \rangle_{\ne} \varphi^{\alpha}
  \, , \, d_0 (\Phi^{\beta}) = \varphi^{\beta} \, .
\end{eqnarray}
It follows from formulas (\ref{eq:2.6}), (\ref{eq:4.3}) and
(\ref{eq:4.4}) that $\C^-$ is $d_0$-invariant.  Hence by the
K\"unneth lemma, (\ref{eq:4.1}) and  (\ref{eq:4.2}) imply
\begin{equation}
  \label{eq:4.5}
  W_k (\fg ,x,f,k) =H (\C^- , d_0) \, .
\end{equation}

Now we can state and prove the main theorem on the structure of
the vertex algebra $W_k (\fg ,x,f)$.

\begin{theorem}
  \label{th:4.1}
Let $\fg$ be a simple finite-dimensional Lie superalgebra with an
invariant bilinear form $(. \, | \, .)$, and let $x,f$ be a pair
of even elements of $\fg$ such that $\ad x$ is diagonalizable with
eigenvalues in $\tfrac{1}{2}\ZZ$ and $[x,f]=-f$.  Suppose that
all eigenvalues of $\ad x$ on $\fg^f$ are non-positive:  $\fg^f
=\oplus_{j \leq 0} \fg^f_j$.  Then

\alphaparenlist
\begin{enumerate}
\item %%a
For each $a \in \fg^f_{-j}(j \geq 0)$ there exists a $d_0$-closed
field $J^{\{ a \}}$ in $\C^-$ of conformal weight $1+j$ (with
respect to $L$) such that $J^{\{ a \}} -J^{(a)}$ is a linear
combination of normal ordered products of the fields $J^{(b)}$,
where $b \in \fg_{-s}$, $0\leq s <j$, the fields $\Phi_{\alpha}$, where
$\alpha \in S_{1/2}$, and the derivatives of these
fields.

\item %%b
  The homology classes of the fields $J^{\{ a_i \}}$, where $
 a_1,a_2, \ldots$ is  a basis of $\fg^f$ compatible with its
 $\tfrac{1}{2}\ZZ$-gradation, strongly generate the vertex
 algebra $W_k (\fg ,x,f)$ and obey the PBW theorem (see
 Remark~\ref{rem:4.2} below).

\item %%c
$H_0 (\C (\fg ,x,f,k), d_0) = W_k (\fg ,x,f)$ and $H_j (\C (\fg
,x,f,k), d_0)=0$ if $j \neq 0$.
\end{enumerate}
\end{theorem}

\begin{proof}

Define an ascending vertex algebra fltration of $\C^-$  by the following
relations:% $(\tfrac{1}{2}\ZZ)^2$-gradation $\C^-=\oplus_{m,n \in
%  \tfrac{1}{2}\ZZ} \C^-_{m,n}$ defined
%
\begin{eqnarray*}
  \deg \vac = 0 \, , \, \deg T =0 \, , \, 
  \deg J^{(v)} =j+\tfrac{1}{2} \hbox{  if  } v \in \fg_{-j} \, \quad
      (j \in \tfrac{1}{2}\ZZ_+) \, , \\
  \deg \Phi_{\alpha} = 0 \, (\alpha \in S_{1/2}) \, ,
% \,
  \deg \varphi^{\alpha} = j-\tfrac{1}{2} \quad
      \hbox{  if  } u_{\alpha} \in \fg_j \,(j> 0) \, .
\end{eqnarray*}

The differential $d_0$ of $\C^-$ induces the following
differential of the associated graded vertex algebra $\Gr \C^-$:
\begin{equation}
  \label{eq:4.6}
   d_1 (J^{(v)})  = \sum_{\beta \in S_+}
    ([f,v]| u_{\beta})\varphi^{\beta} \, ,\,
  d_1 (\varphi^{\alpha}) = 0 \, ,\,\, d_1 (\Phi^{\alpha})
  =\varphi^{\alpha}\, , \,\, d_1 (\vac) =0 \, .
\end{equation}
Since the pair $(x,f)$ is good,  by
(\ref{eq:1.11}), for each $r \geq 1$, $r \in \tfrac{1}{2}\ZZ$, we
can choose elements $u^{\alpha} \in \fg_{1-r}$ $(\alpha \in S_r)$
such that
\begin{equation}
  \label{eq:4.7}
  (f|[u^{\alpha},u_{\beta}]) =\delta_{\alpha \beta}
  \hbox{  for all  } \alpha,\beta \in S_r \, .
\end{equation}
Note that
\begin{equation}
  \label{eq:4.8}
  \fg_{\leq} =\sum_{\alpha \in S_+ \backslash S_{1/2}}
     \CC u^{\alpha} \oplus \fg^f_{\leq} \, .
\end{equation}
Then we have from (\ref{eq:4.6}):
\begin{equation}
  \label{eq:4.9}
  d_1 (J^{(u^{\alpha})})=\varphi^{\alpha} \, , \,\,
  \alpha \in S_+ \backslash S_{1/2}\, .
\end{equation}

Note that $\Gr \C^-$ is the universal enveloping vertex algebra
of the Lie conformal algebra
\begin{equation}
\label{eq:4.11n}
   \tilde{R} =\sum_{v \in \fg_{\leq}} \CC [\partial] J^{(v)}
  + \sum_{\alpha \in S_+} \CC [\partial] \varphi^{\alpha}
  + \sum_{\alpha \in S_{1/2}} \CC [\partial]
  \Phi^{\alpha} + \CC \vac \, ,
\end{equation}
divided by the ideal generated by $\vac - \vac_{\Gr}$ (where all
$\lambda$-brackets vanish, except between the $\Phi^{\alpha}$'s).
It follows from (\ref{eq:4.6}), (\ref{eq:4.8}), (\ref{eq:4.9})
and Lemma~\ref{lem:3.4}(b) that the vertex algebra $H (\Gr \C^-
\,, \, d_1)$ is strongly generated by the elements $J^{(v)}$, $v
\in \fg^f_{\leq}$.

Note that this first term of the spectral sequence is
concentrated in zero charge, and since the corresponding
differential and all higher differentials change charge by $-1$,
we conclude that the spectral sequence converges to $H (\Gr \C^-
\, , \, d_1)$.

 Note also that
 $d_1$ preserves the conformal weights, i.e.,~the
eigenvalues of $L_0$.  Since the eigenspaces of $L_0$ on $\C^-$
are finite-dimensional, the complex $\C^-$ is locally finite.  Hence the
spectral sequence converges to the associated graded vertex
algebra of $H (\C^-\, , \, d_0)$.

Hence, given $a \in \fg^f_{-j}$ $(j \in \ZZ_+)$ the  field
$J^{(a)}$ can be extended to a closed field
$J^{\{ a \}}$ with respect to $d_0$, of the form:
\begin{equation}
  \label{eq:4.11}
  J^{\{ a \}} = J^{(a)} + (\hbox{finite sum of lower degree
    terms})\, .
%%%%%  \, r \in \ZZ \, , \, r \geq 1)\, .
\end{equation}
By the charge considerations,  the lower degree terms do not  involve the fields $\varphi^{\alpha}$, and
hence they are linear combinations of normally ordered products
of the fields $J^{(b)}$, $\Phi_{\alpha}$, and their derivatives.
This proves (a).
The statement~(b) follows from (a) by applying the K\"unneth
lemma (to (\ref{eq:4.2})) and (\ref{eq:4.1}).  The statement~(c) follows from (b).

\end{proof}

\begin{remark}
  \label{rem:4.1}
Since the exact elements of the complex $(\C^- \, , \, d_0)$  have obviously strictly negative charge, 
the vertex algebra $W_k (\fg , x,f)$ is a subalgebra of the
vertex algebra $\C (\fg ,x,f,k)$, which consists of $d_0$-closed
charge $0$ elements of $\C^-$  (i.e. the complex has the so called formality 
property).
Using this, one can compute the elements $J^{\{a\}}$ recursively.
It is easy to see that for a given
$a \in \fg^f_j$ the solution is unique if $j=0$ or $-1/2$.  However, 
for $j \leq -1$
the solution is not unique, except for the following cases:
$j=-1$ and $\fg^f_0=0$; $j$ is a negative half-integer and $\dim \fg_j = 
\dim \fg_{-1/2}$ .
%The equation $(d_1 + d_2)J^{\{ a \}}=0$ is equivalent to the
%following system of equations on the $A_{-j+r,j-r}$ in
%(\ref{eq:4.11}):
%%
%\begin{equation}
%  \label{eq:4.12}
%  d_1 (A_{-j+1,j-1}) =-d_2 (J^{(a)})\, , \,\, d_1 (A_{-j+r,j-r})
%     =d_2 (A_{-j+r-1,j-r+1}) \hbox{  for  } r \geq 2 \,.
%\end{equation}
%
\end{remark}
%
%Of course the fields of conformal weight $1$ and $\tfrac{3}{2}$
%of the vertex algebra $W_k (\fg ,x,f)$ constructed in
%Theorem~\ref{th:2.1} (a) and (d), respectively, are of the form
%given by Theorem~\ref{th:4.1} (a).  It is easy to see that they
%are the only solutions of the system (\ref{eq:4.12}) for a given
%the solution is not unique, and it is more difficult to write
%down a solution explicitly.

\begin{remark}
  \label{rem:4.2}
Write each field $J^{\{ a_i \}}$ in the form  $J^{\{ a_i \}} (z)
=\sum_{n \in \ZZ -\Delta_i} J^{\{i\}}_n z^{-n-\Delta_i}$, where
$\Delta_i$ is the conformal weight.  Then the property that the
$J^{\{ a_i \}}$ strongly generate the vertex algebra $W_k (\fg
,x,f)$ means that the monomials,
\begin{equation}
  \label{eq:4.13}
  (J^{\{i_1\}}_{-m_1})^{b_1} (J^{\{i_2\}}_{-m_2})^{b_2} \ldots
   (J^{\{i_s\}}_{-m_s})^{b_s} \vac \hbox{  where  }b_i \in \ZZ_+ \, ,
   \, b_i \leq 1 \hbox{  if  } a_i \hbox{  is odd, and  }
   m_{i} > \Delta_{i}-1 \, ,
\end{equation}
span this vertex algebra.  The property that they obey the PBW
theorem means that the monomials (\ref{eq:4.13}), where the
sequence of pairs $(i_1,m_1)$, $(i_2,m_2)$, $\ldots$ is
decreasing in the lexicographical order, form a basis of
$W_k (\fg ,x,f)$.

\end{remark}

\section{The structure of vertex algebras $W_k (\fg ,e_{-\theta})$,
  associated to minimal gradations.}
\label{sec:5}

A $\tfrac{1}{2} \ZZ$-gradation of the Lie superalgebra $\fg $ is called
\emph{minimal} if it has the form
\begin{equation}
  \label{eq:5.1}
  \fg = \fg_{-1} \oplus \fg_{-1/2} \oplus \fg_0
         \oplus \fg_{1/2} \oplus \fg_1
\end{equation}
and satisfies the following additional properties:

\alphaparenlist
\begin{enumerate}
\item %%a
  $\fg_1 =\CC e$ and $\fg_{-1}=\CC f$, where $e$ and $f$ are even
  non-zero elements,

\item %%b
  (\ref{eq:5.1}) is the eigenspace decomposition with respect to
  $\ad x$, where $x=[e,f]$.
\end{enumerate}

Since $\{ e,x,f \}$ is an $\sl_2$-triple, the minimal  gradation
is determined, up to conjugation, by the nilpotent element $f$.
Note that $\fg^{\natural}$ is the centralizer of this triple, and   we have:
\begin{equation}
  \label{eq:5.2}
  \fg^{\natural} =\{ a \in \fg_0 | (x|a)=0 \} \, , \quad
  \fg^f = \fg_{-1}+\fg_{-1/2}+ \fg^{\natural} \, .
\end{equation}

Choose a Cartan subalgebra $\fh$ of the even part of $\fg_0$.
Then $\fh$ is a Cartan subalgebra of the even part of  $\fg$
 and $x \in \fh$, so that $\fh^{\natural}: =\{ h \in \fh | (x|h)=0
 \}$ is the Cartan subalgebra of the even part of $\fg^{\natural}$.
In particular, we have
\begin{equation}
  \label{eq:5.3}
  \fh = \fh^{\natural} \oplus \CC x \, .
\end{equation}

Let $\Delta \subset \fh^*$ be the set of roots of  $ \fh$ in $\fg$.  Choose a subset
 of positive roots $\Delta_+ \subset \Delta$ such that $\alpha
 (x) \geq 0$ if $\alpha \in \Delta_+$, and let $\theta$ be the
 highest root.  Hence $e=e_{\theta}$ and $f=e_{-\theta}$ are
 root vectors attached to $\theta$ and $-\theta$.

It follows from the classification of simple finite-dimensional
Lie superalgebras \cite{K1} that such a Lie superalgebra $\fg$
carries a non-degenerate even supersymmetric invariant bilinear
form and is not a Lie algebra iff $\fg$ is isomorphic to $\sl
(m|n)/\delta_{m,n} \CC I$ $(m,n\geq 1, (m,n)\neq(1,1))$ , 
$osp (m|n) =$\break $ spo (n|m)$ $(m \geq
1 \, , \, n \geq 2 \even)$, $D(2,1;a)$, $F(4)$, $G(3)$ or $H(n)$
$(n \geq 6 \even)$.  It is easy to see that in the case of $H(n)$
the gradation (\ref{eq:5.1}) corresponding to $e_{-\theta}$ has
$\dim \fg_{\pm 1} \geq 2$, hence does not satisfy (a).  In all
the remaining cases the even part of $\fg$ is reductive.  Thus,
$\theta$ is the highest root of one of the simple components of
the even part of $\fg$ (which is $\fg$ if $\fg$ is a simple Lie
algebra), and therefore the adjoint orbit of $e$ (which coincides
with that of $f$) in this simple component is the unique non-zero
nilpotent orbit of minimal dimension in this simple component.

Conversely, if $\theta$ is an even highest root of $\fg$ for some
ordering of the set of roots, choosing a root vector $e \in
\fg_0$ and embedding $e$ in an $\sl_2$-triple $\{ e,x,f \}$, we
obtain a minimal $\tfrac{1}{2} \ZZ$-gradation given by the eigenspaces of
$\ad x$.  Thus, up to conjugation, minimal $\tfrac{1}{2} \ZZ$-gradations
of $\fg$ are classified by simple components of the even part of
$\fg$ whose highest root can be made a highest root of
$\fg$ for some ordering of $\Delta$.  It is easy to see, using
the description of $\Delta$ in \cite{K1}, that this is always
possible except when $\fg =osp (3|n)$ and the simple component of
its even part is $so_3$.

We normalize the invariant bilinear form on $\fg$ by the
condition $(\theta | \theta)=2$.  This determines uniquely the
Casimir operator of $\fg$ and hence its eigenvalue $2h\spcheck$ on
$\fg$.  The number $h\spcheck$ is called the dual Coxeter number
of the gradation (\ref{eq:1.1}).  We shall identify $\fh$ with
$\fh^*$ using this form.  Then we have: $x=\theta /2$.

Since $[e,f] = \tfrac{1}{2}\theta$, we obtain $(e|f)=\tfrac{1}{2}$, and since $[a,b] \in \CC e$ for $a,b \in \fg_{1/2}$, we get
\begin{equation}
  \label{eq:5.4}
  [a,b]=2 \langle a,b \rangle_{\ne} e \quad
  (a,b \in \fg_{1/2}) \, .
\end{equation}

Since in general \cite{KW3}
\begin{equation}
  \label{eq:5.5}
  \str_{\fg} (\ad \, a)(\ad \, b)=2h\spcheck (a|b)\, , \quad
  a,b \in \fg \, ,
\end{equation}
and $(x|x)=1/2$, we conclude by letting $a=b=x$ in (\ref{eq:5.5})
that
\begin{equation}
  \label{eq:5.6}
  \sdim \fg_{1/2} = 2h\spcheck -4 \, .
\end{equation}
(Note that this equality means that $\sdim[\fg,e]=2h\spcheck-2$.)

The above discussion gives a complete list of minimal gradations of $\fg$, along with $\fg^{\natural}$,
 $h\spcheck$ and the description of the $\fg^{\natural}$-module
 $\fg_{1/2}$ ($\simeq \fg^*_{-1/2}$),  which is given below
 (cf.~\cite{FL}, \cite{KRW}):
%%%%\newpage

\Romanlist
\begin{enumerate}
\item %%I
 ${\fg}$ is a simple Lie
algebra:
$$
\begin{array}{| c | c | c|  c || c | c | c | c|}
\hline
{\fg} & {\fg}^{\natural} & {\fg}_{1/2}& h\spcheck &
{\fg} & {\fg}^{\natural} & {\fg}_{1/2}& h\spcheck \\
\hline
\sl_n \ (n \geq 3) & g\ell_{n-2} &
\CC^{n-2} \oplus \CC^{n-2 *} & n & F_4 & sp_6 & \Lambda_0^3 \CC^6
& 9\\
\hline
so_n \ (n \geq 5) & \sl_2
\oplus so_{n-4} & \CC^2 \otimes
\CC^{n-4} & n-2 & E_6 & \sl_6 & \Lambda^3
\CC^6  &12 \\
\hline
sp_n \ (n \geq 2) & sp_{n-2}
 &
\CC^{n-2} & \tfrac{n}{2}+1& E_7 & so_{12}
& spin_{12} & 18\\
\hline
G_2 & \sl_2 & S^3 \CC^2 & 4 & E_8 & E_7 &
56 -\dim & 30
\\
\hline
\end{array}
$$

\item %%II
 $\fg$ is not a Lie algebra, but $\fg^{\natural}$ is  and
 $\fg_{1/2}$ is purely odd  $(m \geq   1)$:
$$
\begin{array}{| c | c | c | c || c | c | c | c | }
\hline
{\fg} & {\fg}^{\natural} & {\fg}_{1/2} & h\spcheck
&  {\fg} & {\fg}^{\natural}
& {\fg}_{1/2} & h\spcheck\\
\hline
\sl (2|m) \ (m \neq 2) & g\ell_m &
\CC^m \oplus \CC^{m *} & 2-m & D(2, 1 ;a) & \sl_2 \oplus \sl_2
&  \CC^2 \otimes \CC^2  & 0\\
\hline
\sl (2| 2)/\CC I & \sl_2 & \CC^2
\oplus \CC^2 & 0  & F(4) & so_7 & spin_7  & -2\\
\hline
spo(2 | m )  & so_m &
\CC^m & 2-\tfrac{m}{2} & G(3) & G_2
& 7-\dim & -\tfrac{3}{2}
\\
\hline
osp(4 | m )  & s\ell_2 \oplus
sp_m &
\CC^2 \otimes \CC^m & 2-m &  & & & \\
\hline
\end{array}
$$

\item %%III
${\fg}$  and $\fg^{\natural}$ are not Lie algebras $(m,n \geq 1)$:
$$
\begin{array}{| c | c | c | c | }
\hline
{\fg} & {\fg}^{\natural} & {\fg}_{1/2} & h\spcheck \\
\hline
\sl (m | n) \,  (m \neq n , m>2) & g\ell(m-2 | n) &
\CC^{m-2 | n} \oplus \CC^{m-2 | n  *} & m-n \\
\hline
\sl (m| m)/\CC I (m>2) & \sl (m-2 | m)  &
\CC^{m-2 | m} \oplus \CC^{m-2 | m  *} & 0 \\
\hline
spo (n | m ) \,  (n \geq 4) & spo (n-2 | m) & \CC^{n-2 | m}
& \tfrac{n-m}{2}+1 \\
\hline
osp (m | n ) \, (m \geq 5) & osp
(m-4 | n) \oplus \sl_2 & \CC^{m-4 |
n} \otimes \CC^2  & m-n-2\\
\hline
 F(4) & D ( 2 , 1 ; 2) &
\stackrel{1}{\circ} \leftarrow \!\!\!\otimes \!\!\!\!\rightarrow \circ
\ \  (6 | 4)-\dim & 3
\\
\hline
 G(3) &osp ( 3 | 2 ) &
\stackrel{-3}{\otimes} \Longrightarrow
\stackrel{1}{\circ}
\ \ \ (4 | 4)-\dim &2 \\
\hline
\end{array}
$$
\end{enumerate}

%%%\newpage

Now we turn to a more detailed description of the vertex algebras
$W_k (\fg ,e_{-\theta})$ corresponding to minimal gradations of
$\fg$.  First, note that the central charge of the Virasoro field
$L(z)$ defined by (\ref{eq:2.2}) is given by the following formula
(we use (\ref{eq:5.6}) here):
\begin{equation}
  \label{eq:5.7}
  c(\fg ,e_{-\theta},k) =
     \frac{k \sdim \fg}{k+h\spcheck} -6k + h\spcheck -4\, .
\end{equation}
Note that
\begin{equation}
  \label{eq:5.8}
\sdim \fg = \sdim \fg^{\natural} + 2\sdim \fg_{1/2}+3 = \sdim \fg^{\natural}
+ 4h\spcheck - 5\, .
\end{equation}
Let $\Omega_0$ be the Casimir operator of $\fg_0$ for the
bilinear form $(. \, | \, .)$ restricted to $\fg_0$.  Since the
$\fg_0$-module $\fg_{-1/2}$ (and$\fg_{1/2}$) is either
irreducible or is a direct sum of two contragredient modules,
$\Omega_0$ has only one eigenvalue on $\fg_{-1/2}$ (equal that on
$\fg_{1/2}$), which we   denote  by $h_{1/2}$.

\begin{lemma}
  \label{lem:5.1}
$h_{1/2}=h\spcheck -1$.
\end{lemma}

\begin{proof}
  Let $S_j \subset \Delta$ denote the set of roots of $\fh$ in
  $\fg_j$ and let $\rho_0=\tfrac{1}{2}\sum_{\alpha \in \Delta_+
    \cap S_0}(-1)^{p(\alpha)}\alpha$.  Then we have:
  \begin{equation}
    \label{eq:5.9}
    2\rho =2\rho_0 +\sum_{\alpha \in S_{1/2}}(-1)^{p(\alpha)}\alpha
    +\theta \, .
  \end{equation}
It follows from (\ref{eq:5.4}) and the non-degeneracy of the form
$\langle . \, , \, .\rangle_{\ne}$, that
\begin{equation}
  \label{eq:5.10}
  \alpha \in S_{1/2} \hbox{  iff  }\theta -\alpha \in S_{1/2} \, .
\end{equation}
Let $\mu \in S_{1/2}$.  Then, by (\ref{eq:5.9}), (\ref{eq:5.10})
and (\ref{eq:5.6}) we have:
\begin{eqnarray*}
  2(\mu |\rho) =2(\mu |\rho_0)+\tfrac{1}{2}
    \sum_{\alpha \in S_{1/2}}(-1)^{p(\alpha)}
    ((\mu |\alpha )+(\theta -\mu |\alpha))+(\mu |\theta)\\
    = 2(\mu |\rho_0)+\tfrac{1}{2} \sdim \fg_{1/2}+1=
    2(\mu |\rho_0)+h\spcheck -1\, .
\end{eqnarray*}
Thus, we obtain
\begin{equation}
  \label{eq:5.11}
  2(\mu |\rho -\rho_0)=h\spcheck -1 \, .
\end{equation}
Let $\Omega =\sum^r_{i=1} h^ih_i +2\rho +2\sum_{\alpha
  \in\Delta_+}e_{-\alpha} e_{\alpha}$ (resp. $\Omega_0
=\sum^r_{i=1} h^ih_i+2\rho_0 +2$\break $\sum_{\alpha \in\Delta_+\cap
  S_0}e_{-\alpha}e_{\alpha}$) be the Casimir operator
of $\fg$ (resp. of $\fg_0$), where
$(e_{\alpha}|e_{-\alpha})=1$,
$(h^i|h_j)=\delta_{ij}$.  We have:
\begin{displaymath}
  \Omega = 2 (\rho-\rho_0)+\Omega_0+2\sum_{\alpha \in S_{1/2}}
  e_{-\alpha}e_{\alpha}+2e_{-\theta}e_{\theta}\, ,
\end{displaymath}
hence
\begin{displaymath}
  2h\spcheck e_{\mu}=\Omega e_{\mu}=2(\rho -\rho_0|\mu)
  e_{\mu}+h_{1/2} e_{\mu}+2[e_{\mu -\theta},
  [e_{\theta -\mu}, e_{\mu}]] \, .
\end{displaymath}
Hence, using (\ref{eq:5.11}), we get:
\begin{displaymath}
  h\spcheck e_{\mu}=(h_{1/2}-1)e_{\mu}+2
  [e_{\mu-\theta}, [e_{\theta -\mu} ,e_{\mu}]]\, .
\end{displaymath}
But the second summand on the right is $ae_{\mu}$, where
$a=c^{\theta}_{\theta -\mu ,\mu} c^{\mu}_{\mu-\theta ,\theta}$.  Hence,
it remains to show that $a=1$.  We have:
\begin{eqnarray*}
  a e_{\theta} =[e_{\theta -\mu}, [e_{\mu-\theta}, e_{\theta}]]
  =[[e_{\theta -\mu}, e_{\mu-\theta}],e_{\theta}]
  =(\theta |\theta -\mu) e_{\theta} =e_{\theta}, \hbox{  hence  }
  a=1 \, .
\end{eqnarray*}

\end{proof}

We denote by $h\spcheck_{0,i}$ the dual Coxeter number of the $i^{th}$
simple component $ \fg^{\natural}_i$ of $\fg^{\natural}$ with respect to
the bilinear form $(.|.)$ restricted to $\fg^{\natural}_i$.
We have the following more precise version of
Theorems~\ref{th:2.1} and \ref{th:4.1} in the case of a minimal gradation.

\begin{theorem}
  \label{th:5.1}\alphaparenlist
  \begin{enumerate}
  \item %%%a
    All the fields $J^{\{ a \}} (z)$, $a \in \fg^{\natural}$, and $
   G^{\{ v \}} (z)$, $v \in \fg_{-1/2}$, (see Theorem~\ref{th:2.1}),
   are primary fields of the vertex algebra $W_k (\fg
   ,e_{-\theta})$ of conformal weight $1$ and $\tfrac{3}{2}$,
   respectively.
\item %%b
  The fields $J^{\{ a \}} (z)$, $a \in \fg^{\natural}$, $G^{\{ v
    \}}(z)$ $(v \in \fg_{-1/2})$ and $L(z)$ strongly generate the
  vertex algebra $W_k (\fg ,e_{-\theta})$.

\item %%c
Let $\{ u^{\alpha} \}_{\alpha \in S_0}$ be the dual basis to
$\{u_{\alpha}\}_{\alpha \in S_0}$,
i.e.,~$(u_{\alpha}|u^{\beta})=\delta_{\alpha \beta}$, $\alpha
,\beta \in S_0$.  Then we have the
following equality in $W_k (\fg ,e_{-\theta})$:
\begin{eqnarray*}
  L &=&- \frac{1}{k+h\spcheck} \Big(J^{(f)} + \sum_{\alpha \in
  S_{1/2}} (-1)^{p(\alpha)} \Phi^{\alpha} J^{([f,u_{\alpha}])}\\
&&- \tfrac{1}{2} \sum_{\alpha \in S_0} (-1)^{p(\alpha)} :
J^{(u_{\alpha})} J^{(u^{\alpha})}: - \tfrac{1}{2}
(k+h\spcheck) \sum_{\alpha \in S_{1/2}} : \partial
\Phi^{\alpha} \Phi_{\alpha}: -(k+1)\partial J^{(x)}\Big)\, .
\end{eqnarray*}

\item %%d
  $[{J^{\{ a\}}}{}_{\lambda} J^{\{ b\}}] = J^{\{ [a,b]\}} +
    \lambda
    ((k+\tfrac{1}{2}h\spcheck)(a|b)- \tfrac{1}{4}\kappa_{\fg_0}(a,b))$, and
    $[{J^{\{ a \}}}{}_{\lambda} G^{\{ v\}}]=G^{\{[a,v]\}}$.

\item %%e
$ [{G^{\{ u \}}}{}_{\lambda} G^{\{ v \}}] =-2(k+h\spcheck)
  (e|[u,v])L+(e|[u,v])\sum_{\alpha \in S^{\natural}}
  : J^{\{u^{\alpha}\}} J^{\{u_{\alpha}\}} :$
\vspace{-1ex}
\begin{eqnarray*}
&& +\sum_{\gamma \in S_{1/2}}:J^{\{
    [u,u^{\gamma}]^{\natural}\}} J^{\{ [u_{\gamma},v]^{\natural}\}}:
  +2(k+1) (\partial+ 2\lambda) J^{\{[[e,u],v]^{\natural}\}}\\
  &&+  \lambda \sum_{\gamma \in S_{1/2}}
  J^{\{ [[ u,u^{\gamma}],[u_{\gamma},v]]^{\natural}\}}
        + \tfrac{\lambda^2}{3} c(u,v) \, \\
\noalign{where}\\
      c(u,v) &=&(e|[u,v])( -(k+h\spcheck)c(k) + 
         (k+\tfrac{h\spcheck}{2}) \, \sdim \fg^{\natural} 
         -\tfrac{1}{2} \sum_i h\spcheck_{0,i} \sdim \fg^{\natural}_i)\\
  %%           \sum_{i \in S^f_0}\\
%
             &&+ \sum_{\gamma \in S_{1/2}} c' ([u,u^{\gamma}]^{\natural} \, ,
             \,    [u_{\gamma} , v]^{\natural}) \, , \\
       c' (a,b) &=& k(a|b)+\tfrac{1}{2} \str_{\fg_+} (\ad a)(\ad b)\\
            &=& (k+\tfrac{h\spcheck}{2}) (a|b)-\tfrac{1}{4}
                \kappa_{\fg_0} (a,b)+\tfrac{1}{4} \str_{\fg_+}
                \ad ([a,b])(a,b \in \fg_0)\, .
% (e|[u,v])\Big(-2 (k+h\spcheck) c (\fg
%  ,e_{-\theta},k) + (2k+h\spcheck) \sdim \fg^{\natural} -\sum_i h\spcheck_{0.i}
%  \sdim \fg^{\natural}_i\\
%%
%&& + (2k+h\spcheck -\sum_i h\spcheck_{0,i})\,
%  (2h\spcheck -3)\Big),
\end{eqnarray*}
  \end{enumerate}
\noindent
Here the superscript $\natural$ denotes the orthogonal
  projection of $\fg_0$  on $\fg^{\natural}$,
  $u_{\alpha}$ and $u^{\alpha}$
  (resp.~$u_{\gamma}$ and $u^{\gamma}$) are bases of
  $\fg^{\natural}$ (resp.~$\fg_{1/2}$) such that $(u_{\alpha} |
  u^{\alpha'}) =\delta_{\alpha \alpha'}$ (resp.~$\langle
  u_{\gamma}, u^{\gamma'}\rangle_{\ne} =\delta_{\gamma \gamma'}$).

\end{theorem}

\begin{proof}

The claim (a) for the fields $J^{\{ a \}}$ was proved in
\cite{KRW}, and for the fields $G^{\{ v \}}$ in
Theorem~\ref{th:2.1}(d).

In order to prove (b), it suffices to show, in view of
Theorem~\ref{th:4.1}, that $L$ can be written in the form
const.$\, J^{\{ f \}}$ (described in Theorem~\ref{th:4.1}(a)).
  From \cite{KRW}, formula~(2.4), we have:
  \begin{displaymath}
    d_0 (\varphi_{\theta}) =e+\tfrac{1}{2} \, , \, d_0(f)
     = \sum_{\substack{\alpha \in S_+\\ \gamma \in S}}
     (-1)^{p(\alpha)p(\gamma)} c^{\gamma}_{\alpha ,-\theta}
     :e_{\gamma} \varphi^{\alpha}:+\tfrac{1}{2} k\partial
     \varphi^{\theta} \, .
  \end{displaymath}
Since $c^{\gamma}_{\alpha ,-\theta} \neq 0$ only if $\gamma
<\theta$, we see that $:ef :+: fe:$ is homologous to
$-f + \cdots$.  Here and below $\cdots$ signify a linear combination of normal
ordered products of fields of conformal weight $<2$.  Hence, by
the Sugawara construction, $L^{\fg}:= \tfrac{1}{k+h\spcheck} (:
ef :+:fe :+ \cdots)$ is homologous to $-\tfrac{1}{k+h\spcheck}
f+\cdots$.  In view of Theorem~\ref{th:4.1}, this proves (b).
The statement (c) is obtained by solving the system of equations
(4.12) for  $a=-\tfrac{1}{k+h\spcheck}f$ and inserting
the solution in (\ref{eq:4.11}). We use Lemma \ref{lem:5.1} in this
calculation.  (d)~follows from Theorem~\ref{th:2.1}(c). (e) is
obtained by a very long, but a straightforward $\lambda$-bracket calculation.

\end{proof}

\begin{remark}
  \label{rem:5.1}
Let $u \in \fg_j$, $v \in \fg_{-j}$.  Then we have
\begin{eqnarray}
  \label{eq:5.12}
  \str_{\fg_-} (\ad \, u) (\ad \, v)
     = \str_{\fg_+} (\ad \, u)(\ad \, v)-\str_{\fg_+}
        \ad \, [u,v],\\
    \label{eq:5.13}
  \str_{\fg_+} (\ad \, u) (\ad \, v)
     = \tfrac{1}{2} (\kappa_{\fg} (u,v)-\kappa_{\fg_0} (u,v)
       + \str_{\fg_+} \ad \, [u,v]) \, .
\end{eqnarray}
Formula (\ref{eq:5.12}) follows by a simple computation from the
formulas $\str_{\fg_+} A_+=\sum_{\alpha \in S_+} (-1)^{p(\alpha)}$
$(A_+u_{\alpha}|u^{\alpha})$,  $\str_{\fg_-}A_-=\sum_{\alpha
  \in S_+}(-1)^{p(\alpha)} \,\, (u_{\alpha}|A_- u^{\alpha})$,
where $A_{\pm} \in \End \fg_{\pm}$, and $u_{\alpha}, u^{\alpha}$
are dual bases of $\fg_+$ and $\fg_- : (u_{\alpha}|
u^{\beta})=\delta_{\alpha \beta}$.  Formula (\ref{eq:5.13})
 is immediate from (\ref{eq:5.12}).

\noindent Likewise, we have
$\str_{\fg_0} (\ad \, v) (\ad \, u)=(\Omega_0 v|u)$.  Hence
\begin{equation}
  \label{eq:5.14}
  \str_{\fg_0} (\ad \, v) (\ad \, u) =
  h_{1/2} (v|u) \hbox{  if  } u \in \fg_{1/2} \, , \,
  v \in \fg_{-1/2} \, .
\end{equation}

\end{remark}

Since $[v,u]=[v,u]^{\natural}$ - $(v|u)x$, $u \in \fg_{1/2}$, $v
\in \fg_{-1/2}$, where the superscript $\natural$ denotes the orthogonal
projection of $\fg_0$ on $\fg^{\natural}$,we obtain, using (\ref{eq:5.6}):
\begin{equation}
  \label{eq:5.15}
  \str_{\fg_+} \ad [v,u]=- (v|u) (h\spcheck -1) \, .
\end{equation}
It follows from (\ref{eq:5.13}),  (\ref{eq:5.14}),
(\ref{eq:5.15}) and Lemma \ref{lem:5.1} that in the case of
$W_k (\fg ,e_{-\theta}) $ the last term in the formula for $G^{\{v\}}$
in Theorem~\ref{th:2.1} is equal to
\begin{displaymath}
  -(k+1)\sum_{\beta \in S_{1/2}} (v|u_{\beta})\partial \Phi^{\beta} =
  2(-1)^{p(v)}(k+1)\partial \Phi_{[e,v]} \, .
\end{displaymath}

\begin{remark}
  \label{rem:5.2}
If $\kappa_{\fg_0} (a,b) = 2h\spcheck_0 (a|b)$ for all $a,b \in
[\fg^{\natural} , \fg^{\natural}]$, in particular, if
$[\fg^{\natural}, \fg^{\natural}]$ is simple, we have the
following relation: 
\begin{displaymath}
2+h_{1/2} \sdim \fg_{1/2} =h\spcheck \sdim \fg_0
     -h\spcheck_0 \sdim [\fg^{\natural} ,\fg^{\natural}] \, ,
\end{displaymath}
which is obtained by calculating
$\str_{\fg_+} \Omega_0$ using (\ref{eq:5.12}). This relation allows one
to compute $h\spcheck_0$. Note also that in this case we have
$u,v \in \fg_{-1/2}$:

\alphaparenlist
 \begin{enumerate}
 \item %%a
$\sum_{\gamma \in S_{1/2}} c' ([u,u^{\gamma}]^\natural ,[u_{\gamma},v]^{\natural}) =
(e|[u,v]) (k+\frac{h\spcheck - h\spcheck_0}{2})(\sdim \fg_{1/2}
+1)$,

\item %%b
  $c(u,v) =- (e|[u,v]) \{ (k +h\spcheck) c (k) -
  (k+\frac{h\spcheck -h\spcheck_0}{2}) (\sdim \fg_0 +\sdim \fg_{1/2})\}$.
 \end{enumerate}
It turns out that the dual Coxeter
numbers $h\spcheck_{0,i}$ for $\fg$ from Table~I are equal to the
usual Coxeter numbers of the Lie algebras $\fg^{\natural}_i$ (as
in Table~I), except for $\fg =G_2$, in which case
$\fg^{\natural}=\sl_2$ and $h\spcheck_0 =2/3$.  For Tables~II
and~III the numbers $h\spcheck_{0,i}$ are given in Table~$H_0$.

\vspace{3ex}
%%%%\begin{table*}[htbp]
%%% \caption{$H_0$}
  \centering{Table $H_0$} \small
  \begin{tabular}{@{}c@{\;}|@{\;}c@{\;}|@{\;}c@{\;}|@{\;}c@{\;}|@{\;}
                  c@{\;}|@{\;}c@{\;}|@{\;}c@{\;}|@{\;}c@{\;}|@{\;}
                  c@{\;}|@{\;}c@{\!}}
\hline
   $\fg$ &\hbox{\Large{ $\sl (m|n ) \atop  m \neq n$}} & $\sl
   (m|m)/\CC I$  & \hbox{\Large$spo (n|m) \atop(n,m) \neq (2,4)$}
   & $osp (4| n)$ &\hbox{\Large{ $osp (m|n) \atop m \geq 5$}}
    & $F(4)$ & $G(3)$ & $F(4)$ & $G(3)$\\[2ex]
\hline
   $\fg^{\natural}_i$ & $\sl (m-2|n)$ & $\sl (m-2|m)$
   & $spo (n-2|m)$ &\hbox{\Large{ $\sl_2 \atop sp_n$}}
   &\hbox{\Large{ $\sl_2  \atop osp (m-4|n)$}} & $so_7$
   & $G_2 $ & $ D(2,1;a)$ & $ osp (3|2)$\\[2ex]
\hline
$h\spcheck_{0,i}$ & $m-n-2$ & $ -2$ &  $(n-m)/2$
& \hbox{\Large{$2\atop -1-n/2$}} & \hbox{\Large{$ 2\atop m-n-6$}}&
$-10/3$ & $-3$& $0$& $-2/3$\\[2ex]
\hline
  \end{tabular}

%%%\end{table*}
%

\end{remark}

\vspace{3ex}

Theorem~\ref{th:5.1} implies a ``free field realization'' of all
vertex algebras $W_k (\fg ,e_{-\theta})$.  In order to state the
result, recall the $2$-cocycle $\alpha_k$ on $\fg_0
[t,t^{-1}]$ (cf. (\ref{eq:2.5}) and (\ref{eq:5.5})):
\begin{equation}
\label{eq:5.16}
  \alpha_k (at^m, bt^n) = m\delta_{m,-n} ((k+h\spcheck)(a|b)-\tfrac{1}{2}
\kappa_{\fg_0}(a,b))\, .
\end{equation}
%
%where $\delta_x (a,b) =0$ if $a$ or $b$ lies in $\fg^{\natural}$
%and $\delta_x (x,x)=1$.
Let $\hat{\fg}_0=\fg_0 [t,t^{-1}]+\CC 1$ be the affine Lie
superalgebra corresponding to this cocycle $(a,b \in \fg_0 \,
,\,\,  m,n \in \ZZ)$:
\begin{displaymath}
  [at^m ,bt^n] = [a,b]t^{m+n} +\alpha_k (at^m,bt^n)1 \, ,
\end{displaymath}
and let $V_{\alpha_k} (\fg_0)$ be the corresponding universal
affine vertex algebra.

\begin{theorem}
  \label{th:5.2}
The following formulas define a vertex algebra homomorphism of
$W_k (\fg ,e_{-\theta})$ to $V_{\alpha_k} (\fg_0) \otimes
F(A_{\ne})$:
\begin{eqnarray*}
&&  J^{\{ a \}} \mapsto a + \frac{(-1)^{p(a)}}{2}
      \sum_{\alpha \in S_{1/2}}: \Phi^{\alpha}
      \Phi_{[u_{\alpha},a]}: (a \in \fg^{\natural}), \\
&&      G^{\{ v \}} \mapsto \sum_{\alpha \in S_{1/2}}
         : [v,u_{\alpha}]\Phi^{\alpha}:
         -(k+1)\sum_{\alpha \in S_{1/2}}
         (v|u_{\alpha})  \partial \Phi^{\alpha}\\
&&         -\frac{(-1)^{p(v)}}{3} \sum_{\alpha ,\beta \in S_{1/2}}
         : \Phi^{\alpha}
         \Phi^{\beta}\Phi_{[u_{\beta},[u_{\alpha},v]]}
         : (v \in \fg_{-1/2})\, , \\
&&         L \mapsto \frac{1}{2 (k+h\spcheck)}
         \sum_{\alpha \in S_0} (-1)^{p(\alpha)}:u_{\alpha}
         u^{\alpha}:+ \frac{k+1}{k+h\spcheck}\partial x +\frac{1}{2}
         \sum_{\alpha \in S_{1/2}} : \partial \Phi^{\alpha}
         \Phi_{\alpha}: .
\end{eqnarray*}

\end{theorem}

\begin{proof}
  Let $\hat{\fg}_{\leq}$ be the affine superalgebra associated to
  the Lie superalgebra $\fg_{\leq}$ and the $2$-cocycle $\alpha_k$
  on $\fg_{\leq} [t,t^{-1}]$ extended from $\fg_0 [t,t^{-1}]$ to
  $\fg_\leq  [t,t^{-1}]$ trivially.  Theorems~\ref{th:2.1} and
  \ref{th:5.1} give us a realization of the vertex algebra
  $W_k(\fg ,e_{-\theta})$ as a subalgebra of the vertex algebra $
 V_{\alpha_k} (\fg_{\leq}) \otimes F(A_{\ne})$.  The realization
given by Theorem~\ref{th:5.2} is given by the homomorphism of
$V_{\alpha_k} (\fg_{\leq}) \otimes F(A_{\ne})$ to $V_{\alpha_k}
(\fg_0) \otimes F(A_{\ne})$ induced by the canonical homomorphism
$\fg_{\leq} \to \fg_0$.  (We use also Remark~5.1 to simplify a
coefficient in $G^{\{ v \}}$.)
 \end{proof}

\begin{remark}
\label{rem:5:3}
Replacing $V_{\alpha_k}(\fg_0)$ and $F(A_{\ne})$ in Theorem~\ref{th:5.2} by 
their twisted versions,
we obtain free field realizations of twisted superconformal algebras,
in particular, the Ramond sector (see \cite{KW5} for details).
\end{remark}

\section{Highest weight modules over $W_k (\fg ,x,f)$}
\label{sec:6}

Let $V$ be a vertex algebra with a conformal vector $\nu$, that
is the field $Y(\nu ,z)$, corresponding to $\nu$, is of the form
$Y(\nu ,z)=\sum_{n \in \ZZ} L_n z^{-n-2}$, where the $L_n$ form
 a representation of the Virasoro algebra with central charge
$c$, $L_{-1}$ coincides with the infinitesimal translation
operator and $L_0$ is diagonalizable on $V$ \cite{K4}.  If $a \in
V$ is an eigenvector of $L_0$ with eigenvalue $=$ conformal weight
$\Delta (a)$, one writes $Y (a,z)=\sum_{n \in \ZZ -\Delta (a)} \, a_n
z^{-n-\Delta(a)}$.  Recall \cite{K4} that one has the following
commutator formula for $Y(a,z)$ and $Y(b,w)$:
\begin{equation}
  \label{eq:6.1}
  [a_m,b_n] =\sum_{j \in \ZZ_+} \binom{\Delta (a) + m-1}{j} \,
       (a_{(j)} b)_{m+n} \, ,
\end{equation}
where $\Delta (a)$ is the conformal weight of $a$ and $a_{(j)}b$ is
the $j$\st{th} product in $V$.  In particular,
\begin{equation}
  \label{eq:6.2}
  [L_0 \, ,\,  a_n]=-na_n \, .
\end{equation}

Recall that a $V$-module is a vector space $M$ and a collection
of $\End M$-valued fields $\{ Y^M (a,z)$ $ = \sum_{n \in
  \ZZ-\Delta_{(a)}} a^M_n z^{-n-\Delta_{(a)}} \}_{a \in V}$, such that the
following properties hold:

\Mlist
\begin{enumerate}
\item %%M1
  $Y^M (\vac ,z) = I_M$, i.e., $\vac^M_n = \delta_{n,0}$,

\item %%M2
  $Y^M (L_{-1}a,z) =\tfrac{d}{dz} Y^M (a,z)$, i.e.,
  $(L_{-1}a)^M_n =-(n + \Delta (a)) a^M_n$,

\item %%M3
  $Y^M (a_{(-1)}b,z) =:Y^M (a,z) Y^M(b,z):$,

\item %%M4
  $[a^M_m , b^M_m] =\sum_{j \in \ZZ_+} \binom{\Delta (a) +m-1}{j}
  \, (a_{(j)}b)^M_{m+n}$ (cf.~(\ref{eq:6.1})),

\item %%M5
  $L^M_0$ is diagonalizable on $M$ with eigenvalues bounded below.
\end{enumerate}

Suppose that $V$ is strongly generated by a collection of fields
$\{ J^{\{ i \}} (z) =\sum_{m \in \ZZ-\Delta (i)}$\break $J^{\{ i \}}_m
z^{-m-\Delta (i)} \}_{i \in I}$, where $J^{\{ i \}}$ has conformal
weight $\Delta (i) \in \RR$.  Then by (\ref{eq:6.1}), we have:
\begin{equation}
  \label{eq:6.3}
  [J^{\{ i \}}_m \, , \, J^{\{ j \}}_n] =
  \sum_{\vec{s} , \vec{t}} c^{ij}_{m,n} (\vec{s} ,\vec{t})
  J^{\{ s_1 \}}_{t_1} \,  J^{\{ s_2 \}}_{t_2} \ldots \, ,
\end{equation}
where the sum is finite and for each term of this sum we have:
\begin{equation}
  \label{eq:6.4}
  t_r \in \ZZ -\Delta (s_r) \, , \, \sum_r t_r =m+n
     \hbox{  and  } t_1 \leq t_2 \leq \ldots \, .
\end{equation}

Denote by $\A$ the unital associative superalgebra on generators
$J^{\{ i \}}_m$ $(i \in I , m \in \ZZ-\Delta (i))$ and defining
relations (\ref{eq:6.3}).  Let $\tilde{\A}_-$, $\tilde{\A}_+$ and
$\tilde{\A}_0$ be the subalgebras of $\A$ generated by the $J^{\{
  i \}}_m$ with $m<0$, $m>0$ and $m=0$, respectively.  It follows
from (M4) that any $V$-module $M$ is an $\A$-module.  It follows
from (\ref{eq:6.3}) and (\ref{eq:6.4}) that
\begin{equation}
  \label{eq:6.5}
  \A = \tilde{\A}_- \tilde{\A}_0 \tilde{\A}_+ \, .
\end{equation}
By (\ref{eq:6.2}), $L_0$ lies in the center of $\tilde{\A}_0$,
hence each eigenspace of $L_0$ in $M$ carries a representation of
$\tilde{\A}_0$.

Denote by $\A_0$ the unital associative superalgebra on
generators $J^{\{ i \}}_0$ $(i \in I)$ and the defining relations
(cf.~(\ref{eq:6.3})):
\begin{equation}
  \label{eq:6.6}
  [J^{\{ i \}}_0 \, , \, J^{\{ j \}}_0]=\sum_{\vec{s}}c^{ij}_{00}
  (\vec{s} ,0) J^{\{ s_1 \}}_0 \, J^{\{ s_2\}}_0 \ldots \, .
\end{equation}

It is clear from (\ref{eq:6.3}),  (\ref{eq:6.4}) and
(\ref{eq:6.5}) that the representation of $\tilde{\A}_0$ in the
eigenspace of $L^M_0$ with the minimal eigenvalue induces a
representation of the associative superalgebra $\A_0$.

The proof of the following theorem is immediate from the
above remarks.

\begin{theorem}
  \label{th:6.1}

Let $M$ be an irreducible $V$-module.  Then the representation of $\A_0$ in the eigenspace
of $L^M_0$ with the minimal eigenvalue is irreducible, and this
representation uniquely determines $M$.
\end{theorem}

\begin{example}
  \label{ex:6.1}
  Let $V=W_k (\fg , e_{-\theta})$.  Recall (Theorem~\ref{th:5.1})
  that $V$ is strongly generated by $L$, primary fields $J^{\{ a
    \}}$ $(a \in \fg^{\natural})$ of conformal weight $1$ and
  primary fields $G^{\{ v \}}$ $(v \in \fg_{-1/2})$ of conformal
  weight $3/2$.  The fields $G^{\{ v \}} $ obviously do not
  contribute to $\A_0$, and, by Theorem~\ref{th:2.1}(c), relation
    (\ref{eq:6.6}) is simply $[J^{\{a \}}_0 \, , \, J^{\{ b \}}_0]=
    J^{\{[ a,b ]\}}_0$.  Hence
    \begin{displaymath}
      \A_0 = U (\fg^{\natural}) \otimes \CC [L_0]  \simeq U (\fg_0)\, .
    \end{displaymath}
\end{example}

Recall that, given a restricted $\hat{\fg}$-module $P$ of level
$k$ (i.e.,~$K=k \, I_P$), one constructs the associated $\C (\fg
,x,f,k)$-module
\begin{displaymath}
  \C (P) = P \otimes F (\fg ,x,f)
\end{displaymath}
with the differential $d^P_0= \Res_z \, d^{\C (P)}(z)$.  The
homology $H(P)$ of the complex $(\C (P),\, d^P_0)$ with the
induced charge decomposition (by setting charge$\, P=0$) is a
direct sum of $W_k (\fg ,x,f)$-modules:
\begin{displaymath}
  H(P) = \oplus_{j \in \ZZ} H_j (P) \, .
\end{displaymath}
Thus we get a functor from the category of restricted
$\hat{\fg}$-modules to the category of $\ZZ$-graded $W_k (\fg
,x,f)$-modules \cite{FF2}, \cite{FKW}, \cite{KRW}.

Let $P_0$ be a $\fg$-module on which $x$ is a diagonalizable and
$U(\fg_+)$ is locally finite (for example, a Verma module over
$\fg$).  Extend $P_0$ to a $\fg [t]+\CC D +\CC K$-module by
letting $D$ and $\fg t^n$ for $n \geq 1$ act trivially, and
$K=kI_{P_0}$.  The induced $\hat{\fg}$-module
\begin{equation}
\label{eq:6.7}
  P= U (\hat{\fg}) \otimes_{U (\fg [t]+\CC D + \CC K)} P_0
\end{equation}
is called a \emph{generalized Verma} module over $\hat{\fg}$.
(For example, if $P_0$ is a Verma module over $\fg$, then $P$ is
a Verma module over $\hat{\fg}$.)

\begin{theorem}
  \label{th:6.2}
  Suppose that $(x,f)$ is a good pair (i.e.,~(\ref{eq:1.10})
  holds) and that $P$ is a generalized Verma module over
  $\hat{\fg}$.  Then
  \begin{displaymath}
    H_j(P) =0 \hbox{  if  } j \neq 0 \, .
  \end{displaymath}

\end{theorem}

\begin{proof}
  We have:
  \begin{displaymath}
    \C (P) = \C (\fg ,x,f,k) \otimes P_0
  \end{displaymath}
with the following action of the differential:
\begin{equation}
\label{eq:6.8}
  d^P_0 (a \otimes v) =d_0 (a) \otimes v +
 (\sum_{n \in \ZZ}\,  \sum_{\alpha \in S_+} (-1)^{p(\alpha)}
    \varphi^{\alpha}_{(n-1)} \otimes u_{\alpha (-n)} )
    (a \otimes v) \, ,
\end{equation}
where $a \in \C (\fg ,x,f,k)$, $v \in P_0$.

Recall that we have  decomposition (\ref{eq:4.2}) of the
complex $\C (\fg ,x,f,k)$, hence
\begin{displaymath}
  (\C (P) , d^P_0) = (\C^+ , d_0 )\otimes
(\C^- \otimes P_0 , d^P_0)
\end{displaymath}
is a decomposition of complexes.  Hence by (\ref{eq:4.1}) and the
K\"unneth lemma, we need to prove
\begin{equation}
  \label{eq:6.9}
  H_j (\C^- \otimes P_0 , \, d^P_0) =0
  \hbox{  if  } j \neq 0 \, .
\end{equation}

%Recall that in Section~\ref{sec:4} we introduced a bicomplex
%structure on $(\C^- , \, d_0)$:
%%
%\begin{displaymath}
%  \C^- = \oplus_{m,n \in \tfrac{1}{2}\ZZ} \C^-_{m,n} \, , \,
%     d_0 = d_1 + d_2 \, .
%\end{displaymath}
%
We extend the filtration of $\C^-$, introduced in the proof of
Theorem~4.1, 
 to the complex $(\C^- \otimes
P_0 , \, d^P_0)$ by letting
%
%\begin{displaymath}
%P_0 = \otimes_{j \in
%  \tfrac{1}{2}\ZZ} (P_0)_{j,-j}\, , \hbox{  where  }(P_0)_{j,-j}
%\hbox{  is the
%eigenspace of  } x \hbox{   with the eigenvalue  }j\, ,
%\end{displaymath}
%%
%and letting
%%
%\begin{displaymath}
%  d^P_1 = _1 \otimes 1\, ,\,  d^P_2 = d_2 \otimes 1 + 
%       \sum_{\alpha \in S_+}
%       (-1)^{p(\alpha)} \varphi^{\alpha}_{(-1)} \otimes u_{\alpha (0)} \,.
%\end{displaymath}
%%
%Since $d_1 (\varphi^{\alpha}) =0$, due to Lemma~4.1,
%we have:  $d^P_1 \, d^P_2 + d^P_2 \, d^P_1 =0$.  Also, obviously
%$(d^P_1)^2 =0$, and since $(d^P_0)^2=0$, we conclude that
%$(d^P_2)^2=0$.

%But by Lemma~\ref{lem:4.2}, 
\begin{displaymath}
  \deg v = -j \hbox{ if }v \in P_0 \, \hbox{and} \, 
  \pi (x) v =jv \, .
\end{displaymath}
Then the differential $d^P_0$ induces the differential $d^P_1 =
d_1 \otimes 1$ on $\Gr (\C^- \otimes P_0)$.  Hence it follows
from the proof of Theorem~4.1 that $H_j (\C^- \otimes P_0 \, , \, d^P_1)
= 0$ if $j \neq 0$.  Also the bicomplex $\C^- \otimes P_0$ is
locally finite since $P_0$ is a locally finite
$U(\fg_+)$-module.  Hence the spectral sequence converges to the
homology of $(\C^- \otimes P \, , \, d^P_0)$ and (\ref{eq:6.9})
holds.

\end{proof}

Let $\fh^{\natural}$ be a maximal diagonalizable subalgebra of the
even part of $\fg^{\natural}$, and include $\fh^{\natural}$ in a
Cartan subalgebra $\fh$ of the even part of $\fg_0$.  Let
$\Delta_j\subset \fh^*$ be the set of roots of $\fh$ in $\fg_j$ and let
$\Delta_{0+}$ be a subset of positive roots of $\Delta_0$.  Then
$\fh$ is a Cartan subalgebra of the even part of $\fg$ and
$\Delta = \coprod_j \Delta_j$ is the set of roots of $\fh$ in
$\fg$, $\Delta_+ := \Delta_{0+} \coprod (\coprod_{j>0} \Delta_j)$ being
the subset of positive roots.  Denoting by $\fn_{0+}$, $\fn_{0-}$, $\fn_+$ and
$\fn_-$ the sum of all root spaces corresponding to $\Delta_{0+}$,
$-\Delta_{0+}$, $\Delta_+$ and $-\Delta_+$, we obtain the
triangular decompositions:
\begin{equation}
  \label{eq:6.10}
  \fg = \fn_- \oplus \fh \oplus \fn_+ \, \quad
  \fg_0 = \fn_{0-} \oplus \fh \oplus \fn_{0+} \, .
\end{equation}

Recall that a Verma module over $\fg$ with highest weight $\lambda
\in\fh^*$ is the module
$U(\fg)\otimes_{U(\fh +\fn_+)} \CC v_{\lambda}$, where
$\CC v_{\lambda}$ is a $1$-dimensional $\fh +
\fn_+$-module such that $\fn_+ v_{\lambda}=0$  and $hv_{\lambda}
=\lambda (h) v_{\lambda}$ for all $h \in \fh$.

Let $\hat{\fh}=\fh + \CC K + \CC D$ be the Cartan subalgebra of
$\hat{\fg}$.  We extend the
bilinear form $(. \, |\, . )$ from $\fh$ to  $\hat{\fh}$ by
letting $(\fh | \CC K+\CC D)=0$, $(D|D)=(K|K)=0$,
$(K|D)=(D|K)=1$, and identify $\hat{\fh}$ with $\hat{\fh}^*$
using this form.  Given $k$, we extend $\lambda \in \fh^*$ to
$\hat{\lambda} \in \hat{\fh}^*$ letting $\hat{\lambda}|_{\fh}
=\lambda$, $\lambda (K) =k$, $\lambda (D)=0$.

A Verma module
over $\hat{\fg}$ with highest weight $\hat{\lambda}$ of level $k$ is a generalized Verma module
(\ref{eq:6.7}) for which $P_0$ is a Verma $\fg$-module with
highest weight $\lambda$.  Any quotient of a Verma
$\hat{\fg}$-module with highest weight $\hat{\lambda}$ is called
a highest weight $\hat{\fg}$-module.

Recall that the Euler--Poincar\'e character of the $\ZZ$-graded
$W_k (\fg ,x,f)$-module $H(P)$ is defined by:
\begin{displaymath}
  \ch_{H(P)}(h) = \sum_{j \in \ZZ} (-1)^j {\rm tr}_{H_j(P)}
     q^{L_0} e^{J^{\{ h \}}_0} \, , \, h \in \fh^{\natural} \, .
\end{displaymath}
As usual, we define the character of a $\hat{\fg}$-module $P$ by
$\ch_P (H) ={\rm tr}_P e^H$, where $H \in \hat{\fh}$.    We let, as usual, $\hat{\rho}=h\spcheck
D+\rho$.  Let $P$ be a highest weight $\hat{\fg}$-module with highest
weight $\hat{\lambda}$ of level $k$.  Due to
\cite{KRW} (see Remark 3.1 there), we have the following formula for
$\ch_{H(P)}$ in terms of $\ch_P$:
\begin{eqnarray}
  \label{eq:6.11}
 \lefteqn{\hspace{-5in} \ch_{H(P)} (h) =
    \frac{q^{\frac{(\hat{\lambda}|\hat{\lambda}+2\hat{\rho})}
        {2(k+h\spcheck)}}}{\prod^{\infty}_{j=1}(1-q^j)^{\dim \fh}}
    \hat{R} \, \ch_P\,  (2\pi i (-\tau D-\tau  x)+h)}\\
\nonumber
  \times \prod^{\infty}_{n=1} \,( \prod_{\alpha \in \Delta_{0+}}
        (1-s(\alpha) q^{n-1} e^{-\alpha (h)})^{-s(\alpha)}
        (1-s(\alpha)q^n e^{\alpha (h)})^{-s(\alpha)}
\prod_{\alpha \in \Delta_{1/2}}\!
  (1-s(\alpha)q^{n-\frac{1}{2}}e^{\alpha (h)})^{-s(\alpha)}) \, ,
\end{eqnarray}
where $\hat{R}$ is the Weyl denominator for $\hat{\fg}$,
$s(\alpha)=(-1)^{p(\alpha)}$, $q=e^{2\pi i \tau}$, $h \in
\fh^{\natural}$.

Now we shall define a Verma module over the vertex algebra $W_k (\fg
,x,f)$, assuming that the pair $(x,f)$ is good.  The isomorphism
(\ref{eq:1.12}) and the triangular decomposition (\ref{eq:6.10})
of $\fg_0$ induce the following decomposition of $\fg^f$ as
$\fh^{\natural}$-modules:
\begin{equation}
  \label{eq:6.12}
  \fg^f = \fn_{0-} \oplus \fh \oplus \fn_{0+} \oplus \fg_{1/2} \, ,
\end{equation}
compatible with the gradation (\ref{eq:1.10}).

\begin{example}
  \label{ex:6.2}

In the case of a minimal gradation we have $\fg^{\natural} =
\fn_{0-} \oplus \fh^{\natural} \oplus \fn_{0+}$ and $\fh =
\fh^{\natural} + \CC x$, so that the decomposition
(6.12) is identified with (\ref{eq:5.2}) via the
identification of $x$ with $f$ and $\fg_{1/2}$ with $\fg_{-1/2}$
given by $\ad f$.

\vspace{1ex}

Choose a basis $\{ a_i | i=1,\ldots ,s\}$  of $\fg^f$ compatible with the
decomposition (\ref{eq:6.12}), the root space decomposition with respect  to
$\fh^{\natural}$ of (\ref{eq:6.12}) and the gradation (\ref{eq:1.10}).  Let
$J^{\{ a_i \}} (z) =\sum_{n \in \ZZ - \Delta (i)} J^{\{ i \}}_n z^{-n-\Delta (i)}$ be the
field of $W_k (\fg,x,f)$ corresponding to $a_i$ (given by
Theorem~\ref{th:4.1}a).  Recall that the conformal weight
$\Delta (i) =1+j$ if $a_i \in \fg_{-j}$ $(j \in \tfrac{1}{2}\ZZ_+)$.
  We order the $a_i$ in such a way that $a_1,\ldots ,a_r=x$ form
  a basis of $\fh$.  A $W_k(\fg ,x,f)$-module $M$ is called a \emph{highest weight
  module} with \emph{highest weight} $\lambda = (\lambda (a_1),
\ldots ,\lambda (a_r)) \in \CC^r$, if there
exists a non-zero vector $v$ such that:

\begin{list}{}{}

\item %%HW1
\hspace*{-6ex} (HW1)\qquad $W_k (\fg ,x,f) v =M $,

\item %%HW2
 \hspace*{-6ex} (HW2)\qquad $J^{\{ i \}}_0 v=\lambda (a_i) v$ if $a_i \in \fh$,

\item %%HW3
\hspace*{-6ex}(HW3)\qquad $J^{\{ i \}}_m v=0 $ if $m>0$, or $m=0$ and $a_i \in \fn_{0+}$.
\end{list}

\end{example}

Note that the following vectors span a highest weight $W_k (\fg
,x,f)$-module $M$:
\begin{equation}
  \label{eq:6.13}
  \left( J^{\{ i_1 \}}_{-m_1}\right)^{b_1} \cdots
   \left( J^{\{ i_s \}}_{-m_s}\right)^{b_s} v \, ,
   \hbox{  where  } b_i \in \ZZ_+ \, , \,
   b_i \leq 1 \hbox{  if  }a_i \hbox{  is odd,  } m_i >0 \, ,
   \hbox{  or  }  m_i=0 \hbox{  and  }a_i \in \fn_{0-} \, .
\end{equation}
A highest weight $W_k (\fg ,x,f)$-module $M$ is called a Verma
module if the vectors (\ref{eq:6.13}), where the sequence of
pairs $(i_1,m_1)$, $(i_2,m_2), \ldots$ decreases in
lexicographical order, form a basis of $M$
(cf. Remark~\ref{rem:4.2}).  It is clear that any irreducible
$W_k (\fg ,x,f)$-module is a highest weight module, hence is a
quotient of the Verma module with the same highest weight.

\begin{theorem}
  \label{th:6.3}

  Let $P$ be a Verma $\hat{\fg}$-module with highest weight
  $\hat{\lambda}\in \hat{\fh}^*$, where $\hat{\lambda} |_{\fh}
  =\lambda$, $\hat{\lambda}(K) =k$, $\hat{\lambda}(D) =0$.  Then
  $H(P) = H_0(P)$, and it is a Verma $W_k (\fg ,x,f)$-module with highest
  weight $\lambda_W \in \fh^*$ such that
\begin{equation}
  \label{eq:6.14}
  \lambda_W |_{\fh^{\natural}} =\lambda |_{\fh^{\natural}} \hbox{  and  }
  \lambda_W(x) =
  \frac{(\hat{\lambda}|\hat{\lambda}+2 \hat{\rho})}
     {2 (k+h\spcheck)}-\lambda (x) \, .
\end{equation}
\end{theorem}

\begin{proof}
  It is clear from (\ref{eq:6.8}) that the vector
  $v=v_{\hat{\lambda}} \otimes \vac$, where $v_{\hat{\lambda}}$
  is a highest weight vector of $P$ and $\vac$ is the vacuum
  vector of $F(\fg ,x,f)$, is $d^P_0$-closed in $\C (P)$.
  Furthermore, it is easy to see that $W_k (\fg ,x,f)v$ is a highest weight
  submodule of the $W_k (\fg ,x,f)$-module $H(P)$ with a highest weihgt vector $v$.

Next, it is clear from Theorem~\ref{th:4.1} that $W_k (\fg
,x,f)v$ is a Verma module.  Due to Theorem~\ref{th:6.2}, it
remains to show that the Euler--Poincar\'e character of $H(P)$ coincides with the character of a $W_k (\fg ,x,f)$-Verma module.
Since $\hat{R} \ch_P =e^{\hat{\lambda}}$ by definition of a Verma
$\hat{\fg}$-module $P$ with highest weight $\hat{\lambda}$, we
obtain from (\ref{eq:6.11}) the following formula $(h \in \fh^{\natural})$:
\begin{eqnarray*}
 && \ch_{H(P)}(h) = e^{\lambda (h)}q^{
  \frac{(\hat{\lambda}|\hat{\lambda}+2\hat{\rho})}
  {2(k+h\spcheck)}-(\lambda |x)} \!\!
  \prod^{\infty}_{j=1}(1-q^j)^{-\dim \fh } \\
  &&\times \!\! \prod^{\infty}_{n=1}( \prod_{\alpha \in
    \Delta_{0+}} \!\! (1-s(\alpha)
   q^{n-1} e^{-\alpha (h)})^{-s(\alpha)}
   (1-s(\alpha)q^n e^{\alpha (h)})^{-s(\alpha)}
\!\!\!\! \prod_{\alpha \in \Delta_{1/2}}
  (1-s(\alpha )  q^{n-\tfrac{1}{2}} e^{\alpha
    (h)})^{-s(\alpha)}) \, .
\end{eqnarray*}
But, in view of (6.12) and (6.13), this is precisely the
character of a $W_k (\fg ,x,f)$-Verma module with highest weight
$\lambda_W$ satisfying (\ref{eq:6.14}).

\end{proof}

\begin{corollary}
  \label{cor:6.1}
\alphaparenlist
\begin{enumerate}
\item %%% (a)
For any $\lambda \in \CC^r$ there exists a Verma $W_k (\fg
  ,x,f)$-module with highest weight $\lambda$.

\item %%b
  If $a_i,a_j \in \fh$, then $[J^{\{ i \}}_0, J^{\{j\}}_0]=0$ in
  $\A_0$.
\end{enumerate}

\end{corollary}

\begin{proof}
  (a)~is immediate by Theorem~\ref{th:6.3}.  (b)~follows from (a)
  and (\ref{eq:6.6}) by induction on the conformal weight of
  $J^{\{ a_i \}}$, using that $J^{\{ i \}}_0 v_{\lambda}=\lambda
  (a_i)v$ if $v$ is the highest weight vector (cf.~\cite{FKW}).
\end{proof}

In conclusion of this section, we construct an anti-involution
$\omega$ of the associative superalgebra $\A$ and the
corresponding contravariant form on any highest weight module
over $W_k (\fg ,x,f)$.  We call an \emph{anti-involution} of an
associative superalgebra (resp. Lie superalgebra) a vector superspace
 automorphism $\omega$ such that $\omega ^2 = 1$ and $\omega (ab) = \omega (b) \omega
 (a)$ (resp. $\omega [a,b]] = [\omega (b) ,\omega (a)]$).

First, we construct an anti-involution $\omega$ of the Lie
superalgebra $\fg $ such that the following properties hold:
  $\omega (x) =-x$, $\omega (a)=a$ if $a\in \fh^{\natural}$, $\omega (f) =f$,
   $\omega (\fg_{\alpha})=\fg_{-\alpha}$ if $\alpha \in \Delta_0$,
  $(\omega (a) | \omega (b)) = (b|a)$, $a,b \in \fg$.

Next, we lift $\omega$ to an anti-involution of the affine Lie
superalgebra $\hat{\fg}$ by letting $\omega (a_n) =\omega
(a)_{-n} +2K (a|x) \delta_{0,n}$ (the extra term occurs because
the field $x(z)$ is not primary), $\omega (K) = K$, $\omega (D)
=D$. Recall that we write $a(z)=\sum_{n \in \ZZ - \Delta}a_n z^{-n-\Delta}$,
where $\Delta$ is the conformal weight of $a(z)$.
Furthermore, $\omega$ induces a super vector space
automorphism of $A_{\ne}$ such that $\langle \omega (a),\omega
(b) \rangle_{\ne} = \langle b,a \rangle_{\ne}$.  Also $\omega$
induces a super vector space automorphism of $A$, which in turn
induces an automorphism of $A^*$ by $\omega (b^*) (a)
=-(-1)^{p(a)p(b)} b^* (\omega(a))$, so that we have on
$A_{\ch}$:  $\langle
\omega (a),\omega (b) \rangle_{\ch} =\langle b,a \rangle_{\ch}$.
We extend this $\omega$ to an anti-involution of $\hat{A}_{\ne}$
(resp.~$\hat{A}_{\ch}$)
 by letting $\omega (\Phi_n)=\omega
(\Phi)_{-n}$, $\Phi \in A_{\ne}$ (resp.~$\omega
(\varphi_n)=-(-1)^{p(\varphi)} \omega (\varphi)_{-n}$, $\omega
(\varphi^*_n) =\omega (\varphi^*)_{-n}$, $\varphi \in
A$, $\varphi^* \in A^*$),
$\omega (K) =K$.  Next, $\omega$ induces an
anti-automorphism of the tensor product of $U_k (\hat{\fg})=U
(\hat{\fg})/(K-k)$, $U_1 (\hat{A}_{\ne})/(K-1)$ and $U_1
(\hat{A}_{\ch})/(K-1)$.  The associative superalgebra $\A (\C)$
corresponding to the vertex algebra $\C (\fg ,x,f,k)$ is the
``local completion'' of this tensor product, i.e.,~it is generated
by the Fourier coefficients of normal ordered products of
currents $a(z)$, $a\in \fg$, ghosts $\varphi_{\alpha}(z)$,
$\varphi^{\alpha}(z)$, $\Phi_{\alpha}(z)$ and their derivatives.
We extend the anti-involution $\omega$ to $\A (\C)$ in
the obvious way.

In order to compute the action of $\omega$ on normally ordered
products, we need the following lemma.

\begin{lemma}
  \label{lem:6.1}
Let $U$ be an associative superalgebra and let $\omega$ be its
anti-involution.  Let $a(z) =\sum_{n \in \ZZ -\Delta_a}
a_nz^{-n-\Delta_a}$, $b(z) =\sum_{n \in \ZZ -\Delta_b}
b_nz^{-n-\Delta_b}$ be two $U$-valued formal distributions such that
$\Delta_a$, $\Delta_b \in \tfrac{1}{2} \ZZ$, and
let $: ab : (z) = \sum_{n \in \ZZ -\Delta_a-\Delta_b} : ab :_n
z^{-n-\Delta_a-\Delta_b}$ be their normally ordered product.
Define the formal distribution $\omega (a)$ (and similarly $\omega (b)$)
by $\omega (a)(z) = \sum_{n \in -\Delta_a + \ZZ} \omega (a_{-n}) z^{-n-\Delta_a}$.
Then:
\begin{displaymath}
  \omega (: ab :_n) =:\omega (b) \omega (a) :_{-n}
    +\sum_{k \in \ZZ_+} \binom{-n-\Delta_a+\Delta_b}{k+1} (\omega (b)_{(k)}\omega (a))_{-n}\, .
\end{displaymath}

\end{lemma}

\begin{proof}
  We have for $n \in -\Delta_a -\Delta_b +\ZZ$:
  \begin{displaymath}
    : ab :_n =\sum_{k \in -\Delta_a -\ZZ_+} a_k b_{n-k}
    +(-1)^{p(a)p(b)} \sum_{k \in -\Delta_a +1+\ZZ_+}
    b_{n-k} a_k \, .
  \end{displaymath}
We let $n_{ab} =-n-\Delta_a +\Delta_b$ to simplify notation (note that it is an integer).
Applying $\omega$ to both sides and replacing $k$ by $j+n$, where
$j \in -\Delta_b +\ZZ$, we obtain:
\begin{displaymath}
  \omega (: ab :_n) =\sum_{j+\Delta_b \leq n_{ab}}
  \omega (b)_j \omega (a)_{-j-n} + (-1)^{p(a)p(b)}
  \sum_{j+\Delta_b >n_{ab}} \omega (a)_{-j-n} \omega (b)_j \, .
\end{displaymath}
First, consider the case $n_{ab} <0$.  Then the last equality can
be rewritten as follows:
\begin{displaymath}
  \omega (: ab :_n) =\omega (b) \omega (a) :_{-n}
  -\sum_{n_{ab}<j+\Delta_b \leq 0}
  [\omega (b)_j , \omega (a)_{-j-n}] \, .
\end{displaymath}
The commutator formula~(6.1)  gives
the result in this case, by making use of the following identity that holds
for any negative integer $A$ and non-negative integer $k$:
$\sum^{-1}_{s=A} \binom{s}{k} = -\binom{A}{k+1}$.  The case $n_{ab} \geq 0$ is treated similarly.

\end{proof}

Usng Lemma~6.1, it is straightforward to check the following
formulas:
\begin{eqnarray}
  \label{eq:6.15n}
      && \omega (L_n)=L_{-n} \, , \\[1ex]
  \label{eq:6.16n}
      && \omega (v^{\ch}_n)=\omega (v)^{\ch}_{-n}+
         \delta_{j0}\delta_{n0}(2h\spcheck (x|v)-\str_{\fg_+}{\ad}v)
         \hbox{  if  } v \in \fg_j \, , \\[1ex]
  \label{eq:6.17n}
      && \omega (v^{\ne}_n) = (\omega v)^{\ne}_{-n}
         \hbox{  if  } v \in \fg^{\natural} \, , \\[1ex]
  \label{eq:6.18n}
      && \omega (J^{(v)}_n) = J^{(\omega (v))}_{-n} +
          \delta_{j0}\delta_{n0} (2(k+h\spcheck) (x|v)
          -\str_{\fg_+}\adv) \hbox{  if  }v\in \fg_{j} \, , \\[1ex]
  \label{eq:6.19n}
      && \omega (J^{\{ v \}}_n) = J^{\{ \omega (v) \}}_{-n}
         \hbox{  if  }v \in \fg^{\natural} \, .
%
%  \label{eq:6.20n}
%      && \omega (G^{\{v \}}_n) = G^{\{ \omega (v) \}}_{-n}
%          \hbox{  if  } v \in \fg^f_{1/2}\, .
%%
\end{eqnarray}

It is also straightforward to check that $\omega (d_n)=d_{-n}$,
so that $d_0$ is fixed by
$\omega$. Hence $\omega$ induces an anti-involution of the
associative superalgebra $\A$ corresponding to the vertex
algebra $W_k (\fg ,x,f)$, which we again denote by $\omega$.
It follows from (6.18) that, for $v \in \fg^{f}_{-1/2}$, we have:
 $\omega (G^{\{ v \}})=J^{(\omega (v))} + \cdots$, where $\cdots$
 denote a field from the subalgebra strongly generated by the
 $J^{(a)}$, $a \in \fg_0$, and the $\Phi_{\alpha}$.  Since the
 fields $G^{\{ \omega (v)\}}$ and $\omega (G^{\{ v \}})$ are
 $d_0$-closed, it follows from Remark~\ref{rem:4.1} that $\omega
 (G^{\{ v \}}) = G^{\{ \omega (v)\}}$, hence we obtain
 \begin{equation}
   \label{eq:6.20n}
    \omega (G^{\{v \}}_n) = G^{\{ \omega (v) \}}_{-n}
          \hbox{  if  } v \in \fg^f_{-1/2}\, .
 \end{equation}

Now let $M$ be a Verma  module over $W_k (\fg ,x,f)$ with
highest weight vector $v_{\lambda}$.  Any vector $v \in M$ can be
written uniquely  in the form $v=\langle v \rangle
v_{\lambda}+v'$, where $\langle v \rangle \in \CC$ is called the
\emph{expectation value} of $v$, and $v'$ is a linear combination
of  weight vectors $v_{\mu}$ with  $\mu \neq \lambda$.  The basic
property of the expectation value, which follows from the
decomposition (\ref{eq:6.5}) is
\begin{equation}
  \label{eq:6.15}
  \langle \omega (a) v_{\lambda}\rangle =
  \langle av_{\lambda}\rangle\, , \, \quad a \in \A \, .
\end{equation}
We define
the \emph{contravariant bilinear form} $B (. \, , \, .)$ on $M$ by the
formula:
\begin{displaymath}
  B (av_{\lambda},bv_{\lambda})
  = \langle \omega (a)bv_{\lambda} \rangle, \hbox{  where  }
  a,b \in \A \, .
\end{displaymath}
By the definition, this bilinear form is contravariant:
\begin{equation}
  \label{eq:6.16}
  B (au,v)= B(u,\omega (a) v)\, , \quad u,v \in M\, , \, a \in \A\, ,
\end{equation}
we have
\begin{equation}
  \label{eq:6.17}  B (v_{\lambda}, v_{\lambda})=1 \, ,
\end{equation}
and it follows from (\ref{eq:6.15}) that it is symmetric:
\begin{equation}
  \label{eq:6.18}
B(u,v) = B(v,u) \, , \quad u,v \in M \, .
\end{equation}
The properties (\ref{eq:6.16})---(\ref{eq:6.18}) determine the
form  $B(. \,,\,  . )$ uniquely.

As usual the maximal submodule of the Verma module $M$ coincides
with the kernel of the form $B (. \,,\, . )$.

\section{Highest weight modules over vertex algebras $W_k (\fg
  ,e_{-\theta})$ and the determinant formula.}
\label{sec:7}

\begin{theorem}
  \label{th:7.1}

\alphaparenlist

\begin{enumerate}

\item %a
The map that associates to an irreducible $W_k (\fg
,e_{-\theta})$-module $M$ the lowest eigenvalue $h$ of $L_0$
and the representation of the Lie superalgebra $\fg^{\natural}$
in the corresponding eigenspace $M_{h}$, given by $a \mapsto
J^{\{ a \}}_0$, $a \in \fg^{\natural}$, is a bijection between
irreducible modules of $W_k (\fg ,e_{-\theta})$ and the set of
pairs $(h,\pi)$, where $\pi$ is an irreducible representation of
$\fg^{\natural}$.

\item %%b
  All Verma $W_k (\fg ,e_{-\theta})$-modules are of the form $H_0
  (P)$, where $P$ is a Verma $\hat{\fg}$-module with highest
  weight $\hat{\lambda}$.  The highest weight $\lambda_W$ of such
  a $W_k (\fg ,e_{-\theta})$-Verma module is completely
  determined by (\ref{eq:6.14}), the number $\lambda_W (x)$ being
  the lowest eigenvalue of $L_0$.

\item %%c
  The Euler--Poincar\'e character $\ch_{H(P)}$ is non-zero if and
  only if $t^{-1}e_{\theta}$ is not locally nilpotent in the
$\hat{\fg}$-module
$P$.

\end{enumerate}

\end{theorem}

\begin{proof}
  (a) and (b) follow from Theorems~\ref{th:6.1}, \ref{th:6.3},
    Example~6.1 and (\ref{eq:6.3}).  (c) follows from \cite{KRW},
    Theorem~3.2.
\end{proof}

Consider now a Verma module $M$ over the vertex algebra $W_k (\fg
,e_{-\theta})$.  Its highest weight $\Lambda_W$ is a pair
$\Lambda,h$, where $\Lambda \in \fh^{\natural *}$ and $h \in \CC$
is the lowest (i.e.,~$\Re h$ is minimal) eigenvalue of $L_0$.  We
have:
\begin{displaymath}
  J^{\{ a \}}_0 v_{(\Lambda ,h)}=\Lambda (a) v_{(\Lambda ,h)} \,,
  \quad a \in \fh^{\natural} \, , \, L_0 v_{(\Lambda ,h)}
  = hv_{(\Lambda ,h)} \, ,
\end{displaymath}
where $v_{(\Lambda ,h)}$ is the highest weight vector of $M$.
With respect to the commuting pair $(\fh^{\natural} ,L_0)$, $M$
decomposes into a direct sum of weight spaces:
\begin{equation}
  \label{eq:7.1}
  M=\oplus_{\substack{\lambda \in \fh^{\natural *}\\ m \in
     \tfrac{1}{2} \ZZ_+}}
  M_{(\lambda , h+m)} \, ,
\end{equation}
so that $M_{(\Lambda ,h)} =\CC v_{(\Lambda ,h)}$.  It is clear
that these weight spaces are mutually orthogonal with respect to
the contravariant form $B (. \, , \, .)$.

Denote by $\natural$: $\alpha \mapsto \alpha^{\natural}$ the projection
of $\fh$ on $\fh^{\natural}$
with respect to the decomposition (\ref{eq:5.3}).  Then we have
for $\alpha \in \fh$:
\begin{equation}
  \label{eq:7.2}
  \alpha = (\alpha | x)\theta +\alpha^{\natural}\, .
\end{equation}

Let $\Delta^{\natural} \subset \fh^{\natural *}$ be the set of
roots of $\fg^{\natural}$ with respect to the Cartan subalgebra
$\fh^{\natural}$ and let $\Delta^{\natural}_+$ be the subset of
positive roots compatible with that of $\fg$. It follows from
(\ref{eq:5.10}) that the ``rho'' for $\Delta^{\natural}_+$, i.e. the half
of the difference between the sums of the sets of even and odd roots
from $\Delta^{\natural}_+$, coincides with $\rho^{\natural}$.
Let $\Delta' \subset \fh^{\natural *}$ be
the set of weights of $\fg^{\natural}$ in $\fg_{1/2}$.

We define the set of roots $\Delta_W$ of the vertex algebra $W_k
(\fg ,x,f)$ as the disjoint union of three sets of pairs $(\alpha
,m) \in \fh^{\natural *} \times \tfrac{1}{2} \ZZ$:
\begin{eqnarray*}
  \Delta^{\natural}_W = \{ (\alpha ,m )|
      \alpha \in \Delta^{\natural} \, , \, m \in \ZZ \} \, ,
 \Delta'_W = \{ (\alpha ,m )|
      \alpha \in \Delta' \, , \, m \in \tfrac{1}{2} +\ZZ \} \, ,
\Delta^{im}_W = \{ (0 ,m )|  m \in \ZZ \} \, ,
\end{eqnarray*}
the multiplicity of a root being $1$ for the first two sets
(except for the case $\fg = sl(2|2)/\CC I$ discussed in  Section \ref{sec:8.4}) and
being $r=\dim \fh^{\natural}+1$ for the third set.  Define the
subset of real positive roots $\Delta^{+ re}_W$ (resp. all
positive roots $\Delta^+_W$) as the disjoint
union of the first two (resp. all three) of the following subsets:
\begin{eqnarray*}
   \Delta^{\natural +}_W &=&  \{ (\alpha ,m )|
       \alpha \in \Delta^{\natural} \, , \, m \in \NN \}
       \cup \{ (\alpha ,0)|\alpha \in \Delta^{\natural}_+\} \, , \\
    \Delta^{\prime +}_W &=&    \{ (\alpha ,m )|  \alpha \in \Delta' \, , \,
    m \in \tfrac{1}{2} +\ZZ_+ \} \, , \,
    \Delta^{im+}_W
    =\{ (0,m) | m \in \NN \} \, .
\end{eqnarray*}
A root $(\alpha ,m)$ is called odd if the corresponding to $\alpha$
root or weight vector is odd.  We define the corresponding
partition function $P_W (\eta )$, $\eta \in \fh^{\natural *}
\times\tfrac{1}{2} \ZZ$, as the number of ways $\eta$ can be represented in
the form (counting multiplicities of roots):
\begin{displaymath}
  \sum_{\alpha \in \Delta^+_W} k_{\alpha} \alpha \, , \,
  \hbox{ where  } k_{\alpha} \in \ZZ_+ \, , \,
  \hbox{ and  } k_{\alpha}\leq 1 \hbox{  if  }\alpha
  \hbox{  is odd.}
\end{displaymath}

%We denote by $P_{W,\gamma}(\eta)$ the number of such partitions
%that do not involve the given root $\gamma \in \Delta^+_W$.

We denote by $\det_{(\eta ,s)}(k,h, \Lambda)$ the determinant of
the contravariant form $B (. \, , \, .)$ restricted to the weight
subspace $M_{(\Lambda -\eta , h+s)}$, $\eta \in \ZZ_+
 \Delta^+_W$, $s \in \tfrac{1}{2} \ZZ_+$, of the Verma module $M$ with highest
weight $(\Lambda ,h)$.  This is a function of the level $k$ (or,
in view of (\ref{eq:5.7}), of the central charge $c$), the lowest
eigenvalue $h$ of $L_0$ and of $\Lambda \in \fh^{\natural *}$.

In the determinant formula given below we use the normalization
of the bilinear form $(. \, | \, .)$ on $\fg$ such that $(\theta
| \theta)=2$.  The corresponding value of $h\spcheck$ is given in
Tables~I--III.  The form $(. \, | \, .)$ restricted to
$\fh^{\natural}$ is non-degenerate, and we identify
$\fh^{\natural}$ with $\fh^{\natural *}$ using this form.
%We denote by $\rho^{\natural} \in \fh^{\natural *}$ the half of the
%difference between the sum of all even roots from
%$\Delta^{\natural}_+$ and the sum of all odd roots from
%$\Delta^{\natural}_+$. Finally, we let $\epsilon =2$ if $0 \in \Delta'$
%and $\epsilon =1$ otherwise.

\begin{theorem}
  \label{th:7.2}
Let $\hat{\eta} \in \fh^{\natural *} \times \tfrac{1}{2}\ZZ$.  Then, up to a
non-zero constant factor, depending only on the choice of a basis of
the weight space $M_{(\Lambda ,h)-\hat{\eta}}$, the determinant
$\det_{\hat{\eta}} (k,h,\Lambda)$ is given by the following
formula:
\begin{eqnarray*}
 && (k+h\spcheck)^{(r-1)\sum_{m,n \in \NN} P_W (\hat{\eta}-(0,mn))}
    \prod_{m,n \in \NN}
 (h-h_{n,m}(k,\Lambda))^
{P_W (\hat{\eta}-(0, mn))}\\
   &&  \times \prod_{m,n \in \NN} \prod_{\beta \in \Delta^{\natural}_+}
     \, \varphi_{n,m,-\beta} (k,\Lambda)^{(-1)^{p(\beta)(n+1)}P_W
     (\hat{\eta}-n(-\beta ,m))}
      \varphi_{n,m-1,\beta}(k,\Lambda)^{(-1)^{p(\beta)(n+1)}
      P_W (\hat{\eta}-n (\beta ,m-1))} \\
  && \times  \prod_{m \in \tfrac{1}{2}+\ZZ_+} \,
   \prod_{n \in \NN}
   \prod_{\gamma \in   \Delta'}
   (h-h_{n,m,\gamma} (k,\Lambda))^{(-1)^{p(\gamma)(n+1)}P_W
     (\hat{\eta}- n(\gamma ,m))} \, ,
\end{eqnarray*}
where
\begin{eqnarray*}
  \varphi_{n,m,\beta}(k,\Lambda) &=& (\Lambda +
  \rho^{\natural}|\beta) + m (k+h\spcheck) -\tfrac{n}{2}
  (\beta | \beta) \, , \\
  h_{n,m,\gamma }(k,\Lambda) &=& \frac{1}{4(k+h\spcheck)}
  ((2(\Lambda + \rho^{\natural}|\gamma)+2m(k+h\spcheck)-n(\gamma|\gamma))^2
  -(k+1)^2 + 2(\Lambda |\Lambda +2 \rho^{\natural})  ) \, ,  \\
  h_{n,m} (k,\Lambda) &=& \frac{1}{4(k+h\spcheck)}
(( m (k+h\spcheck)-n)^2 - (k+1)^2 + 2(\Lambda |\Lambda
  +2\rho^{\natural}))\, .
\end{eqnarray*}

\end{theorem}

Theorem~\ref{th:7.2} follows from
Theorem~\ref{th:7.1}b and some  lemmas on Verma modules over
$\hat{\fg}$ stated below.  Recall that a weight vector $v_{\mu}$ of a
$\hat{\fg}$-module (resp. the weight $\mu$) is called
\emph{singular} if $v_{\mu} \neq 0$ and $v_{\mu}$ is annihilated
by $\hat{\fn}_+ :=\fn_+ +\sum_{n \geq 1}\fg t^n$.  Recall that
the set of roots of $\hat{\fg}$ is $\hat{\Delta}= \{ \alpha +mK |
\alpha \in \Delta \cup \{ 0 \} \, , \, m \neq 0$ if $\alpha \in 0
\}$.  A root $\alpha +mK$ of $\hat{\fg}$ is positive if either
$\alpha \in \Delta_+$ and $m \in \ZZ_+$ or $\alpha \in -\Delta_+
\cup \{ 0 \}$ and $m \in \NN$, it is even iff $\alpha$ is even,
and it has multiplicity $1$ unless $\alpha =0$, in which case the
multiplicity is $r= \dim \fh$.

\begin{lemma}
  \label{lem:7.1}\cite{KK}, \cite{K2}.  Let $P$ be a Verma
  $\hat{\fg}$-module with highest weight $\hat{\lambda}$.  Let
  $\tilde{\alpha}$ be a positive root of $\hat{\fg}$ and let $n$
  be a positive integer.  Suppose that $(\hat{\lambda}
  +\hat{\rho}| \tilde{\alpha}) = \tfrac{n}{2}(\tilde{\alpha} |
  \tilde{\alpha})$.  Then $\hat{\lambda}-n\tilde{\alpha}$ is a
  singular weight of $P$  in the following
  cases:

\romanparenlist
\begin{enumerate}
\item %%i
  $\tilde{\alpha} = \alpha + mK$ is an even real
  (i.e.,~$\alpha$ is an even root) positive root,
%  \Delta_+$, $m \in \NN$ if $\alpha \in -\Delta_+ $)
  such that $\tfrac{1}{2}\alpha$ is not an odd root,

\item %%ii
$\tilde{\alpha} = \alpha +mK$ is an odd positive root such that
$2\alpha$ is an even root and $n$ is odd,

\item %%iii
  $\tilde{\alpha} = \alpha +mK$ is an odd positive root such that $2
  \alpha$ is not a root and $n =1$.

\end{enumerate}
\end{lemma}

\begin{remark}
  \label{rem:7.1}
This lemma follows from the determinant formula for the
contravariant form on a weight space with weight $\lambda -\eta$
of a Verma module with highest weight $\lambda$ over any contragredient Lie superalgebra $\fg
(A,\tau)$ associated to a symmetrizable matrix $A$ and the
anti-involution $\omega (e_i)=(f_i)$, $\omega (f_i)=e_i$, $\omega
(h_i)=h_i$ \cite{K2}:
\begin{displaymath}
\hbox{det}_{\eta} (\lambda) =\prod_{\alpha \in \Delta_+}
 \, \prod_{n \in \NN} \,\,((\lambda + \rho
 |\alpha)-\tfrac{n}{2}(\alpha |
 \alpha))^{(-1)^{p(\alpha)(n+1)}P(\eta -n \alpha)\dim
   \fg_{\alpha}}
\end{displaymath}
where $P$
 is the partition function for $\fg (A,\tau)$.  (Unfortunately
 there is a misprint \cite{K2} in the exponent of this formula:
 $n+1$ is missing there.)  The determinant formula for
 $\hat{\fg}$, which is a special case of the above formula, is as
 follows:
 \begin{eqnarray*}
   \hbox{det}_{\hat{\eta}} (\hat{\lambda}) =
     (k+h\spcheck)^{\sum_{m,n \in \NN}P (\hat{\eta}-mnK)}
     \prod_{n \in \NN} \prod_{\tilde{\alpha}}
     ((\hat{\lambda}+\hat{\rho}|\tilde{\alpha})-\tfrac{n}{2}
     (\tilde{\alpha}|\tilde{\alpha}))^{P(\hat{\eta}-n\tilde{\alpha})}\\
     \times \prod_{n\in 1+2\ZZ_+} \prod_{\tilde{\beta}}
     (( \hat{\lambda}+\hat{\rho} |\tilde{\beta})-
     \tfrac{n}{2}(\tilde{\beta}|\tilde{\beta}))^{P(\hat{\eta}-n\tilde{\beta})}
     \prod_{\tilde{\gamma}}  (\hat{\lambda}+ \hat{\rho}|
     \tilde{\gamma})^{P_{\tilde{\gamma}}(\hat{\eta}-\tilde{\gamma})}\, ,
 \end{eqnarray*}
where $\tilde{\alpha}$, $\tilde{\beta}$ and $\tilde{\gamma}$ run
over the positive roots described in (i), (ii) and (iii),
respectively, and $P_{\tilde{\gamma}}$ denotes the number of
partitions not involving $\tilde{\gamma}$.  Of course, both
determinant formulas are given up to a non-zero factor depending
on the choice of basis of the weight space.  The latter
determinant formula implies that for a generic $\hat{\lambda}$ on
any hyperplane $k=a$, where $a \neq -h\spcheck$, the Verma
$\hat{\fg}$-module with highest weight $\hat{\lambda}$ is irreducible.

\end{remark}

\begin{lemma}
  \label{lem:7.2}
Let $P$ be a Verma $\hat{\fg}$-module with highest weight vector
$v_{\hat{\lambda}}$, and let $v_{\hat{\lambda}-n\tilde{\alpha}}$
be a singular vector corresponding to one of the singular weights
described by Lemma~\ref{lem:7.1}, cases (i)--(iii).  Then

\alphaparenlist
\begin{enumerate}
\item %%a
in cases (i) and (ii), the map $U(\hat{\fn}_-)\to P$ defined by
$u \mapsto u (v_{\hat{\lambda}-n\tilde{\alpha}})$, is injective,

\item %%b
in case (iii), the map $U^{-\tilde{\alpha}} (\hat{\fn}_-)\to P$,
defined by $u \mapsto u (v_{\hat{\lambda}-\tilde{\alpha}})$, is
injective, where $U^{-\tilde{\alpha}}(\hat{\fn}_-)$ denotes the
span of all PBW monomials in negative root vectors, except for
that corresponding to $-\tilde{\alpha}$.
\end{enumerate}
\end{lemma}

\begin{proof}
  In all cases (i)--(iii) we have:
  $v_{\hat{\lambda}-n\tilde{\alpha}}
  =(e^n_{-\tilde{\alpha}}+\cdots)v_{\hat{\lambda}}$, where
  $e_{-\tilde{\alpha}}$ is a root vector attached to
  $-\tilde{\alpha}$ and $\cdots$ signify a linear combination of
  products of root vectors $e_{-\tilde{\beta}}$, where $0<\tilde{\beta}<\tilde{\alpha}$.
\end{proof}

A weight vector of a $W_k (\fg ,e_{-\theta})$-module is called
singular if it is annihilated by all ``raising'' operators
$L_n$, $J^{\{ a \}}_n$, $G^{\{ v \}}_n$ for $n>0$ and $J^{\{ a
  \}}_0$ for $a \in \fn_{0+}$ (and it is non-zero).

\begin{lemma}
  \label{lem:7.3}
Let $P$ be a Verma module over $\hat{\fg}$ with highest weight
$\hat{\lambda}=\lambda + kD$ ($\lambda \in \fg$) of level $k$,
and let $\vac$ be the vacuum vector
of $F (\fg ,x,f)$.

\alphaparenlist
\begin{enumerate}
\item %%a
Let $u$ be a singular vector of the $\hat{\fg}$-module $P$ with
weight $\hat{\lambda}-n\tilde{\alpha}$, where
$\tilde{\alpha}=\alpha +mK$ is as in Lemma~\ref{lem:7.1}
(i)--(iii) and $(\alpha |x)=0$, $1/2$ or $1$.  Then the homology
class of the vector $u \otimes \vac$ is a singular vector of the
$W_k (\fg ,e_{-\theta})$-module $H(P)$.
\item %%b
If $u$ is a weight vector of $P$ with weight $\hat{\mu}$, then
the weight of the vector $u \otimes \vac$ of the $W_k (\fg,
e_{-\theta})$-module $H(P)$ is:
\begin{displaymath}
  (\mu \, , \, \frac{(\hat{\lambda}|\hat{\lambda}+2\hat{\rho})}
    {2(k+h\spcheck)}-\hat{\mu}(x+D))\, , \, \hbox{  where  }
    \mu = \hat{\mu}|_{\fh}\, .
\end{displaymath}

\item %%c
The condition
$(\hat{\lambda}+\hat{\rho}|mK+\alpha)=\tfrac{n}{2}(\alpha |
\alpha)$, $\alpha \in \Delta$, $m,n \in \ZZ$, on $\hat{\lambda}$ is
equivalent to the following condition on the highest weight
$(\Lambda ,h)$ of the $W_k(\fg ,e_{-\theta})$-Verma module
$H(P)$:
%, where $\alpha'$ denotes the restriction of $\alpha$ to $\fg^{\natural}$:
%
\begin{eqnarray*}
  \varphi_{n,m,\alpha^{\natural}} (k,\Lambda) = 0 \quad \hbox{  if  }
     (x|\alpha)=0 \, , \\
     h_{n,m+1/2,\alpha^{\natural}}(k,\Lambda)=h \quad \hbox{  if  }
     (x|\alpha)=1/2  \, , \\
     h_{n,m+1}(k,\Lambda)=h \quad \hbox{  if  }
     \alpha=\theta \, .
\end{eqnarray*}

\end{enumerate}
\end{lemma}

\begin{proof}
  It is clear that if $u$ is a singular vector of the
  $\hat{\fg}$-module $P$, then the vector $u \otimes \vac$ of $\C
  (P)$ is $d^P_0$-closed.  It is also clear that this vector is
  annihilated by all raising operators.  It follows from the
  computation of $H(P)$ in Section~\ref{sec:6} that if $u$ is a
  singular vector of the type considered, the vector $u \otimes
  \vac$ is not $d^P_0$-exact, proving (a).

Statement (b) follows from the following expression for $L_0$ :
\begin{equation}
\label{eq:7.3}
  L_0 = \frac{\hat{\Omega}}{2(k+h\spcheck)}
  -D-x+(\hbox{ ghost terms})\, ,
\end{equation}
where $\hat{\Omega}$ is the Casimir operator for $\hat{\fg}$ \cite{K3},
hence has eigenvalue $(\hat{\lambda}|\hat{\lambda}+2\hat{\rho})$
on $P$.

Statement (c) is derived from (b) by a straightforward
calculation.  By (\ref{eq:7.2}) we have:
\begin{equation}
  \label{eq:7.4}
  (\hat{\lambda}+\hat{\rho}|\hat{\lambda}+\hat{\rho})
  = \tfrac{1}{2} (\lambda +\rho|\theta)^2 +
  | \lambda^{\natural}+ \rho^{\natural}|^2 \, .
\end{equation}
Letting $\hat{\lambda}=0$ in (\ref{eq:7.4}), we get
\begin{equation}
  \label{eq:7.5}
  (\hat{\rho}|\hat{\rho})=\tfrac{1}{2}(h\spcheck -1)^2
     + (\rho^{\natural} | \rho^{\natural}) \, .
\end{equation}
Using (b), (\ref{eq:7.4}) and  (\ref{eq:7.5}), we find $h$, the
minimal eigenvalue of $L_0$:
\begin{equation}
  \label{eq:7.6}
  h=\frac{1}{4 (k+h\spcheck)}((\lambda + \rho |\theta)
     -(k+h\spcheck))^2 -(k+1)^2  +2
     (\lambda^{\natural}| \lambda^{\natural}+2\rho^{\natural})\, .
\end{equation}
Furthermore, we have by (\ref{eq:7.2}):
\begin{equation}
  \label{eq:7.7}
  (\hat{\lambda}+\hat{\rho}|mK+\alpha) =m(k+h\spcheck)
  +(\alpha |x)(\lambda +\rho |\theta)+
  (\lambda^{\natural}-\rho^{\natural} |\alpha^{\natural})\, .
\end{equation}
In particular, we have, provided that $(\alpha |x)\neq 0$:
\begin{displaymath}
  (\lambda +\rho |\theta)=\tfrac{1}{(\alpha|x)} ((\hat{\lambda}+\hat{\rho}
  |mK+\alpha)-m(k+h\spcheck)-(\lambda^{\natural} +\rho^{\natural}
  |\alpha^{\natural})) \, .
\end{displaymath}
Plugging this in (\ref{eq:7.6}), we get, provided that $(\alpha
|x) \neq 0$:
\begin{eqnarray}
  \label{eq:7.8}
  h &=& \frac{1}{4(k+h\spcheck)(\alpha |x)^2}
  ((m+(\alpha |x))(k+h\spcheck)-(\hat{\lambda}+\hat{\rho}
  |mK+\alpha)+(\lambda^{\natural}+\rho^{\natural}|
  \alpha^{\natural}))^2\\
\nonumber
&& -(\alpha |x)^2  (k+1)^2
  +\frac{(\lambda^{\natural}|\lambda^{\natural}+2\rho^{\natural})}
  {2(k+h\spcheck)} \, .
\end{eqnarray}
By (\ref{eq:7.7}) and (\ref{eq:7.8}), the condition
$(\hat{\lambda}+\hat{\rho} |mK+\alpha)=\tfrac{n}{2}(\alpha
|\alpha)$ is equivalent to:
\begin{eqnarray*}
  \varphi_{n,m,\alpha^{\natural}} (k,\lambda^{\natural}) =0
        \hbox{  if  } (\alpha |x)=0 \, , \\
  h=h_{n,m+1}(k,\lambda^{\natural})
         \hbox{  if  }\alpha  =\theta\, ,\\
  h=h_{n,m+1/2,\alpha^{\natural}} (k,\lambda^{\natural})
  \hbox{  if  } (\alpha |x)=1/2 \, .
\end{eqnarray*}

\end{proof}

\begin{proof}[Proof of Theorem 7.2]
  The proof follows the traditional lines, cf. e.g. \cite{KR}.
 Choose an ordered basis $\{ u_{\alpha}|\alpha \in
  \Delta^{\natural}\} \coprod \{ u^i_0 |i=1,\ldots ,r-1 \}$ of
  $\fg^{\natural}$ consisting of root vectors and a basis of
  $\fh^{\natural}$, such that $(\omega
  (u_{\alpha})|u_{\beta})=\delta_{\alpha \beta}$ and $(u^i_0 |
  u^j_0) =\delta_{ij}$.  Choose an ordered basis $\{
  v_{\gamma}|\gamma \in \Delta' \}$ of $\fg_{-1/2}$ consisting
  of weight vectors such that $\langle \omega (v_{\beta}),v_{\gamma}
  \rangle_{\ne}=\delta_{\beta \gamma}$.

Due to Theorem~\ref{th:5.1}, the vertex algebra is strongly
generated by the fields of the following three types:
\begin{eqnarray*}
 &&  L(z) = \sum_{m \in \ZZ} L_m z^{-m-2} \, , \\
  && J^{\{ u_{\alpha}\}}(z) = \sum_{m \in \ZZ}
     J^{\{ u_{\alpha}\}}_m z^{-m-1}\, , \, J^{\{ u^i_0 \}}
     =\sum_{m \in \ZZ} J^{\{ u^i_0 \}}_m z^{-m-1} \, , \\
  &&   \frac{1}{(k+h\spcheck)^{1/2}}G^{\{ v_{\gamma}\}} =
     \sum_{m \in \tfrac{1}{2}+\ZZ} G^{\{v_{\gamma}\}}_m
     z^{-m-3/2} \, .
\end{eqnarray*}
We shall need the following commutation relations:
\begin{eqnarray}
  \label{eq:7.9}
  [L_m,L_{-m}] &=& 2mL_0 +\frac{m^3-m}{12}c(k),
\hbox{where}\,  c(k)\, \hbox {is given by (\ref{eq:5.7}),}\,\\
  \label{eq:7.10}
  [J^{\{ u \}}_m, J^{\{u'\}}_{-m}] &=& J^{\{ [u,u']\}}_0
  +m \Big(k+\tfrac{1}{2}\big(h\spcheck - \sum_i h\spcheck_{0,i}\big)\Big)(u|u')
%  -\tfrac{1}{4}\kappa_{\fg_0}(u,u'))
  \hbox{  (see Theorem~\ref{th:5.1}d),}\\
  \label{eq:7.11}
  [G^{\{ v \}}_m , G^{\{ v'\}}_{-m}] &=& - 2 (e|[v,v'])  L_0
  +\frac{1}{k+h\spcheck}A (v,v',m)\, ,
\end{eqnarray}
where $A(v,v',m)$ is a (possibly infinite) linear combination
over $\CC$ of monomials of the form $J^{\{ u_{\alpha}\}}_{-n}
J^{\{ u_{\beta} \}}_n $ with $n>0$ plus a polynomial of degree at
most $2$ in $J^{\{ u_{\alpha}\}}_0$ and $k$, not involving $k^2$
(see Theorem~\ref{th:5.1}e).

Let $M$ be a Verma module over $W_k (\fg ,e_{-\theta})$ with
highest weight vector $v_{\Lambda ,h}$.  Recall that the
following vectors form a basis of $M$ compatible with its weight
space decomposition (\ref{eq:7.1}):
\begin{equation}
  \label{eq:7.12}
  (G^{\{ v_{\gamma_1}\}}_{-r_1})^{c_{1}}
   (G^{\{ v_{\gamma_2}\}}_{-r_2})^{c_{2}} \ldots
   (J^{\{ u_{\alpha_1}\}}_{-n_1})^{b_{1}}
   (J^{\{ u_{\alpha_2}\}}_{-n_2})^{b_{2}} \ldots
   (L_{-m_1})^{a_1}(L_{-m_2})^{a_2} \ldots
   v_{(\Lambda ,h)}\, ,
\end{equation}
where $a_i, b_i, c_i \in \ZZ_+$ and $b_i$ (resp. $c_i$) $\leq 1$
if $\alpha_i$ (resp. $\gamma_i$) is odd; $m_i,n_i,r_i >0$ or $n_i
=0$ and $\alpha_i$ is a negative root; and each sequence $m_1,
m_2, \ldots$; $(\alpha_1,n_1)$, $(\alpha_2 ,n_2), \ldots$;
$(\gamma_1,r_1)$, $(\gamma_2,r_2), \ldots$ is strictly
decreasing.  Note that in the weight space decomposition
(\ref{eq:7.1}) the vector (\ref{eq:7.12}) has weight
\begin{equation}
  \label{eq:7.13}
  (\Lambda -\sum n_i\alpha_i-\sum r_i\gamma_i\, , \, h+\sum m_i
  +\sum n_i +\sum r_i\,) .
\end{equation}

The function $\det_{\hat{\eta}}(k,h,\Lambda)$ is a rational
function in $k$ and a polynomial function in $h$ and $\Lambda$.
First, we compute the total degree of this function (by letting
the degree of $k$, $h$ and $\Lambda$ be $1$ and defining the
degree of a rational function as the difference between the
degrees of the numerator and the denominator).

Let $\{ R_i \}$ be the basis of the weight space $M_{(\Lambda
  ,h)-\hat{\eta}}$ consisting of monomials (\ref{eq:7.12}) such
that (\ref{eq:7.13}) is equal to $(\Lambda ,h)-\hat{\eta}$.  Then
\begin{displaymath}
  \hbox{det}_{\hat{\eta}} (k,h,\Lambda)=\det (B(R_i,R_j))_{i,j}\,.
\end{displaymath}
It is clear from (\ref{eq:7.9})--(\ref{eq:7.11}) that the leading
term of $\det_{\hat{\eta}}$ comes from the diagonal of the matrix
$(B(R_i,R_j))_{i,j}$, and that the leading term of each diagonal
term $B(R_i,R_i)$ is the product of leading terms of factors of
(\ref{eq:7.12}), i.e.,~the following three types of factors:
\begin{displaymath}
B(L^n_{-m}v_{(\Lambda ,h)}\,, \, L^n_{-m}v_{(\Lambda ,h)}),\,
   B(J^{\{u_{\alpha}\}n}_{-m} v_{(\Lambda ,h)},
   J^{\{ u_{\alpha}\}n}_{-m} v_{(\Lambda ,h)}) \, , \,
   B(G^{\{v_{\gamma}\}n}_{-m} v_{(\Lambda ,h)}\, , \,
   G^{\{ v_{\gamma}\}n}_{-m} v_{(\Lambda ,h)})\, .
\end{displaymath}
It is clear from  (\ref{eq:7.9})--(\ref{eq:7.11}) that each of
these factors contributes $n$ to the total degree of
$\det_{\hat{\eta}}$.  Hence the contribution to the total degree
of all first type factors is equal to
\begin{displaymath}
\sum_{n \in \NN} \sum_{m
  \in \NN} n (P_W (\hat{\eta}-n(0,m))
-P_W(\hat{\eta}-(n+1)(0,m)))=\sum_{n \in
  \NN}\sum_{m \in \NN} P_W (\hat{\eta}-n(0,m))\, .
\end{displaymath}
Likewise, the
second (and resp. third) type factors with $u$ (resp.~$v$) even
contribute to the total degree the number
\begin{displaymath}
  \sum_{n \in \NN} \sum_{\alpha \in \Delta^{\natural +}_{W,\even}}
  P_W (\hat{\eta}-n\alpha) \quad\hbox{(resp.  } \sum_{\gamma \in
    \Delta^{\prime +}_{W,\even}} P_W (\hat{\eta}-n\gamma)
  \hbox{).}
\end{displaymath}
If $u$ (resp. $v$) is odd, then the second (resp. third) type
factors contribute
\begin{displaymath}
  \sum_{\alpha \in \Delta^{\natural +}_{W,\odd}}P_{W,\alpha}
  (\hat{\eta}-\alpha) \quad (\hbox{resp. }
  \sum_{\gamma \in \Delta^{\prime +}_{W,\odd}}P_{W,\gamma}
  (\hat{\eta}-\gamma)) \, ,
\end{displaymath}
where $P_{W,\alpha}$ denotes the number of partitions by elements
from $\Delta^+_W\backslash \{ \alpha \}$.  But if $\alpha$ is an
odd root, we have:
\begin{displaymath}
  P_W (\hat{\eta}) =P_{W,\alpha}(\hat{\eta})+
  P_{W,\alpha} (\hat{\eta}-\alpha) \, ,
\end{displaymath}
therefore
\begin{displaymath}
  P_{W,\alpha} (\hat{\eta}-\alpha)=
  \sum_{n \in \NN}(-1)^{n+1}P_W (\hat{\eta}-n\alpha)\, .
\end{displaymath}
Thus, the total degree of $\det_{\hat{\eta}}$ is equal to
\begin{equation}
  \label{eq:7.14}
  T=\sum_{\alpha \in \Delta^+_W} \sum_{n \in \NN}
  (-1)^{p(\alpha)(n+1)}P_W(\hat{\eta}-n\alpha)\, .
\end{equation}
Of course, in (\ref{eq:7.14}) and the previous formulas the
$\alpha$ in the summations are counted with their multiplicities,
namely $\alpha \in \Delta^{im+}_W$ is counted $r$ times.

The same arguments show that the total degree in $h$ and
$\Lambda$ of the leading term of $\det_{\hat{\eta}}$ is equal to
\begin{displaymath}
  T'=T-(r-1)\sum_{m,n \in \NN}P_W (\hat{\eta}-(0,mn))\, .
\end{displaymath}
But Lemmas~\ref{lem:7.1}, \ref{lem:7.2} and \ref{lem:7.3} give us
a factor of $\det_{\hat{\eta}}$ whose leading term has total
degree in $h$ and $\Lambda$ at least $T'$  (since singular
vectors that are not highest weight vectors lie in the maximal
submodule of $H(P)$). (The corresponding to $m\delta +\alpha$
root is $(\alpha^{\natural},m+\alpha (x))$.) 
It follows that, up to a factor, which is a
rational function in $k$ with poles only at $k=-h\spcheck$,
$\det_{\hat{\eta}}$ is given by Theorem~\ref{th:7.2}.

In order to compute this factor, note that the free field
realization given by Theorem~\ref{th:5.2} is irreducible for a
generic Verma $V_{\alpha_k} (\fg_0)$-module, provided that for
each simple component of $\fg_0$ (including the commutative ones)
the level plus the dual Coxeter number $h\spcheck_0$ of this component
is non-zero.  But for a simple component its level is
$k+h\spcheck -h\spcheck_0$, hence
$(k+h\spcheck-h\spcheck_0)+h\spcheck_0=k+h\spcheck$ for all
components.  Thus, on any hyperplane $k+h\spcheck =a$, $a \neq
0$, the generic Verma $V_{\alpha_k} (\fg_0)$-module is
irreducible.  Hence the factor may vanish only if $k+h\spcheck
=0$, hence the factor is a power of $k+h\spcheck$.

\end{proof}

\begin {remark}
\label{rem:7.2}
In the case when $W(\fg, e_{-\theta})$ is from Table II, i.e.
$\fg_{1/2}$ is purely odd, the determinant formula, given by Theorem~
\ref{th:7.2} can be a bit simplified, using the fact that
all elements from $\Delta_{-1/2}$ different from $-\theta /2$
are isotropic. Let $\epsilon =2$ if $0 \in \Delta'$ and $\epsilon =1$
otherwise. Then the determinant $\det_{\hat{\eta}}(k,h,\Lambda)$
is given by the following formula:
\begin{eqnarray*}
 && (k+h\spcheck)^{(r-1)\sum_{m,n \in \NN} P_W (\hat{\eta}-(0,mn))}
    \prod_{\substack {m,n \in \epsilon^{-1} \NN \\ m-n \in \ZZ}}
 (h-h_{n,\epsilon m}(k,\Lambda))^
{P_W (\hat{\eta}-(0,\epsilon mn))}\\
   &&  \times \prod_{m,n \in \NN} \prod_{\beta \in \Delta^{\natural}_+}
     \, \varphi_{n,m,-\beta} (k,\Lambda)^{P_W
     (\hat{\eta}-n(-\beta ,m))}
      \varphi_{n,m-1,\beta}(k,\Lambda)^{
      P_W (\hat{\eta}-n (\beta ,m-1))} \\
  && \times  \prod_{m \in \tfrac{1}{2}+\ZZ_+} \,
  \prod_{\gamma \in   \Delta'\backslash\{0\}}
   (h-h_{m,\gamma} (k,\Lambda))^{P_{W,(\gamma,m)}
     (\hat{\eta}- (\gamma ,m))} \, ,
\end{eqnarray*}
where
\begin{eqnarray*}
  \varphi_{n,m,\beta}(k,\Lambda) &=& (\Lambda +
  \rho^{\natural}|\beta) + m (k+h\spcheck) -\tfrac{n}{2}
  (\beta | \beta) \, , \\
  h_{m,\gamma }(k,\Lambda) &=& \frac{1}{4(k+h\spcheck)}
  ((2(\Lambda + \rho^{\natural}|\gamma)+2m(k+h\spcheck))^2
  -(k+1)^2 + 2(\Lambda |\Lambda +2 \rho^{\natural})  ) \, ,  \\
  h_{n,m} (k,\Lambda) &=& \frac{1}{4(k+h\spcheck)} (( m
  (k+h\spcheck)-n)^2 - (k+1)^2 + 2(\Lambda |\Lambda
  +2\rho^{\natural}))\, .
\end{eqnarray*}
\end{remark}

\section{Examples}
\label{sec:8}

\subsection{Virasoro algebra.}\qquad
\label{sec:8.1}

Recall that the Virasoro algebra is $W_k (\sl_2 , e_{-\theta})$.
In this case $\dim \fg_0=1$, $\fg_{1/2}=0$ and $h\spcheck =2$.
We let $\fg_0=\CC a$, where $(a|a)=1$.  Then $x=a/\sqrt{2}$ since
$(x|x)=1/2$.  The Virasoro central charge is given by
(\ref{eq:5.7}):
\begin{displaymath}
  c=1-\frac{6 (k+1)^2}{k+2} \, .
\end{displaymath}
The free field realization is given by Theorem~\ref{th:5.2}:
\begin{displaymath}
  L=\frac{1}{2(k+2)} :aa:+\frac{k+1}{\sqrt{2}(k+2)} \partial a \, ,
\end{displaymath}
where $a$ is a free boson:  $[a_{\lambda} a]=\lambda (k+2)$ \, .
We can remove the singularity at $k=-2$ by letting
$b=\frac{1}{(k+2)^{1/2}}a$, so that $[b_{\lambda}b]=\lambda$, and
\begin{displaymath}
  L=\tfrac{1}{2}:bb:+\left( \tfrac{1-c}{12}\right)^{1/2} \partial
  b\, .
\end{displaymath}
This is the free field realization of the Virasoro
algebra, which goes back to Virasoro and Fairlie.

The determinant formula given by Remark~\ref{rem:7.2} looks as
follows:
\begin{displaymath}
  \hbox{det}_N (k,h) = \prod_{m,n \in \NN} (h-h_{n,m}(k))^{p(N-mn)}\, ,N \in \ZZ_+\, ,
\end{displaymath}
where $p$ is the classical partition function and
\begin{displaymath}
  h_{n,m}(k) = \frac{1}{4(k+2)} ((m(k+2)-n)^2-(k+1)^2) \, .
\end{displaymath}
This is the Kac determinant formula \cite{K2}, \cite{KR}. Note that it was observed
already in \cite{FF2} that under the quantum reduction the Kac-Kazhdan
equations for $\hat{sl_2}$ are transformed to Kac equations for the Virasoro
algebra.

\subsection{Neveu--Schwarz algebra.}\qquad
\label{sec:8.2}

Recall that the Neveu--Schwarz algebra is $W_k (spo (2|1),
e_{-\theta})$.  In this case $\fg_0$ is the $1$-dimensional Lie
algebra, $\fg_{1/2}$ is $1$-dimensional and purely odd, and
$h\spcheck =3/2$.  We use the notation of \cite{KRW}, Section~6,
except for a different normalization of the invariant bilinear
form, namely $(a|b)=\str \, ab$.  Then for $h=\theta$ we have
$(h|h)=2$, and we have a free neutral fermion $\Phi$ such that $[\Phi
_{\lambda} \Phi]=1$.  The vertex algebra $W_k (spo (2|1),e_{-\theta})$ is
strongly generated by two fields:
\begin{eqnarray*}
  J^{\{ f_{\alpha}\}} &=& J^{(f_{\alpha})} +\tfrac{1}{2}
     :\Phi J^{(h)}:+(k+1)\partial \Phi \, , \\
  J^{\{ f_{2\alpha}\}} &=& J^{( f_{2\alpha})}
      +:\Phi J^{(f_{\alpha})}: -\tfrac{1}{4}:
      J^{(h)} J^{(h)} : -\tfrac{k+1}{2}\partial J^{(h)}
      + \tfrac{2k+3}{4}:\Phi \partial \Phi: \, .
\end{eqnarray*}

The fields $L=-\tfrac{1}{k+3/2} J^{\{ f_{2\alpha}\}}$ and
$G=\tfrac{1}{(k+3/2)^{1/2}} J^{\{f_{\alpha}\}}$ satisfy the
relations of the Neveu--Schwarz algebra:
\begin{displaymath}
  [L_{\lambda}L] = (\partial +2\lambda) L
    +\tfrac{\lambda^3}{12}c \, , \, [L_{\lambda}G]=
    (\partial +\tfrac{3}{2}\lambda )G \, , \,
    [G_{\lambda}G]=L+\tfrac{\lambda^2}{6}c \, ,
\end{displaymath}
where the central charge is (see (\ref{eq:5.7})):
\begin{displaymath}
  c=\tfrac{3}{2} -\tfrac{12 (k+1)^2}{2k+3} \, .
\end{displaymath}
As in Theorem~\ref{th:6.2}, we project to the tensor product of
the affine vertex algebra associated to $\fg_0$ and
$F^{\ne}(\fg_{1/2})$ to obtain the free field realization.  In
order to remove the singularity at $k=-3/2$, we let
$b=J^{(h)}/(2k+3)^{1/2}$, so that $[b_{\lambda}b]=\lambda$, and
let $\gamma=(k+1)/(2k+3)^{1/2}$.  As before,
$[\Phi_{\lambda}\Phi]=1$.  Then we get:
\begin{displaymath}
  L=\tfrac{1}{2}:bb:+\gamma\partial b-\tfrac{1}{2}:
  \Phi \Phi : \, , \, G=\tfrac{1}{\sqrt{2}}:b\Phi:+\sqrt{2}\gamma \partial
\Phi,
\end{displaymath}
the central charge being $c=3/2-12 \gamma^2$.  This is the
free field realization of the Neveu--Schwarz algebra
which goes back to Neveu--Schwarz \cite{NS} and Thorn \cite{T} (cf.~\cite{K4}).

The determinant formula given by Remark~\ref{rem:7.2} looks as
follows (cf.~\cite{K2}, \cite{KW0}):
\begin{displaymath}
  \hbox{det}_N (k,h) = \prod_{\substack{m,n \in \tfrac{1}{2}\NN \\
      m-n\in\ZZ}}(h-h^{NS}_{n,m}(k))^{\tilde{p}(N-2mn)}
     \, , \, N \in \tfrac{1}{2}\ZZ_+ \, ,
\end{displaymath}
where $\tilde{p}(s)$ denotes the number of partitions of $s
\in \tfrac{1}{2}\ZZ_+ $ in a sum of parts from $\tfrac{1}{2}\NN$,
the non-integer parts being distinct, and
\begin{displaymath}
  h^{NS}_{n,m}(k) = \tfrac{1}{2(2k+3)}
  ((m(2k+3)-n)^2 -(k+1)^2)\, .
\end{displaymath}

\subsection{$N=2$ superconformal algebra.}\qquad
\label{sec:8.3}

Recall that the $N=2$ superconformal algebra is $W_k (\sl (2|1),
e_{-\theta})$ \cite{KRW}.  In this case $\fg_0$ is the $2$-dimensional
commutative Lie algebra, $\fg_{1/2}$ is $2$-dimensional and
purely odd, and $h\spcheck =1$.  The Virasoro central charge is
given by (\ref{eq:5.7}):
\begin{displaymath}
  c=-3(2k+1) \, .
\end{displaymath}
Recall that the simple root vectors of $\fg$ are $h_1, h_2$ with scalar
products $(h_1 |h_1) = (h_2 |h_2)=0$, $(h_1 |h_2)=1$.  We have:
$\fh =\CC h_1 +\CC h_2$, $S_{1/2}=\{ h_1,h_2 \}$, $S_0
=\emptyset$, $\theta =h_1 +h_2$.  Let $\alpha
=\tfrac{1}{2}(h_1-h_2)$.  Then $\fh^{\natural} =\CC \alpha$,
$\rho^{\natural} =0$, $h^{\natural}_1 =-h^{\natural}_2=\alpha$.
We have two free neutral fermions $\Phi_1$, $\Phi_2$ with the
$\lambda$-brackets:
\begin{displaymath}
  [\Phi_1 {}_{\lambda} \Phi_2]=-1 \, , \,
  [\Phi_i {}_{\lambda} \Phi_i]=0 \, .
\end{displaymath}

The construction of the vertex algebra in question given by
Theorem~\ref{th:5.1}, is explicitly written down in \cite{KRW},
Section~7 (see~(7.1) and formulas preceeding it).  To obtain the
free field realization we project on the affine vertex algebra
associated to $\fg_0$ to get
$\gamma =(-k-1)^{1/2}$:
\begin{eqnarray*}
  J &=& (h_1-h_2) + : \Phi_1 \Phi_2 : \, ,
  G^+ = -\tfrac{1}{\gamma}:\Phi_2 h_1:+\gamma \partial \Phi_2 \, , \,
      G^- =-\tfrac{1}{\gamma}:\Phi_1h_2:+\gamma \partial \Phi_1 \, , \\
  L &=& -\tfrac{1}{\gamma^2}:h_1h_2:+\tfrac{1}{2}\partial (h_1+h_2)
      -\tfrac{1}{2}(:\Phi_1 \partial \Phi_2:
      +:\Phi_2\partial\Phi_1:) \, .
\end{eqnarray*}

Note that the cubic terms in $G^{\pm}$ disappear since $\dim
\fg_{1/2}<3$.  The filed $L$ is a Virasoro filed with central
charge $c$, the fields $J$ (resp. $G^{\pm}$) are primary of
conformal weight $1$ (resp.~$3/2$), and the remaining
$\lambda$-brackets are as follows:
\begin{eqnarray*}
  [J_{\lambda} J ]= \tfrac{\lambda}{3}c \, , \,
  [G^{\pm}{}_{\lambda} G^{\pm} ]=0 \, , \,
  [J_{\lambda} G^{\pm}] =\pm G^{\pm} \, , \,
  [{G^+}{}_{\lambda} G^- ] = L + (\tfrac{1}{2}\partial+\lambda)
  J+\tfrac{\lambda^2}{6}c \, ,
\end{eqnarray*}
which is the $N=2$ superconformal algebra.  We have:  $\omega
(J_n)=J_{-n}$, $\omega (G^{\pm}_n)=G^{\mp}_{-n}$.

In order to make the formulas more symmetric and to remove the
singularity at $k=-1$, we let:
\begin{displaymath}
  b^+ = -\tfrac{1}{\gamma}h_1 \, , \,
      b^-=\tfrac{1}{\gamma}h_2 \, , \,
      \psi^+=\Phi_2 \, , \, \psi^-=\Phi_1\, ,
\end{displaymath}
so that $[{b^{\pm}}_{\lambda}b^{\mp}]=\lambda$,
$[{b^{\pm}}{}_{\lambda}b^{\pm}]=0$,
$[{\psi^{\pm}}{}_{\lambda}\psi^{\mp}]=1$,
${\psi^{\pm}}{}_{\lambda}\psi^{\pm}=0$.  Then we get the free field
realization by Kato--Matsuda \cite{KM} (cf.~\cite{K4}):
\begin{eqnarray*}
 &&  J = -\gamma (b^++b^-)+:\psi^+\psi^-: \, , \,
 G^+ = :\psi^+b^+:+\gamma\partial\psi^+ \, , \,
      G^- = :\psi^-b^-:-\gamma\partial \psi^- \, , \, \\
 &&  L = :b^+b^-:+\tfrac{\gamma}{2}\partial
      (b^--b^+)+\tfrac{1}{2}
      (:\psi^-\partial\psi^+:+:\psi^+\partial\psi^-:) \, ,
\end{eqnarray*}
the central charge being $c=3+6\gamma^2$.

Using the non-Dynkin gradation of $\sl (2|1)$, discussed in
\cite{KRW} as well, we get another ``free field'' realization of
the $N=2$ superconformal algebra.  In this case $\fg_0 \simeq
g\ell (1|1)$, where $\CC^{1|1}=\CC \epsilon_1 + \CC \epsilon_2$,
$\epsilon_1$ even, $\epsilon_2$ odd, we take the bilinear form
$(a|b)=\str \, ab$ on $\fg_0$, and the following basis:
\begin{displaymath}
  e= E_{12}\, , \, f=E_{21} \, , \, h_1=E_{11}+E_{22}\, , \,
  h_2=E_{11}\, .
\end{displaymath}
Consider the associated affine Lie superalgebra $\hat{\fg}_0$ of
level~$k$.  Then we have the following ``free field'' realization
in $\hat{\fg}_0$ with $c=-6k+3$ (cf.~\cite{KRW}):
\begin{eqnarray*}
  J &=& h_1-h_2 \, , \, G^+=-\tfrac{1}{k}f \, , \,
       G^- =:h_2 e:-(k-1)\partial e \, , \\
  L &=& -\tfrac{1}{k} (:ef :-:h_1h_2 :)+
        \tfrac{1}{2}\partial (h_1 + h_2)\, .
\end{eqnarray*}

In order to write down the determinant formula, let $\Lambda =-j
\alpha$, where $j\in \CC$, so that
\begin{displaymath}
  (\Lambda | \Lambda)=-j^2/2 \, , \,
  (\Lambda |h_1-h_2)=j \, , \,
  (\Lambda |\alpha)=j/2 \, .
\end{displaymath}
The set of  positive roots $\Delta^+_W$ consists of the set
of even positive roots $\{ (0,m)|m \in \NN \}$ and the set of odd
positive roots $\{ (\pm \alpha ,m) |m \in \tfrac{1}{2} +\ZZ_+
\}$, all of multiplicity $1$.  The partition function $P_{W}
(\hat{\eta})$ is equal to the number of partitions of
$\hat{\eta}=(m \alpha ,n)$ in a sum of roots from $\Delta^+_W$,
the odd roots being distinct.

The highest weight $(\Lambda ,h)$ of a Verma module over $N=2$
superconformal algebra is determined by $h=$ the lowest
eigenvalue of $L_0$ and $j=$ the eigenvalue of
$J_{0}^0$ on the highest weight vector.

The factors that occur in Remark~\ref{rem:7.2} are $k+1$,
$h-h_{n,m} (k,j)$ and $h-h_{n,\pm \alpha}(k,j)$, where
\begin{eqnarray*}
  h_{n,m}(k,j) &=& \frac{1}{4(k+1)} ((m(k+1)-n)^2
        -(k+1)^2 -j^2)\,  , \\
  h_{m,\pm \alpha}(k,j) &=& \frac{1}{4(k+1)}
     ((\pm j+2m(k+1))^2 -(k+1)^2 -j^2)
     =(m^2-\tfrac{1}{4})(k+1)\pm jm \, .
\end{eqnarray*}
Hence Remark~\ref{rem:7.2} gives the following determinant
formula (cf.~\cite{KM}):
\begin{eqnarray*}
  \hbox{det}_{\hat{\eta}} (k,h,j) &=& \prod_{m,n \in \NN}
     \Big(4(k+1)h-(m(k+1)-n)^2 +(k+1)^2+j^2\Big)^
     {P_W (\hat{\eta}-(0,mn))}\\
&& \times \prod_{m\in \tfrac{1}{2}+\ZZ_+}
   (h-((m^2-\tfrac{1}{4})(k+1)+jm))^{
   P_{W,(\alpha ,m)}(\hat{\eta}-(\alpha ,m))}\\
   &&\times (h-((m^2-\tfrac{1}{4})(k+1)-jm))^{P_{W,(-\alpha ,m)}
     (\hat{\eta}-(-\alpha ,m))} \, .
\end{eqnarray*}

\subsection{$N=4$ superconformal algebra.} \qquad
\label{sec:8.4}

As we shall see, the $N=4$ superconformal algebra is $W_k (\fg
,e_{-\theta})$, where $\fg =\sl (2|2)/\CC I$.  In this case
$\fg^{\natural}=\sl_2$, $\fg_{1/2}$ is $4$-dimensional purely
odd, and as an $\sl_2$-module it is a direct sum of two
$2$-dimensional irreducible $\sl_2$-modules.  The dual Coxeter
number $h\spcheck =0$, hence by (\ref{eq:5.7}), the Virasoro
central charge is:
\begin{displaymath}
  c=-6 (k+1) \, .
\end{displaymath}
The simple roots of $\fg$ are $\alpha_1$, $\alpha_2$, $\alpha_3$,
where $\alpha_1$ and $\alpha_3$ are odd and $\alpha_2$ is even,
and the non-zero scalar products between them are as follows:
\begin{displaymath}
  (\alpha_1 | \alpha_2)=1 \, , \quad (\alpha_2 |\alpha_3)=1 \,
  , \quad (\alpha_2 |\alpha_2)=-2 \, .
\end{displaymath}
The subalgebra $\fg^{\natural}=\sl_2$ corresponds to the root
$\alpha_2$, so that its dual Coxeter number $h\spcheck_0=-2$, and
$\fh^{\natural}=\CC \alpha_2$.

We choose root vectors $e_{\alpha}$ and $e_{-\alpha}$ $(\alpha
\in \Delta_+)$ such that $(e_{\alpha}| e_{-\alpha})=-1$, so that
$[e_{\alpha \lambda}e_{-\alpha}]=-\alpha -\lambda K$.  We use the
following notations:  $e_i=e_{\alpha_i}$, $f_i=e_{-\alpha_i}$,
$e_{ij}=e_{\alpha_i + \alpha_j}$,
$f_{ij}=e_{-\alpha_i-\alpha_j}$, etc., $h_i=-\alpha_i$,
$h_{ij}=-\alpha_i-\alpha_j$, etc.  We have:  $\theta
=\alpha_1+\alpha_2+\alpha_3$, so that $x=-\tfrac{1}{2} h_{1 2
3}$, and we  take $f=f_{1 2 3}$.  We have $4$ free neutral
fermions:
\begin{displaymath}
  \Phi_1 =\Phi_{\alpha_1} \, , \, \Phi_3 =\Phi_{\alpha_3} \, , \,
  \Phi_{1  2}=\Phi_{\alpha_1 +\alpha_2} \, , \,
  \Phi_{2  3} = \Phi_{\alpha_2 +\alpha_3} \, ,
\end{displaymath}
whose $\lambda$-brackets are:
\begin{displaymath}
  [{\Phi_{\alpha}}{}_{\lambda}\Phi_{\beta}] =1 \hbox{  if  } \alpha +\beta
  =\theta \, , \hbox{  and  }=0 \hbox{  otherwise,}
\end{displaymath}
so that $\Phi^{\alpha}=\Phi_{\theta -\alpha}$.

Then we have the following $d_0$-closed fields, which
strongly generate $W_k (\fg ,e_{-\theta})$  (by
Theorem~\ref{th:4.1}):
\begin{eqnarray*}
  J^0 &=&J^{(h_2)}-:\Phi_1 \Phi_{2 3}:-:\Phi_3\Phi_{12}:\, ,\,
    J^+ =J^{(e_2)}-:\Phi_{12}\Phi_{23}:\, ,\, J^- =
    J^{(f_2)}-:\Phi_1\Phi_3:\, , \, \\
  G^+ &=& \tfrac{1}{\sqrt{k}}(J^{(f_1)} +:\Phi_{23}J^{(h_1)}:
    -:\Phi_3 J^{(e_2)}:-(k+1)\partial \Phi_{23}+:
    \Phi_3\Phi_{12}\Phi_{23}:)\, , \\
  G^- &=& \tfrac{1}{\sqrt{k}} (J^{(f_{12})} -:\Phi_{23}J^{(f_2)}
     :+:\Phi_3 J^{(h_{12})}:-(k+1)\partial \Phi_3 +:
     \Phi_1\Phi_3\Phi_{23}:)\, , \\
  \bar{G}^+ &=& \tfrac{1}{\sqrt{k}}(J^{(f_3)}+:\Phi_{12}
      J^{(h_3)}:+:\Phi_1 J^{(e_2)}:-(k+1)\partial \Phi_{12}
      -: \Phi_1 \Phi_{12}\Phi_{23}:) \, , \\
  \bar{G}^- &=& \tfrac{1}{\sqrt{k}} (J^{(f_{23})} +: \Phi_{12}
     J^{(f_2)}:+: \Phi_1 J^{(h_{23})}:-(k+1)\partial
     \Phi_1-:\Phi_1\Phi_3\Phi_{12}:) \, , \\
  L&=&-\tfrac{1}{k}(J^{(f)}+\sum_{\alpha \in S_{1/2}}: \Phi_{\alpha}
     J^{(f_{\alpha})}:) + \tfrac{1}{4k}(:J^{(h_{123})}J^{(h_{123})}
     :-: J^{(h_2)}J^{(h_2)}:)\\
    && -\tfrac{1}{2k} (:J^{(e_2)}J^{(f_2)}:+:J^{(f_2)}J^{(e_2)}:)
     -\tfrac{k+1}{2k}\partial J^{(h_{123})} +\tfrac{1}{2}
     \sum_{\alpha \in S_{1/2}}:\Phi_{\alpha}\partial
     \Phi_{\theta-\alpha}:\, .
\end{eqnarray*}

The field $L$ is a Virasoro field with central charge $c$, the
fields $J^0$, $J^{\pm}$ (resp.~$G^{\pm}$, $\bar{G}^{\pm}$) being
primary of conformal weight $1$ (resp.~$3/2$).  The remaining
non-zero $\lambda$-brackets are as follows:
\begin{eqnarray*}
  [{J^0}{}_{\lambda} J^{\pm}] &=& \pm 2 J^{\pm}\, , \,
     [{J^0}{}_{\lambda} J^0]=\tfrac{\lambda}{3} c \, , \,
     [{J^+}{}_{\lambda} J^-]=J^0 +\tfrac{\lambda}{6} c \, , \\
{}  [{J^0}{}_{\lambda} G^{\pm}] &=& \pm G^{\pm} \, , \,
    [{J^0}{}_{\lambda}\bar{G}^{\pm}] =\pm \bar{G}^{\pm} \, , \\
{}  [{J^+}{}_{\lambda}G^-] &=& G^+, [J^-{}_{\lambda}G^+]=G^-\,,\,
     [{J^+}{}_{\lambda}\bar{G}^-] =-\bar{G}^+ \, , \,
     [{J^-}{}_{\lambda}\bar{G}^+]=-\bar{G}^- \, , \\
{}  [{G^{\pm}}{}_{\lambda}\bar{G}^{\pm}] &=&(\partial+2\lambda)
     J^{\pm} \, , \, [{G^{\pm}}{}_{\lambda} \bar{G}^{\mp}]=L
     \pm \tfrac{1}{2} (\partial +2\lambda)J^0
     +\tfrac{\lambda^2}{6} c\, .
\end{eqnarray*}
These are the $\lambda$-brackets of the $N=4$ superconformal
algebra (cf. \cite{K4}).  We have: $\omega (J^0_n)=J^0_{-n}$,
$\omega (J^{\pm}_n) =J^{\mp}_{-n}$, $\omega (G^{\pm}_n)=\bar{G}^{\mp}_{-n}$.

By Theorem~\ref{th:5.2}, the free field realization is given in
the vertex algebra
\begin{displaymath}
  V_{-k-2}(\sl_2)\otimes B \otimes F^{\ne}\, ,
\end{displaymath}
where $V_{-k-2}(\sl_2)$ is the universal affine vertex algebra
associated to $\sl_2$, with the standard basis $E,H,F$
and the standard bilinear form $(a|b)=\tr ab$, of level $-k-2$
$(=-(k+h\spcheck -h\spcheck_0))$, $B$ is the vertex algebra strongly
generated by the free boson  $\theta (z)$ and $F^{\ne}$ is the vertex
algebra strongly generated by the free fermions $\Phi_1$,
$\Phi_3$, $\Phi_{12}$, $\Phi_{23}$.  Explicit formulas are
obtained from the above formulas for the $d_0$-closed fields
by projecting on this vertex algebra:
\begin{eqnarray*}
  J^0 &=& H -:\Phi_1\Phi_{23} :-:\Phi_3 \Phi_{12} :\, , \,
     J^+=E -:\Phi_{12}\Phi_{23}:\, , \,
     J^-=F -:\Phi_1\Phi_3:\, , \\
  G^+ &=& \tfrac{1}{\sqrt{k}} (\tfrac{1}{2}:\Phi_{23}
     (\theta -H):-:\Phi_3E:-(k+1)\partial\Phi_{23}
     +:\Phi_3\Phi_{12}\Phi_{23}:)\, , \\
  G^- &=& \tfrac{1}{\sqrt{k}}(\tfrac{1}{2}:\Phi_3(\theta +H)
     :-:\Phi_{23}F:-(k+1)\partial \Phi_3 +:
     \Phi_1\Phi_3\Phi_{23}):\, , \\
  \bar{G}^+ &=& \tfrac{1}{\sqrt{k}}(\tfrac{1}{2}:\Phi_{12}
     (\theta -H):+:\Phi_1 E:-(k+1)\partial \Phi_{12}
     -:\Phi_1\Phi_{12}\Phi_{23}:)\, , \\
  \bar{G}^- &=& \tfrac{1}{\sqrt{k}}(\tfrac{1}{2}:\Phi_1
     (\theta +H): +:\Phi_{12}F:-(k+1)\partial \Phi_1 \, \,
     -:\Phi_1\Phi_3\Phi_{12}:)\, , \\
  L &=& \tfrac{1}{4k}(:\theta \theta : -:HH :-2:EF:-2:FE:)
      -\tfrac{k+1}{2k} \partial \theta \\
    && +\tfrac{1}{2}(:\Phi_1
      \partial \Phi_{23}:+:\Phi_3 \partial \Phi_{12}
      :+: \Phi_{12}\partial \Phi_3 :+:\Phi_{23}
      \partial \Phi_1:)\, .
\end{eqnarray*}
Comparing with the character formulas of \cite{ET}, we see that,
taking a unitary representation of $\hat{\sl_2}$ and that of $B$
such that $l<h$, where $l$ is the isospin and $h$ is the lowest eigenvalue
of $L_0$, the above formulas give an explicit construction of all massive
unitary representations of the $N=4$ superconformal algebra. In the case of
equality, i.e.,~the massless unitary representations, this construction
is finitely reducible.

Of course, the genuine free field realization is obtained by
making use of the free field realization of $\hat{\sl_2}$
given in \cite{W}.

Let $\alpha =\alpha_2 |_{\fh^{\natural}}$.  Then
$\Delta^{\natural}=\{ \pm \alpha \}$, where $\pm \alpha$ are even
of multiplicity~$1$, and $\Delta'=\{ \pm \alpha /2 \}$, where
$\pm \alpha/2$ are odd of multiplicity~$2$.  The set of positive
roots $\Delta^+_W$ consists of the set of even roots
\begin{displaymath}
  \{ (\pm \alpha ,m) |m \in \NN \} \cup \{(\alpha ,0)\}
     \cup \{ (0,m)|m \in \NN \} \, ,
\end{displaymath}
where the multiplicity of roots from the first two subsets is~$1$ and from the third is $2$, and the set of odd roots
\begin{displaymath}
  \{ (\pm \alpha  /2,m)|m \in \tfrac{1}{2}+\ZZ_+ \} \, ,
\end{displaymath}
all having multiplicity~$2$.

In order to write down the determinant formula, let $\Lambda
=\tfrac{j}{2}\alpha$, where $j \in \CC$.  The highest weight
$(\Lambda ,h) \equiv (j,h)$ of a Verma module over $N=4$
superconformal algebra is determined by $h=$~the lowest
eigenvalue of $L_0$ and $j=$~the eigenvalue of $J_{0}^0$ on the
highest weight vector.  We have:
\begin{eqnarray*}
  h_{n,m} (k,j) &=& \tfrac{1}{4k}((mk-n)^2-(k+1)^2-j(j+2))\, , \\
  \varphi_{n,m,\pm \alpha}(k,j) &=& mk + n \mp (j+1)\, , \\
  h_{m,\pm \alpha /2}(k,j) &=& (m^2-\tfrac{1}{4})k\mp m(j+1)
     -\tfrac{1}{2}\, .
\end{eqnarray*}
Hence Remark~\ref{rem:7.2} gives the following determinant
formula (conjectured in \cite{KeR}):
\begin{eqnarray*}
&&  {\rm det}_{\hat{\eta}} (k,h,j ) = \prod_{m,n \in \NN}
   (4kh -(mk-n)^2 +(k+1)^2+j(j+2))^{P_W(\hat{\eta}-(0,mn))}\\
&& \times \prod_{m,n \in \NN} \varphi_{n,m,-\alpha}
   (k,j)^{P_W (\hat{\eta}-n(-\alpha ,m))}
   \varphi_{n,m-1,\alpha}(k,j)^{P_W(\hat{\eta} -n(\alpha ,m-1))}\\
&& \times \prod_{m \in \tfrac{1}{2}+\ZZ_+}
   \Big(\big(h-h_{m,-\alpha /2}
   (k,j)\big)^{2P_{W,(-\alpha/2,m)}(\hat{\eta}-(-\alpha/2,m))}
   (h-h_{m,\alpha/2} (k,j))^{2P_{W,(\alpha/2,m)}
     (\hat{\eta}-(\alpha /2,m))}\Big)\, .
\end{eqnarray*}

\subsection{$N=3$ superconformal algebra.} \qquad
\label{sec:8.5}

As we shall see, the $N=3$ superconformal algebra is $W_k (\fg
,e_{-\theta})$, where $\fg =osp (3|2)$, tensored with one free
fermion.  The simple roots of $\fg$ are $\alpha_1$, $\alpha_2$,
where $\alpha_1$ is odd and $\alpha_2$ is even, with scalar
products:
\begin{displaymath}
  (\alpha_1 | \alpha_1)=0 \, , \,  (\alpha_1 | \alpha_2)=1/2 \, ,  \,
      (\alpha_2 | \alpha_2)=-1/2 \, , \,
\end{displaymath}
with respect to the bilinear form $(a|b)=-\str \, ab $.  The
subalgebra $\fg^{\natural} \cong sl_2$ corresponds to the root
$\alpha_2$ so that its dual Coxeter number $h\spcheck_0=-1/2$ and
$\fh^{\natural}=\CC \alpha_2$.  Furthermore, $S_{1/2}=\{
\alpha_1, \alpha_1 + \alpha_2 , \alpha_1 + 2\alpha_2\}$
consists of all odd positive roots of $\fg$.  All positive even
roots are $\alpha_2$
and $\theta = 2(\alpha_1+\alpha_2)$.  For each positive
root $\alpha$ we choose root vectors $e_{\alpha}$ and
$e_{-\alpha}$ such that $(e_{\alpha}| e_{-\alpha})=1$.

We use the following notation:
\begin{displaymath}
  h_{mn}=m\alpha_1+n\alpha_2 \, , \, e_{mn}=e_{h_{mn}}\, , \,
  f_{mn}=e_{-h_{mn}} \, .
\end{displaymath}
We have:  $\theta =h_{22}$, so that $x=h_{11}$ and $f=f_{22}$.
We choose the positive root vectors such that the
non-zero brackets are as follows:
\begin{displaymath}
  [e_{10}, e_{01}]=-e_{11}\, ,\,
  [e_{10}, e_{12}]=e_{22}\, ,\,
  [e_{01}, e_{11}]=e_{12} \, , \,
  [e_{11}, e_{11}]=-e_{22} \, .
\end{displaymath}
Then $\langle e_{10},e_{12}\rangle_{\ne}=1$,
$\langle e_{11},e_{11}\rangle_{\ne}=-1$, hence all non-zero
$\lambda$-brackets between the neutral fermions $\Phi_{10}$,
$\Phi_{12}$ and $\Phi_{11}$ are as follows:
\begin{displaymath}
  [\Phi_{10} {}_{\lambda} \Phi_{12}] =1 \, , \,
  [\Phi_{11} {}_{\lambda} \Phi_{11}]=-1 \, , \, \hbox{ and  }
  \Phi^{10}=\Phi_{12} \, , \, \Phi^{12}=\Phi_{10} \, ,  \,
  \Phi^{11}=-\Phi_{11}\, .
\end{displaymath}

Since $h\spcheck =1/2$ and $h\spcheck_0=-1/2$, we have by
(\ref{eq:2.5}):
  \begin{displaymath}
    [J^{(a)}{}_{\lambda} J^{(b)}]=J^{([a,b])}
       +\lambda (k+1)(a|b) \, , \, a,b \in \fg_0 \, .
  \end{displaymath}

The $d_0$-closed fields, provided by Theorem~\ref{th:4.1}, which
strongly generate $W_k (\fg ,e_{-\theta})$ are as follows:
\begin{eqnarray*}
  J^0 &=& -4J^{(h_{01})}- 2 :\Phi_{10}\Phi_{12}:
         \, , \, J^+ = -2J^{(e_{01})}-2:\Phi_{12}\Phi_{11}:
         \, , \, J^-=2J^{(f_{01})}-:\Phi_{10}\Phi_{11}:\, , \, \\
  G^+ &=& J^{(f_{10})}+:\Phi_{12} J^{(h_{10})}:
         -\tfrac{1}{2}:\Phi_{11}J^{(e_{01})}:
         + (k+1)\partial \Phi_{12}\, , \\
  G^0 &=& J^{(f_{11})} -: \Phi_{12}J^{(f_{01})}:
        + \tfrac{1}{2}:\Phi_{10} J^{(e_{01})}:-:
        \Phi_{11}J^{(h_{11})}:-(k+1)\partial \Phi_{11}\, ,\\
  G^- &=& J^{(f_{12})} +:\Phi_{10}J^{(h_{12})}:
          + :\Phi_{11} J^{(f_{01})}+(k+1)\partial
          \Phi_{10}\, ,\\
   L &=& -\tfrac{1}{k+1/2}(J^{(f)}+:\Phi_{12}J^{(f_{12})}
         :+: \Phi_{11}J^{(f_{11})}
         :+:\Phi_{10}J^{(f_{10})})\\
     && +\tfrac{1}{k+1/2}(:(J^{(h_{11})})^2 :-
         :(J^{(h_{01})})^2 : + \tfrac{1}{2}:
         J^{(e_{01})}J^{(f_{01})}:
        + \tfrac{1}{2} : J^{(f_{01})} J^{(e_{01})}:
        +(k+1)\partial J^{(h_{11})})\\
     && - (:\Phi_{12} \partial \Phi_{10}:+:
         \Phi_{10}\partial \Phi_{12}:-:
         \Phi_{11}\partial \Phi_{11}:)\, .
\end{eqnarray*}

The field $L$ is a Virasoro field with central charge
\begin{displaymath}
  c=-6k-7/2 \, ,
\end{displaymath}
the fields $J^0$, $J^{\pm}$ (resp.~$G^0$, $G^{\pm}$) being
primary of conformal weight~$1$ (resp.~$3/2$).  The remaining
non-zero $\lambda$-brackets are as follows:
\begin{eqnarray*}
  [{J^0}{}_{\lambda} J^{\pm}] &=& \pm 2 J^{\pm} \, , \,
     [{J^0}{}_{\lambda} J^0] = \tfrac{4\lambda}{3}
     (c+\tfrac{1}{2})\, , \, [{J^+}{}_{\lambda}J^-]=J^0+
     \tfrac{2\lambda}{3} (c+\tfrac{1}{2})\, , \\
{}  [{J^0}{}_{\lambda} G^{\pm}] &=& \pm2 G^{\pm} \, , \,
     [{J^+}{}_{\lambda}G^-] =-2G^0 \, , \, [{J^-}{}_{\lambda}G^+]
     = - G^0 \, , \, [{J^+}{}_{\lambda}G^0]=-2G^+ \, , \,
     [{J^-}{}_{\lambda} G^0]=- G^- \, , \\
{}  [{G^0}{}_{\lambda}G^0] &=& \tfrac{1}{6}(c+\tfrac{1}{2})L-\tfrac{1}{16} :J^0J^0:
    + \tfrac{\lambda ^2}{36}(c+\tfrac{1}{2})(c-1) \,  , \,
     [{G^-}{}_{\lambda}G^-] =-\tfrac{1}{4}:J^-J^- :\, , \\
{}  [{G^+}{}_{\lambda}G^+] &=&-\tfrac{1}{16}:J^+J^+:\, , \, [{G^+}{}_{\lambda} G^0]
       = -\tfrac{1}{16} : J^0J^+:+\tfrac{1}{4}
      (\tfrac{1}{2}-\tfrac{1}{6} (c+\tfrac{1}{2}))
      (\partial+2 \lambda) J^+ - \tfrac{\lambda}{8} J^+ \, , \,  \\
{}  [{G^-}{}_{\lambda}G^0] &=& \tfrac{1}{8}:J^0J^- :+ \tfrac{1}{2}
      (\tfrac{1}{2} - \tfrac{1}{6} (c+\tfrac{1}{2}))
      (\partial +2\lambda) J^- -\tfrac{\lambda}{4} J^- \, , \\
{}  [{G^+}{}_{\lambda}G^-] &=& -\tfrac{1}{6} (c+\tfrac{1}{2})L
      +\tfrac{1}{8} : J^- J^+ :+\tfrac{1}{4}
      (\tfrac{1}{2}-\tfrac{1}{6}(c+\tfrac{1}{2}))
      (\partial +2\lambda)J^0 -\tfrac{\lambda}{8} J^0 -
      \tfrac{\lambda^2}{36} (c+\tfrac{1}{2})(c-1)\, .
\end{eqnarray*}

In order to remove the quadratic terms in $\lambda$-brackets we
introduce a free fermion $\Phi$ with the $\lambda$-bracket
$[\Phi_{\lambda}\Phi]= -(k+\tfrac{1}{2})$, and modify the fields $L$ and $G$ as
follows:
\begin{eqnarray*}
  \tilde{L} &=& L-\tfrac{1}{2k+1}:\partial \Phi \Phi :\, , \,
  \tilde{G}^+ =\tfrac{i}{\sqrt{k+1/2}}G^+ -\tfrac{1}{4k+2}:J^+\Phi : \, , \,\\
  \tilde{G}^- &=& \tfrac{-i}{\sqrt{k+1/2}}G^- -\tfrac{1}{2k+1}:J^- \Phi : \, , \,
  \tilde{G}^0 =\tfrac{-i}{\sqrt{k+1/2}}G^0 +\tfrac{1}{4k+2}:J^0 \Phi : \, .
\end{eqnarray*}
Then $\tilde{L}$ is a Virasoro field with central charge
\begin{displaymath}
  \tilde{c} =c+ \tfrac{1}{2} =-6k-3 \, ,
\end{displaymath}
and we obtain the $N=3$ superconformal algebra, strongly
generated by $\tilde{L}$, the primary fields $J^{\pm}$ and $J^0$
of conformal weight $1$, the primary fields $\tilde{G}^{\pm}$ and
$\tilde{G}^0$ of conformal weight $3/2$ and the primary field
$\Phi$ of conformal weight $1/2$.  The remaining changed
$\lambda$-brackets are as follows:
\begin{eqnarray*}
   [{J^0}{}_{\lambda}\tilde{G}^0] &=& -2\lambda
       \Phi \, , \, [{J^+}{}_{\lambda}\tilde{G}^-]
       =-2\tilde{G}^0 +2\lambda \Phi \, , \,
 [{J^-}{}_{\lambda}\tilde{G}^+] =
        \tilde{G}^0 +\lambda \Phi \, , \,
        [{\tilde{G}^{\pm}}{}_{\lambda}\tilde{G}^{\pm}]=0\, , \\
{} [{\tilde{G}^+}{}_{\lambda} \tilde{G}^-] &=& \tilde{L} +\tfrac{1}{4}
        (\partial +2\lambda)J^0 +
        \tfrac{\lambda^2}{6}\tilde{c} \, ,
        [{\tilde{G}^+}{}_{\lambda}\tilde{G}^0] =
        \tfrac{1}{4}(\partial +2\lambda)J^+ \, ,
         [{\tilde{G}^0}{}_{\lambda}\tilde{G}^0]=\tilde{L}+\tfrac{\lambda^2}{6} \tilde{c} \, , \\
{} [{\tilde{G}^-}{}_{\lambda} \tilde{G}^0] &=&-\tfrac{1}{2}
        (\partial +2\lambda) J^- \, , \,
        [{\tilde{G}^+}{}_{\lambda} \Phi ] = \tfrac{1}{4}J^+ \, , \,
        [{\tilde{G}^-}{}_{\lambda}\Phi]  =\tfrac{1}{2} J^- \, , \,
        [{\tilde{G}^0}{}_{\lambda}\Phi] =-\tfrac{1}{4} J^0 \, .
\end{eqnarray*}
We have:  $\omega (J^0_n)=J^0_{-n}$, $\omega (J^+_n)=2J^-_{-n}$,
$\omega (G^0_n)=G^0_{-n}$, $\omega (G^{\pm}_n) = G^{\mp}_{-n}$,
$\omega (\Phi_n)=\Phi_{-n}$.

The free field realization of the $N=3$ superconformal algebra is
obtained from the formulas for $\tilde{L}$, $J$'s and
$\tilde{G}$'s by removing the terms containing $J^{(f)}$,
$J^{(f_{10})}$, $J^{(f_{11})}$ and  $J^{(f_{12})}$, and
replacing  $J^{(u)}$ by $u$ for $u \in \fg_0 = \fh + \CC e_{0
  1}+\CC f_{01}$.

Let $\alpha = \alpha_2 |_{\fh^{\natural}}$.  Then
$\Delta^{\natural}_+ = \{ \alpha \}$ and $\Delta'=\{ \pm \alpha ,
0 \}$.  The set of positive roots $\Delta^+_{N=3}$ of the $N=3$
superconformal algebra consists of the
set of even roots
\begin{displaymath}
  \{ (\pm \alpha ,m)|m \in \NN \} \cup \{ (\alpha ,0)\} \cup
  \{ (0,m)|m \in \NN \}\, ,
\end{displaymath}
where the multiplicity of a root from the first two subsets is
$1$ and from the third is $2$, and the set of odd roots
\begin{displaymath}
  \{ (\pm \alpha ,m) |m \in \tfrac{1}{2} + \ZZ_+ \}
  \cup \{ (0,m)| m \in \tfrac{1}{2} +\ZZ_+ \} \, ,
\end{displaymath}
where the multiplicity of a root from the first subset is $1$ and
from the second is $2$ (due to the added free fermion $\Phi$).

In order to write down the determinant formula, let $\Lambda
=\tfrac{j}{2}\alpha$, where $j \in \CC$.  The highest weight
$(\Lambda , h) \equiv (j , h)$ of the Verma module over $N=3$
superconformal algebra is determined by $h=$~the lowest
eigenvalue of $L_0$ and $j=$~the eigenvalue of $J^0_0$ on the
highest weight vector.  We have:
\begin{eqnarray*}
  h_{n,2m} (k,j) &=& \tfrac{1}{2(2k+1)}((m(2k+1)-n)^2 -
     (k+1)^2 -\tfrac{1}{4} j(j+2))\, , \\
  \varphi_{n,m,\pm \alpha}(k,j) &=& \tfrac{1}{4}
     (2m(2k+1)+n \mp (j+1))\, , \\
  h_{m,\pm \alpha} (k,j) &=& (m^2 -\tfrac{1}{4})k
     + \tfrac{1}{2} (m^2-\tfrac{3}{4}\mp m (j+1))\, .
\end{eqnarray*}
Hence Remark~\ref{rem:7.2} gives the following determinant
formula (cf.~\cite{M}):
\begin{eqnarray*}
  &&  {\rm det}_{\hat{\eta}} (k,h,j) =
    (k+\tfrac{1}{2})^{P_{N=3,(0,1/2)}(\hat{\eta}-(0,1/2))}\\
  && \times \prod_{\substack{m,n \in \tfrac{1}{2}\NN \\m-n\in \ZZ}}
     (h-h_{n,2m} (k,j))^{P_{N=3} (\hat{\eta}-(0,2mn))}\\
  && \times \prod_{m,n \in \NN} \varphi_{n,m,-\alpha}
     (k,j)^{P_{N=3} (\hat{\eta}-n(-\alpha ,m))}\varphi_{n,m-1,\alpha}
     (k,j)^{P_{N=3} (\hat{\eta}-n (\alpha ,m-1))}\\
  && \times \prod_{m \in \tfrac{1}{2}+ \ZZ_+}
      (h-h_{m,-\alpha} (k,j))^{P_{N=3,(-\alpha ,m)}
        (\hat{\eta}-(-\alpha, m))}
      (h-h_{m,\alpha}(k,j))^{P_{N=3,(\alpha ,m)}(\hat{\eta}-(\alpha, m))}
\end{eqnarray*}

\subsection{Big $N=4$ superconformal algebra.} \qquad
\label{sec:8.6}

In this subsection, $\fg = D (2,1;a)$, where $a \in \CC
\backslash \{ -1 ,0 \}$, is the family of exceptional
$17$-dimensional Lie superalgebras.  The big $N=4$ superconformal
algebra (see \cite{KL}, \cite{S}, \cite{STP}) is obtained from
$W_k(\fg, e_{-\theta})$ by tensoring it with four free fermions
and a free boson (cf. \cite{GS}). Recall that $\fg$ is a
contragredient Lie superalgebra with the following matrix of
scalar products of simple roots $\alpha_i$ $(i=1,2,3)$ \cite{K1}:
\begin{eqnarray*}
  ((\alpha_i | \alpha_j))^3_{i,j=1} =-\tfrac{1}{a+1}\left(
    \begin{array}{ccc}
      2 & -1 & 0 \\
      -1 & 0 & -a\\
      0 & -a & 2a
    \end{array}\right) \, .
\end{eqnarray*}
As in Section~\ref{sec:8.5}, we use the following notation:
\begin{displaymath}
  h_{mnp} = m \alpha_1 +n\alpha_2 +p\alpha_3 \, , \,
  e_{mnp} =e_{h_{mnp}} \, , \, f_{mnp}=e_{-h_{mnp}} \, .
\end{displaymath}
The set of even (resp.~odd) positive roots is:
\begin{displaymath}
  h_{100}\, , \, h_{001}\, , \, h_{1 2 1}
  \hbox{( resp.  }h_{0 1 0} \, , \, h_{1 10 }\, , \,
  h_{0 1 1} \, , \, h_{1 1 1}) \, .
\end{displaymath}
The highest root $\theta =h_{1 2 1}$, and the matrix of
scalar products is normalized in such a way that $(\theta |
\theta )=2$.

We choose positive root vectors $e_{\alpha}$ such that all
non-zero brackets (up to the order) between them are as follows:
$ [e_{10 0}, e_{0 1 0}] =e_{1 1 0}$\, , \,
     $[e_{1 0 0}, e_{0 1 1}] =e_{111}$\, ,
 $[e_{01 0}, e_{0 0 1}] =e_{0 1 1}$\, , \,
     $[e_{0 1 0}, e_{1 1 1}] =e_{121}$\, ,
  $[e_{00 1}, e_{1 1 0}] =-e_{1 1 1}$\, , \,
     $[e_{1 1 0}, e_{0 1 1}] =-e_{121}$\, ,
and we choose negative root vectors $f_{\alpha}$ such that
$[e_{\alpha},f_{\alpha}]=\alpha$.

For the quantum reduction we take the following $\sl_2$-triple:
$f=f_{1 2 1}$, $e=\tfrac{1}{2}e_{1 2 1}$, $x=\tfrac{1}{2}
h_{1 2 1}$, so that $\fh^{\natural}=\CC
\alpha_1+\CC\alpha_3$, and $\fg^{\natural} =\fh^{\natural}+\CC
e_{\alpha_1}+\CC f_{\alpha_1} +\CC e_{\alpha_3}+\CC f_{\alpha_3}$
is isomorphic to $\sl_2 \oplus \sl_2$.  The subspace $\fg_{1/2}$
is spanned by four odd elements:  $e_{0 \, 1\, 0}$, $e_{1 1
  1}$ , $e_{110}$, $e_{011}$, and all non-zero values of
the (symmetric) bilinear form $\langle . \, , \, . \rangle_{\ne}$
on $\fg_{1/2}$ are:
\begin{displaymath}
  \langle e_{010}\, ,\,  e_{111}\rangle_{\ne}=1 \, , \,
  \langle e_{110} \, , \, e_{011}\rangle_{\ne}=-1\, .
\end{displaymath}
Hence the corresponding free neutral fermions satisfy the
following non-zero $\lambda$-brackets:
\begin{eqnarray*}
&&  [{\Phi_{010}}{}_{\lambda} \Phi_{111}] =1 \, , \,
    [{\Phi_{110}}{}_{\lambda}\Phi_{011}]=-1\, , \\
\noalign{\hbox{and  }}
&&   \Phi^{010}=\Phi_{111}\, , \,
      \Phi^{111}=\Phi_{010}\, , \,
      \Phi^{110}=-\Phi_{011} \, , \,
      \Phi^{011}=-\Phi_{110}\, .
\end{eqnarray*}

The $d_0$-closed fields, provided by Theorem~\ref{th:4.1}, which
strongly generate $W_k (\fg ,e_{-\theta})$ are as follows:
\begin{eqnarray*}
 J^0 &=&-(a+1) J^{(h_{100})} -(:\Phi_{010}\Phi_{111}:
        -:\Phi_{110} \Phi_{0 11} :)\, , \\
  J^{\prime\, 0} &=& -\tfrac{a+1}{a}J^{(h_{001})}
                -(:\Phi_{010}\Phi_{111}: - :
                \Phi_{110} \Phi_{011}:)\, , \\
  J^+ &=& J^{(e_{100})} +:\Phi_{110}\Phi_{111}
          :\, , \, J^{\prime +}=J^{(e_{001})}-:
          \Phi_{011} \Phi_{111} :\, ,\\
  J^- &=& -(a+1)J^{(f_{100})}-: \Phi_{010}\Phi_{011}:\, , \,
          J^{\prime -} =-\tfrac{a+1}{a}J^{(f_{001})} +
          :\Phi_{010} \Phi_{110}:\, ,
\end{eqnarray*}
\begin{eqnarray*}
  G^{++} &=& J^{(f_{010})} +:\Phi_{111}
           J^{(h_{010})}:+\tfrac{1}{a+1}:\Phi_{011}J^{(e_{100})}
           :-\tfrac{a}{a+1}: \Phi_{110} J^{(e_{001})}:\\
           && + \tfrac{a-1}{a+1}:\Phi_{110} \Phi_{011}\Phi_{111}
             :+(k+1) \partial \Phi_{111}\, , \\
  G^{--} &=&  J^{(f_{111})}+:\Phi_{010}
            J^{(h_{111})}:-: \Phi_{110} J^{(f_{100})}
            :+: \Phi_{011} J^{(f_{001})}:\\
           && -\tfrac{a-1}{a+1}:\Phi_{010}\Phi_{110}\Phi_{011}
             :+(k+1)\partial \Phi_{010}\, , \\
  G^{-+} &=& J^{(f_{110})}-:\Phi_{011}
            J^{(h_{110})}:+: \Phi_{111}
            J^{(f_{100})} :+\tfrac{a}{a+1}: \Phi_{010}
            J^{(e_{001})}:\\
            && -\tfrac{a-1}{a+1}:\Phi_{010}\Phi_{011}\Phi_{111}
              :-(k+1)\partial \Phi_{011}\, , \\
  G^{+-} &=& J^{(f_{011})}-:\Phi_{110}
            J^{(h_{011})} :-\tfrac{1}{a+1}:\Phi_{010}
            J^{(e_{100})}:-:\Phi_{111}
            J^{(f_{001})}:\\
            && +\tfrac{a-1}{a+1}:\Phi_{010}\Phi_{110}\Phi_{111}
              :-(k+1)\partial \Phi_{110}\, ,\\
  L &=& -\tfrac{1}{k} J^{(f)}-\tfrac{1}{k}
          \sum_{\alpha \in S_{1/2}} :\Phi_{\alpha}J^{(f_{\alpha})}:\\
      &&  +\tfrac{1}{4k} (:(J^{(x)})^2:-(a+1):(J^{(h_{100})})^2:-\tfrac{a+1}{a}
           :(J^{(h_{001})})^2:)+\tfrac{k+1}{k}\partial J^{(x)}\\
%
%          &&  +\tfrac{1}{4(a+1)} (:J^{(\alpha_1)} J^{(\alpha_1)}
%             :+\tfrac{1}{a}:J^{(\alpha_3)}J^{(\alpha_3)}:)
%             -(k+1)\partial J^{(x)}\\
           && +\tfrac{1}{2k} (: J^{(e_{100})}J^{(f_{100})}:
             +:J^{(f_{100})}J^{(e_{100})}:+: J^{(e_{001})}
             J^{(f_{001})}:+:
             J^{(f_{001})}J^{(e_{001})}:)\\
          && -\tfrac{1}{2} (: \Phi_{111}\partial
             \Phi_{010} :+: \Phi_{010}\partial
             \Phi_{111}:-:\Phi_{110}\partial
             \Phi_{011}:-: \Phi_{011}
             \partial \Phi_{110}:) \, .
\end{eqnarray*}

The fields $J$'s (resp.~$G$'s) are  primary with respect to $L$
of conformal weight $1$ (resp.~$3/2$).  The non-zero
$\lambda$-brackets between these fields are as follows:
\begin{eqnarray*}
  [{J^0}{}_{\lambda}J^0] &=& -2\lambda ((a+1)k+1) \, , \,
     [{J^{\prime 0}}{}_{\lambda}J^{\prime 0}] =-2 \lambda
       (\frac{a+1}{a}k+1) \, , \\
{} [{J^+}{}_{\lambda}J^-] &=& J^0 -\lambda ((a+1)k+1)\, , \,
     [{J^{\prime +}}{}_{\lambda} J^{\prime -}]=
     J^{\prime 0}-\lambda \ (\frac{a+1}{a}k+1 ) \, , \\
{} [{J^0}{}_{\lambda}J^{\pm}] &=& \pm 2 J^{\pm}\, , \,
     [{J^{\prime 0}}{}_{\lambda} J^{\prime \pm}]=\pm
     2J^{\prime \pm}\, , \\
{} [{J^0}{}_{\lambda} G^{\pm\pm}] &=& \pm G^{\pm\pm}\, , \,
     [{J^0}{}_{\lambda}G^{\pm\mp}] =\pm G^{\pm\mp}\, , \,
   [{J^{\prime 0}}{}_{\lambda} G^{\pm\pm}] = \pm G^{\pm\pm}
     \, , \, [{J^{\prime 0}}{}_{\lambda} G^{\pm\mp}]
     =\mp G^{\pm\mp}\, , \\
{} [{J^+}{}_{\lambda}G^{- -}]&=& -G^{+-}\, , \,
     [{J^+}{}_{\lambda}G^{-+}] = -G^{++}\, , \,
     [{J^{\prime +}}{}_{\lambda} G^{--}] =G^{-+}\, , \,
     [{J^{\prime +}}{}_{\lambda}G^{+-}]=G^{++}\, , \\
{} [{J^-}{}_{\lambda} G^{++}] &=& -G^{-+}\, , \,
     [{J^-}{}_{\lambda}G^{+-}] = -G^{--}\, , \,
     [{J^{\prime -}}{}_{\lambda}G^{++}] = G^{+-}\, , \,
     [{J^{\prime -}}{}_{\lambda}G^{-+}]=G^{--}\, , \\
{} [{G^{++}}{}_{\lambda}G^{++}] &=& \frac{2a}{(a+1)^2}:
     J^+J^+ : \, , \, [{G^{--}}{}_{\lambda}G^{--}] =
     \frac{2}{(a+1)^2} : J^-J^- : \, , \\
{} [{G^{-+}}{}_{\lambda}G^{-+}] &=& -\frac{2a}{(a+1)^2}:
     J^-J^{\prime +}: \, , \, [{G^{+-}}{}_{\lambda}G^{+-}]
     =-\frac{2a}{(a+1)^2}:J^+J^{\prime -}:\, , \\
{} [{G^{++}}{}_{\lambda}G^{--}] &=& kL+\frac{1}{4}
     (\frac{1}{a+1}:J^0J^0 :+\frac{a}{a+1}:
  J^{\prime 0}J^{\prime 0}:
   -\frac{1}{(a+1)^2}:( J^0+aJ^{\prime 0})^2:) \\
    && +\frac{a}{(a+1)^2}(:J^+J^- :+: J^{\prime +}J^{\prime -}:)
       -\frac{1}{2(a+1)}\partial (J^0+aJ^{\prime 0})\\
    && +\frac{k+1}{2(a+1)}(\partial +2\lambda)(J^0+aJ^{\prime 0})
       -\frac{\lambda}{(a+1)^2} (J^0+a^2J^{\prime 0})-
       \lambda^2 (k(k+1)+\frac{a}{(a+1)^2})\, , \\
{} [{G^{-+}}{}_{\lambda}G^{+-}] &=& -kL +\frac{1}{4}
       (-\frac{1}{a+1}: J^0J^0:-\frac{a}{a+1}:
       J^{\prime 0}J^{\prime 0}:+\frac{1}{(a+1)^2}:
       (J^0-aJ^{\prime 0})^2 :)\\
    && -\frac{a}{(a+1)^2}(:J^+J^-:+:J^{\prime +}J^{\prime -}:)
       +\frac{1}{2(a+1)}\partial (J^0+aJ^{\prime 0})-
       \frac{1}{(a+1)^2}\partial J^0 \\
    && +\frac{k+1}{2(a+1)}(\partial +2\lambda)(J^0-aJ^{\prime 0})
       -\frac{\lambda}{(a+1)^2} (J^0-a^2J^{\prime 0}) +
       \lambda^2  (k(k+1)+\frac{a}{(a+1)^2}) \, ,
\end{eqnarray*}
\begin{eqnarray*}
{} [{G^{++}}{}_{\lambda}G^{-+}] &=& \frac{a}{(a+1)^2}:J^0J^{\prime+}
       :+\frac{a}{a+1}(\frac{a}{a+1}+k+1)(\partial +2\lambda)
       J^{\prime +}\, , \\
{} [{G^{++}}{}_{\lambda}G^{+-}] &=& -\frac{a}{(a+1)^2}:
       J^{\prime 0}J^+ : -\frac{1}{a+1}(k+1+\frac{1}{a+1})
       (\partial +2\lambda)J^+ \, , \\
{} [{G^{--}}{}_{\lambda}G^{-+}] &=& \frac{a}{(a+1)^2}:
       J^{\prime 0}J^- :+\frac{1}{a+1}(k+1-\frac{1}{a+1})
       (\partial +2\lambda)J^- \, , \\
{} [{G^{--}}{}_{\lambda}G^{+-}] &=& -\frac{a}{(a+1)^2}:
       J^0J^{\prime -} :-\frac{a}{a+1}(k+1-\frac{a}{a+1})
       (\partial +2\lambda)J^{\prime -}\, .
\end{eqnarray*}

Since $h\spcheck =0$ and $\sdim \fg =1$, by (\ref{eq:5.7}), the
Virasoro central charge is equal
\begin{displaymath}
  c=-6k-3 \, .
\end{displaymath}

The big $N=4$ superconformal algebra is obtained from $W_k
(D(2,1;a),e_{-\theta})$ by tensoring the latter vertex algebra by
the vertex algebra strongly generated by four (odd) free fermions
$\sigma^{--}$, $\sigma^{++}$, $\sigma^{+-}$, $\sigma^{-+}$ with
non-zero $\lambda$-brackets $[{\sigma^{--}}{}_{\lambda}
\sigma^{++}]=k$, $[{\sigma^{+-}}{}_{\lambda}
\sigma^{-+}]=k$, and one (even) free boson $\xi$ with
$\lambda$-bracket $[\xi_{\lambda} \xi]=\lambda k$.

Then the five fields $\sigma^{--}$, $\sigma^{++}$, $\sigma^{+-}$,
$\sigma^{-+}$, $\xi$ along with the following modifications of
the six fields $J^0$, $J^{\prime 0}$, $J^{\pm}$, $J^{\prime \pm}$, the four
fields $G^{++}$, $G^{--}$, $G^{+-}$, $G^{-+}$ and the field $L$,
close in the big $N=4$ superconformal algebra with Virasoro central charge
$\tilde{c} = -6k$ (cf.~\cite{IKL}):
\begin{eqnarray*}
  \tilde{J}^0 &=& J^0 -\tfrac{1}{k}: \sigma^{--} \sigma^{++}:
          + \tfrac{1}{k}: \sigma^{+-} \sigma^{-+}: \, , \,
          \tilde{J}^{\prime 0} = J^{\prime 0} +\tfrac{1}{k}: \sigma^{--}\sigma^{++}:
          + \tfrac{1}{k}:\sigma^{+-} \sigma^{-+}:\, ,\\
   \tilde{J}^+ &=& J^+ +\tfrac{a}{k}: \sigma^{+-}\sigma^{++}: \, , \,
          \tilde{J}^- = J^- + \tfrac{1}{ak}
          :\sigma^{--}\sigma^{-+}: \, , \\
   \tilde{J}^{\prime +} &=& J^{\prime +} +
          \tfrac{1}{k}:\sigma^{-+}\sigma^{++}: \, , \,
          \tilde{J}^{\prime -} = J^{\prime -}+
          \tfrac{1}{k}:\sigma^{--}\sigma^{+-}: \, , \\
   \tilde{G}^{++} &=& \tfrac{1}{\sqrt{k}}G^{++}-\tfrac{1}{k(a+1)}:J^+ \sigma^{-+}:
          + \tfrac{a}{k(a+1)}: J^{\prime +}\sigma^{+-}:
          + \tfrac{a}{2k(a+1)}:J^0 \sigma^{++}:\\
      && -\tfrac{a}{2k(a+1)} :J^{\prime 0}\sigma^{++}:
          +\tfrac{1}{k} \left( \tfrac{a}{2}\right)^{1/2}:\xi \sigma^{++}:
          +\tfrac{a}{k^2 (a+1)}:\sigma^{++}\sigma^{+-}\sigma^{-+}:\, ,\\
    \tilde{G}^{--} &=&\tfrac{1}{\sqrt{k}} G^{--} + \tfrac{a}{k(a+1)}: J^- \sigma^{+-}:
          -\tfrac{1}{k(a+1)}: J^{\prime -} \sigma^{-+}:
          -\tfrac{1}{2k(a+1)}: J^0 \sigma^{--}:\\
      && + \tfrac{1}{2k(a+1)}: J^{\prime 0}\sigma^{--}:
          +\tfrac{1}{k}\left( \tfrac{1}{2a}\right)^{1/2}:\xi \sigma^{--}:
          -\tfrac{1}{k^2(a+1)}:\sigma^{--}\sigma^{+-}\sigma^{-+}:\, ,\\
    \tilde{G}^{+-} &=& -\tfrac{1}{\sqrt{k}}G^{+-} +\tfrac{1}{k(a+1)}:J^+\sigma^{--}:
          +\tfrac{a}{k(a+1)}: J^{\prime -}\sigma^{++}:
          +\tfrac{a}{2k(a+1)}: J^0 \sigma^{+-}: \,  \\
      && +\tfrac{a}{2k(a+1)}: J^{\prime 0} \sigma^{+-}:
          + \tfrac{1}{k}\left( \tfrac{a}{2}\right)^{1/2}:\xi \sigma^{+-}:
          -\tfrac{a}{k^2 (a+1)}:\sigma^{+-}\sigma^{--}\sigma^{++}:\, ,\\
    \tilde{G}^{-+} &=&\tfrac{1}{\sqrt{k}}G^{-+}-\tfrac{1}{k(a+1)}: J^{\prime +}
          \sigma^{--}:-\tfrac{a}{k(a+1)}: J^- \sigma^{++}:
          -\tfrac{1}{2k(a+1)}:J^0 \sigma^{-+}: \,  \\
      && -\tfrac{1}{2k(a+1)}:J^{\prime 0}\sigma^{-+}:+\tfrac{1}{k}
          \left( \tfrac{1}{2a}\right)^{1/2}:\xi \sigma^{-+}:
          +\tfrac{1}{k^2 (a+1)}:\sigma^{-+}\sigma^{--}\sigma^{++}:\,,\\
    \tilde{L} &=& L + \tfrac{1}{2k}(: \partial \sigma^{--}\sigma^{++}:
           +:\partial \sigma^{++} \sigma^{--}:
           +:\partial\sigma^{-+}\sigma^{+-}:
           +:\partial \sigma^{+-}\sigma^{-+}:+ : \xi^2 :)\, .
\end{eqnarray*}

The fields $\sigma$'s, $\xi$, $\tilde{J}$'s and $\tilde{G}$'s are
primary with respect to $\tilde{L}$ of conformal weight $1/2$,
$1$,  $1$ and $3/2$, respectively.
The $\lambda$-brackets for the pairs
$(\tilde{J},
\tilde{J}')$ are zero, and the non-zero $\lambda$-brackets for the pairs
$(\tilde{J}, \tilde{G})$, $(\tilde{J}', \tilde{G})$,
$(\tilde{J}, \tilde{J}) $, and $(\tilde{J}',\tilde{J}')$ are as follows:
\begin{eqnarray*}
  [\tilde{J}^0 {}_{\lambda} \tilde{G}^{++}]
     &=& \tilde{G}^{++}-\lambda a \sigma^{++}\, , \,
         [\tilde{J}^0 {}_{\lambda} \tilde{G}^{--}]
         = -\tilde{G}^{--}+\lambda \sigma^{--} \, , \,
         [\tilde{J}^0 {}_{\lambda} \tilde{G}^{-+}]
         = -\tilde{G}^{-+}+\lambda \sigma^{-+}\, , \\
{} [\tilde{J}^0 {}_{\lambda} \tilde{G}^{+-}]
     &=& \tilde{G}^{+-}-\lambda a\sigma^{+-} \, , \,
         [\tilde{J}^+ {_{\lambda}} \tilde{G}^{--}]
         = \tilde{G}^{+-} -\lambda a \sigma^{+-}\, , \,
         [\tilde{J}^+ {}_{\lambda} \tilde{G}^{-+}]
         =-\tilde{G}^{++}+\lambda a \sigma^{++}\, , \, \\
{} [\tilde{J}^- {}_{\lambda} \tilde{G}^{++}]
     &=& -\tilde{G}^{-+}+\lambda \sigma^{-+} \, , \,
         [\tilde{J}^- {}_{\lambda} \tilde{G}^{+-}]
         =\tilde{G}^{--}-\lambda \sigma^{--} \, , \, \\
{} [\tilde{J}^{\prime 0} {}_{\lambda} \tilde{G}^{++}]
     &=& \tilde{G}^{++}+\lambda \sigma^{++} \, , \,
         [\tilde{J}^{\prime 0}{}_{\lambda} \tilde{G}^{--}]
         = -\tilde{G}^{--} -\tfrac{\lambda}{a}\sigma^{--}\, , \,
         [\tilde{J}^{\prime 0}{}_{\lambda} \tilde{G}^{-+}]
         = \tilde{G}^{-+}+\tfrac{\lambda}{a}\sigma^{-+}\, , \\
{}  [\tilde{J}^{\prime 0}{}_{\lambda}\tilde{G}^{+-}]
      &=& -\tilde{G}^{+-}-\lambda \sigma^{+-}\, , \,
          [\tilde{J}^{\prime +}{}_{\lambda} \tilde{G}^{--}]
          = \tilde{G}^{-+}+\tfrac{\lambda}{a}\sigma^{-+}\, , \,
          [\tilde{J}^{\prime +} {}_{\lambda} \tilde{G}^{+-}]
          = -\tilde{G}^{++}-\lambda \sigma^{++}\, , \\
{}  [\tilde{J}^{\prime -}{}_{\lambda} \tilde{G}^{++}]
       &=& -\tilde{G}^{+-}-\lambda \sigma^{+-}\, , \,
          [\tilde{J}^{\prime -}{}_{\lambda} \tilde{G}^{--}]
          = \tilde{G}^{--}+\tfrac{\lambda}{a}\sigma^{--}\, ,\\
{}  [\tilde{J}^0{}_{\lambda} \tilde{J}^0]
       &=& \lambda \frac{\tilde{c}(a+1)}{3} \, , \,
          [{{\tilde{J}^0}}{}_{\lambda} \tilde{J}^{\pm}]
          = \pm 2 \tilde{J}^{\pm} \, ,\,
          [{\tilde{J}^+}{}_{\lambda} \tilde{J}^-]
          = \tilde{J}^0 +\lambda \frac{\tilde{c} (a+1)}{6}\, , \,
          [{\tilde{J}^{\prime 0}}{}_{\lambda} \tilde{J}^{\prime 0}]
          =\lambda \frac{\tilde{c}(a+1)}{3a} \, ,\\
{} [{\tilde{J}^{\prime 0}}{}_{\lambda} \tilde{J}^{\prime \pm}]
       &=& \pm 2 \tilde{J}^{\prime \pm} \, , \,
          [{\tilde{J}^{\prime +}}{}_{\lambda} \tilde{J}^{\prime -}]
          = \tilde{J}^{\prime 0}+\lambda \frac{\tilde{c}(a+1)}{6a}\,.
\end{eqnarray*}
The non-zero $\lambda$-brackets for the pairs $(\tilde{J}
,\sigma)$ are as follows:
\begin{eqnarray*}
   [{\tilde{J}^0}{}_{\lambda} \sigma^{\pm\pm}] &=&
        \pm \sigma^{\pm\pm} \, , \,
        [{\tilde{J}^{\prime 0}}{}_{\lambda} \sigma^{\pm \mp}] =
        \pm \sigma^{\pm \mp} \, , \\[1ex]
{}  [{\tilde{J}^+}{}_{\lambda} \sigma^{\scriptmm\mp}] &=&
        \pm a \sigma^{\scriptpp \mp} \, , \,
        [{\tilde{J}^-}{}_{\lambda} \sigma^{\scriptpp\pm}]
        =\mp \tfrac{1}{a} \sigma^{\scriptmm \pm} \, , \\[1ex]
{}  [{\tilde{J}^{\prime 0}}{}_{\lambda} \sigma^{\pm\pm}] &=&
        \pm \sigma^{\pm\pm} \, , \,
        [{\tilde{J}^{\prime 0}}{}_{\lambda} \sigma^{\pm \mp}]
        =\mp \sigma^{\pm \mp} \, , \\[1ex]
{}  [{\tilde{J}^{\prime +}{}_{\lambda}} \sigma^{\mp\scriptmm}] &=&
        \pm \sigma^{\mp \scriptpp} \, , \,
        [{\tilde{J}^{\prime -}}{}_{\lambda} \sigma^{\pm \scriptpp}]
        = \mp \sigma^{\pm \scriptmm}\, .
\end{eqnarray*}
The non-zero $\lambda$-brackets for the pairs $(\tilde{G} ,
\sigma)$ and $(\tilde{G} , \xi)$ are as follows:
\begin{eqnarray*}
     [{\tilde{G}^{\,\scriptpp \pm}}{}_{\lambda} \sigma^{\scriptmm \mp}] &=&
         \frac{a}{2(a+1)}(\tilde{J}^0 \mp \tilde{J}^{\prime 0})
         + \left( \tfrac{a}{2}\right)^{1/2}\xi \, , \,
%         [{\tilde{G}^{++}}{}_{\lambda}\sigma^{+-}]=
%         -\frac{1}{a+1} \tilde{J}^+ \, ,
\\[1ex]
{}   [{\tilde{G}^{\,\scriptpp\pm}}{}_{\lambda} \sigma^{\scriptmm \pm}] &=&
         \frac{a}{a+1} \tilde{J}^{\prime \pm} \, , \,
         [{\tilde{G}^{\scriptmm \mp}}{}_{\lambda}\sigma^{\scriptpp\mp}]
         = \pm \frac{1}{2(a+1)}
         (\tilde{J}^{\prime 0} \mp \tilde{J}^0)+
         \left( \frac{1}{2a}\right)^{1/2} \xi \, , \\[1ex]
{}   [{\tilde{G}^{\scriptmm \mp}}{}_{\lambda} \sigma^{\scriptpp \mp}] &=&
         -\frac{1}{a+1} \tilde{J}^{\prime \mp} \, , \,
         [{\tilde{G}^{\scriptmm \mp}}{}_{\lambda}\sigma^{\scriptmm\pm}]
         = \pm \frac{a}{a+1}\tilde{J}^- \, , \,
         [{\tilde{G}^{\scriptpp \mp}}{}_{\lambda} \sigma^{\scriptpp \pm}]
         =-\frac{1}{a+1} \tilde{J}^+ \, , \\[1ex]
{}   [{\tilde{G}^{\scriptpp \pm}}{}_{\lambda}\xi] &=& (\partial + \lambda)
        \left( \frac{a}{2}\right)^{1/2} \sigma^{\scriptpp \pm} \, , \,
        [{\tilde{G}^{\scriptmm \mp}}{}_{\lambda} \xi] =(\partial + \lambda)
        \left( \frac{1}{2a}\right)^{1/2}\sigma^{\scriptmm \mp} \, .
\end{eqnarray*}
Finally, the non-zero $\lambda$-brackets between the $\tilde{G}$'s
are as follows:
\begin{eqnarray*}
      [{\tilde{G}^{\pm \scriptpp}}{}_{\lambda}
          \tilde{G}^{\mp \scriptmm}] &=&
         \tilde{L} + \frac{1}{(a+1)} (\partial +2\lambda)
         (\pm \tilde{J}^0 +a\tilde{J}^{\prime 0}) +
         \frac{\lambda^2}{6}\tilde{c}\, , \\[1ex]
{}    [{\tilde{G}^{\pm\pm}}{}_{\lambda} \tilde{G}^{\mp \pm}] &=&
         \mp \frac{a}{a+1} (\partial +2\lambda)
         \tilde{J}^{\prime \pm} \, , \,
         [{\tilde{G}^{\pm\pm}}{}_{\lambda} \tilde{G}^{\pm \mp}]
         = \mp \frac{1}{a+1}(\partial +2\lambda)\tilde{J}^{\pm}\,.
\end{eqnarray*}
We have: $\omega (J^0_n)=J^0_{-n}$, $\omega
(J^{\pm}_n)=J^{\mp}_{-n}$, $\omega (J^{\prime 0}_{-n})=J^{\prime
  0}_{-n}$, $\omega (J^{\prime \pm}_n)= J^{\prime \mp}_{-n}$,
$\omega (G^{++}_n)=G^{--}_{-n}$, $\omega (G^{+-}_n)=G^{-+}_{-n}$,
$\omega (\sigma^{--}_n)=-a\sigma^{++}_{-n}$, $\omega
(\sigma^{-+}_n)=-a\sigma^{+-}_{-n}$, $\omega (\xi_n)=-\xi_{-n}$.

Let $\alpha =\alpha_1 |_{\fh^{\natural}}$ and $\alpha'=\alpha_3
|_{\fh^{\natural}}$.  Then $\Delta^{\natural}_+= \{ \alpha
,\alpha' \}$ and $\Delta'=\{ \pm \tfrac{1}{2}(\alpha + \alpha')$,
$\pm \tfrac{1}{2}(\alpha -\alpha')\}$.  The set of positive roots
$\Delta^+_W$ consists of the set of even roots:
\begin{displaymath}
  \{ (\pm \alpha ,m) \, , \, (\pm \alpha' ,m)|m \in \NN \}
  \cup \{ (\alpha ,0) \, , \, (\alpha' ,0)\} \cup
  (0,m) |m \in \NN \}\, ,
\end{displaymath}
where the multiplicity of a root from the first two subsets is
$1$ and from the third is $3$, and the set of odd roots:
\begin{displaymath}
  \{ (\pm \tfrac{1}{2}(\alpha + \alpha'),m) \, , \,
 (\pm   \tfrac{1}{2}(\alpha -\alpha'),m)|m \in \tfrac{1}{2}
 +\ZZ_+ \} \, ,
\end{displaymath}
all of multiplicity $1$.

In order to write down the determinant formula, let $\Lambda
=\tfrac{1}{2}(j\alpha +j'\alpha')$.  The highest weight $(\Lambda
,h)$ of the Verma module over $W_k (D(2,1;a),e_{-\theta})$ is
determined by the triple $(j,j',h)$, where $h=$~the lowest
eigenvalue of $L_0$ and $j$ ad $j'$ are the eigenvalues of
$J^0_0$ and $J^{\prime 0}_0$ on the highest weight vector.  We
have:
\begin{eqnarray*}
  h_{n,m} (k,j,j') &=& \tfrac{1}{4k}((mk-n)^2-(k+1)^2
     -\tfrac{(j+1)^2 +a(j'+1)^2}{a+1}+1) \, , \\[1ex]
  \varphi_{n,m,\pm \alpha} (k,j,j') &=& \tfrac{1}{a+1}
     ((a+1)mk +n \mp (j+1))\, , \\[1ex]
  \varphi_{n,m,\pm \alpha'} (k,j,j') &=& \tfrac{1}{a+1}
     ((a+1)mk +an \mp a(j'+1)) \, , \\[1ex]
  h_{m,\pm\tfrac{1}{2}(\alpha + \alpha')} (k,j,j') &=&
     \tfrac{-a(j-j')^2}{4(a+1)^2k} +(m^2-\tfrac{1}{4})
     k \mp \tfrac{m(j+1+a(j'+1))}{a+1}-\tfrac{1}{2}\, , \\[1ex]
  h_{m,\pm \tfrac{1}{2}(\alpha -\alpha')} (k,j,j') &=&
     \tfrac{-a(j+j'+2)^2}{4(a+1)^2k}+(m^2-\tfrac{1}{4})k \mp
     \tfrac{m(j+1-a(j'+1))}{a+1}-\tfrac{1}{2}\, .
\end{eqnarray*}
Hence Remark~\ref{rem:7.2} gives the following determinant
formula for $W_k (D(2,1;a) ,\, e_{-\theta})$:
\begin{eqnarray*}
  && {\rm det}_{\hat{\eta}} (k,h,j,j') =
       k^{2\sum_{m,n\in \NN} P_W (\hat{\eta}-(0,mn))}
 \prod_{m,n \in \NN} (h-h_{n,m}
       (k,j,j'))^{P_W (\hat{\eta}-(0,mn))}\\[1ex]
  && \times \prod_{\substack{m,n \in \NN \\ \beta = \alpha,\alpha'}}
       \varphi_{n,m,-\beta} (k,j,j')^{P_W(\hat{\eta}-n(-\beta,m))}
       \varphi_{n,m-1,\beta}(k,j,j')^{P_W (\hat{\eta}-n (\beta ,m-1))}\\[1ex]
  && \times \prod_{\substack{m\in \tfrac{1}{2}+\ZZ_+ \\
            \gamma =\pm \tfrac{1}{2}(\alpha +\alpha'),
          \pm \tfrac{1}{2}(\alpha -\alpha')}}
      (h-h_{m,\gamma}(k,j,j'))^{P_{W,(\gamma ,m)}
      (\hat{\eta}-(\gamma ,m))}\, .
\end{eqnarray*}

The determinant formula for the big $N=4$ superconformal algebra
is obtained by a simple modification of the above formula.  First
the multiplicities of the following roots increase by~$1$:
$(0,m)$ (due to the added free boson) and $(\pm
\tfrac{1}{2}(\alpha +\alpha'),m)$, $(\pm \tfrac{1}{2}(\alpha
-\alpha'),m)$ (due to the added free fermions).  This leads to
the obvious changes of exponents.  Second, one should add the
obvious power of the eigenvalue of the zero's mode of the added
free boson $\xi$.

\end{document}